\documentclass{raa}

\usepackage{graphicx, times}
\usepackage{natbib}
\usepackage{amssymb,amsmath}
\usepackage{bm}

\begin{document}
\title{Made-to-measure galaxy modelling utilising absorption line strength data}

\setcounter{page}{1}          

\author{R. J. Long
      \inst{1,2}}

\institute{National Astronomical Observatories, Chinese Academy of Sciences, A 20 Datun Rd, Chaoyang District, Beijing 100012, China; {\it rjl@nao.cas.cn}\\
        \and
           Jodrell Bank Centre for Astrophysics, School of Physics and Astronomy, The University of Manchester, Oxford Road, Manchester M13 9PL, UK\\}

\date{Received~~2016 month day; accepted~~2016~~month day}

\abstract{
We enhance the Syer \& Tremaine made-to-measure (M2M) particle method of stellar dynamical modelling to model simultaneously both kinematic data and absorption line strength data thus creating a `chemo-M2M' modelling scheme.  We apply the enhanced method to four galaxies (NGC 1248, NGC 3838, NGC 4452, NGC 4551) observed using the SAURON integral-field spectrograph as part of the ATLAS$^{\rm{3D}}$ programme.  We are able to reproduce successfully the 2D line strength data achieving mean $\chi^2$ per bin values of $\approx 1$ with $>95\%$ of particles having converged weights. Because M2M uses a 3D particle system, we are also able to examine the underlying 3D line strength distributions.  The extent to which these distributions are plausible representations of real galaxies requires further consideration.  Overall we consider the modelling exercise to be a promising first step in developing a `chemo-M2M' modelling system and in understanding some of the issues to be addressed.  Whilst the made-to-measure techniques developed have been applied to absorption line strength data, they are in fact general and may be of value in modelling other aspects of galaxies.
}

\authorrunning{R. J. Long}            
\titlerunning{M2M galaxy modelling with line strength data}  

\maketitle


\section{Introduction}\label{sec:introduction}
The creation and evolution of galaxies is a significant research topic involving, for example, an understanding of galaxy mergers and in situ star formation.  A galaxy's construction and evolution are imprinted in its kinematics and chemistry but require significant analysis to identify the contributing componentry.  IFU-based galaxy surveys, providing spatially dense spectral data cubes from which complementary measurements of kinematics and spectral line strength can be obtained, offer major opportunities to investigate the underlying componentry and its assembly, provided appropriate techniques can be developed.  Included  are surveys such as ATLAS$^{\rm{3D}}$ \citep{AtlasI}, SAMI \citep{SAMI2015} and MaNGA \citep{Bundy2015}.  

The made-to-measure (M2M) method proposed by \citet{Syer1996} for modelling stellar dynamical systems has been used on a variety of research projects by a number of different research groups (for example, \citealt{DL2007}, \citealt{Morganti2012}, \citealt{Portail2015};  \citealt{Dehnen2009}; \citealt{Long2010, Long2012}, \citealt{Zhu2014}; \citealt{Hunt2013}; \citealt{Malvido2015}).  With projects 
\begin{itemize}
\item covering elliptical and lenticular galaxies, dwarf spheroidals, and the Milky Way, 
\item using IFU data, long slit data, velocity measurements of individual stars, globular clusters and planetary nebulae, 
\item employing self-gravitation of particles (with particle weights affecting the gravitational mass of particles) and potentials from N-body simulations, multi-Gaussian expansions as well as theoretical formulae, and
\item utilising both inertial and rotating frames, 
\end{itemize}
the M2M method has demonstrated itself to be very flexible and well able to handle a wide variety of applications.  Whereas previous papers utilised the M2M method to model stellar kinematic data, the method is capable of being used to model other types of data.   The current short investigation applies the M2M method to model simultaneously spectral absorption line strength data and kinematic data.  We believe that this is the first time in which line strength data has been employed in M2M modelling.

Our objectives in performing this investigation are four-fold,
\begin{enumerate}
\item to extend the M2M method to model absorption line strength data as well as kinematic data and to create a software implementation of the revised method,
\item to apply the method to a selection of external galaxies and confirm that the usual criteria for a successful M2M run can be met (particle weight convergence and observable reproduction), 
\item to understand the limitations of the extensions and to identify areas for future work, and 
\item to promote further research by bringing the potential chemo-modelling capabilities of the M2M method to the attention of the astrophysical community.
\end{enumerate}

The structure of the paper broadly follows the objectives. In Sections \ref{sec:m2m} and \ref{sec:external}, we describe the M2M method together with our enhancements, and apply the enhanced method to four galaxies taken from the ATLAS$^{\rm{3D}}$ survey. In Sections \ref{sec:discuss} and \ref{sec:conclusions}, we discuss our results and draw conclusions identifying areas for further examination.

\section{The M2M Method}\label{sec:m2m}
Our description of the M2M method is based on \citet{Long2012}.  In Sections \ref{sec:theory} and \ref{sec:enhtheory}, we do not describe the totality of the method but only sufficient so that it is clear how the line strength extensions have been incorporated.

In brief, the M2M method is concerned with modelling stellar systems with test particles orbiting in a gravitational potential.  Weights are associated with the particles and their values adjusted as the particles are orbited so that, by using these weights, observational measurements of a real system are reproduced, with any measurement errors being taken into account.  The weights themselves are expected to converge individually constant values.  For modelling observed kinematics (Section \ref{sec:theory}), the particle weights are related to the luminosity of the stellar system.  For modelling line strength data (Section \ref{sec:enhtheory}), additional particle weights are introduced with one additional weight being required for each absorption line to be modelled.

\subsection{Basic Theory - Luminosity Weights}\label{sec:theory}
For a system of $N$ particles, orbiting in a gravitational potential, with luminosity weights $w_i$, the key equation which leads to the luminosity weight adaption equations is
\begin{equation}
	F_{\rm{lm}}(\bm{w}) = -\frac{1}{2} \chi_{\rm{lm}} ^2 + \mu_{\rm{lm}} S_{\rm{lm}} + \frac{1}{\epsilon_{\rm{lm}}}\frac{d}{dt}S_{\rm{lm}}
\label{eqn:Flm}
\end{equation}
where $\chi_{\rm{lm}} ^2$ and $S_{\rm{lm}}$ are all functions of the particle weights $\bm{w} = (w_1, \cdot \cdot \cdot , w_N)$; $t$ is time; and $\mu_{\rm{lm}}$ and $\epsilon_{\rm{lm}}$ are positive parameters. 
 
The $\chi_{\rm{lm}} ^2$ term arises from assuming that the probability of the model reproducing a single observation can be represented by a Gaussian distribution and then constructing a log likelihood function covering all observations. For $K$ different observables, we take $\chi_{\rm{lm}} ^2$ in the form
\begin{equation}
	\chi_{\rm{lm}} ^2 = \sum _k ^K \lambda _k \chi _k ^2
\label{eqn:chilambda}
\end{equation}
where $\lambda _k$ are small, positive parameters whose role in balancing numerically the weight adaption equation is explained in \citet{Long2012}. As pointed out in \citet{Long2013}, $\chi_{\rm{lm}} ^2$, being a linear combination of $\chi ^2$ functions, is not itself a $\chi ^2$ function.  The individual $\chi ^2 _k$ are defined by
\begin{equation}
	\chi ^2 _k = \sum _j ^{J_k} \Delta _{k,j} ^2
\end{equation}
and
\begin{equation}
	\Delta _{k,j} = \frac{y_{k,j}(\bm{w}) - Y_{k,j}}{\sigma_{k,j}}
\end{equation}
where $Y_{k,j}$ is the measured value of observable $k$ at position $j$ with error $\sigma_{k,j}$. The term $y_{k,j}(\bm{w})$ is the model equivalent of $Y_{k,j}$ and is given by
\begin{equation}
	y_{k,j}(\bm{w}) = \sum _i ^N w_i K_{k,j}(\bm{r}_i, \bm{v}_i) \delta(i \in k,j)
	\label{eqn:modelobs}
\end{equation}
where $K_{k,j}(\bm{r}_i, \bm{v}_i)$ is the kernel for observable $k$ evaluated at position $j$ for a particle with position $\bm{r}_i$ and velocity $\bm{v}_i$.  The selection function $\delta(i \in k,j)$ signifies that only particles $i$ which contribute to observable $k$ at position $j$ should be included in the calculation of $y_{k,j}$.

For the entropy function $S_{\rm{lm}}$ we employ the \citet{Morganti2012} function
\begin{equation}
	S_{\rm{lm}}(\bm{w}) = - \sum _i ^N w_i \left [ \ln (\frac{w_i}{m_i}) - 1 \right ]
\end{equation}
where $m_i$ is taken as the initial value of a particle luminosity weight (in practice, we take $m_i = 1/N$).  $S_{\rm{lm}}$ is used for regularisation purposes with the amount of regularisation being controlled by the parameter $\mu_{\rm{lm}}$. The derivative term $dS_{\rm{lm}}/dt$ acts as the constraint $dS_{\rm{lm}}/dt = 0$ and reflects that over time we require the particle weights, and thus $S_{\rm{lm}}$, to be constant.

The equations for luminosity weight adaption over time arise from maximising $F_{\rm{lm}}(\bm{w})$ with respect to the particle luminosity weights ($\partial F_{\rm{lm}} / \partial w_i = 0,\;\; \forall i$).  This gives equations of the form
\begin{equation}
	\frac{d}{dt} w_i = - \epsilon_{\rm{lm}} w_i \left [ \frac{\partial}{\partial w_i} \left ( \frac{1}{2} \chi_{\rm{lm}} ^2 \right ) - \mu_{\rm{lm}} \frac{\partial S_{\rm{lm}}}{\partial w_i} \right ].
	\label{eqn:wtevoln}
\end{equation}
The overall rate of adaption is controlled by $\epsilon_{\rm{lm}}$.  

Exponential (temporal) smoothing, as described in \citet{Syer1996} and \citet{Long2012}, is used to suppress noise as the numbers of particles contributing to the model observables vary.  This smoothing is parametrised by a small positive parameter $\alpha_{\rm{lm}}$.

\subsection{Extensions for modelling Spectral Line Strength Data}\label{sec:enhtheory}
The equations for modelling absorption line strength data are very similar to those for modelling kinematic data except that we now introduce further sets of weights, one for each absorption line being modelled.  Line strength data are linear in superposition and thus the calculation of model observables from a particle model is straightforward.  Use of such data also reduces considerably the need for any post-observational processing required before M2M modelling can take place.  As a consequence we do not attempt to model non-linear or derived quantities, such metallicity or age, directly.  

There are strong parallels between how mean line-of-sight velocity data and line strength data are modelled so we first make that comparison.
Using equation (\ref{eqn:modelobs}), model mean velocity values $\bar{v}_{j}(\bm{w})$ are calculated as a luminosity-weighted mean using 
\begin{equation}
	\bar{v}_{j}(\bm{w}) = \frac{\sum _i  w_i v_i \delta(i \in j)}{\sum _i  w_i \delta(i \in j)},
\end{equation}
where $v_i$ is the line-of-sight velocity for particle $i$.  If now we associate with each particle a line strength value $a_i$, then the model luminosity-weighted mean line-of-sight line strength values $\bar{a}_{j}(\bm{w})$ have the same form,
\begin{equation}
	\bar{a}_{j}(\bm{w}) = \frac{\sum _i  w_i a_i \delta(i \in j)}{\sum _i  w_i \delta(i \in j)},
\label{eqn:meanspec}
\end{equation}
and can be compared with the observed line strength data as part of the M2M process.  

For model velocity calculations, particle velocities change as particles are orbited with their luminosity weights changing as in equation (\ref{eqn:wtevoln}).  Particle line strength values are not linked to orbiting in the same intrinsic way that particle velocities are, so we need an alternative modification mechanism.  It is at this point that the similarity with velocity calculations reduces.  We could choose to use constant particle line strength values and rely on adjusting the luminosity weights, but this would require significant advance knowledge of the final solution.  We thus need a means of adjusting the particle line strength values such that the model mean line-of-sight values match the observed data values.  If we treat the particle line strength values as M2M particle weights then the weight adaption theory in the previous section can be applied to the particle line strength weights and we arrive at a line strength equivalent to equation (\ref{eqn:wtevoln}).  We can now interpret equation (\ref{eqn:meanspec}) not only as giving the model line strength values but also as a luminosity-weighted mean of line strength weights.  Given that the M2M method is already able to model line-of-sight velocity data successfully, and given our line strength approach is similar, we have every confidence that we will be able to model line strength data and, as will be seen, this is indeed the case.

The spectral line equivalent to equation (\ref{eqn:Flm}) is
\begin{equation}
	F_{\rm{sp}}(\bm{a}) = -\frac{1}{2} \chi_{\rm{sp}} ^2 + \frac{1}{\epsilon_{\rm{sp}}}\frac{d}{dt}S_{\rm{sp}}
\label{eqn:Fsp}
\end{equation}
where $\chi _{\rm{sp}} ^2$ and $S_{\rm{sp}}$ are functions of the particle line strength weights $\bm{a} = (a_1, \cdot \cdot \cdot , a_N)$; $t$ is time; and $\epsilon_{\rm{sp}}$ is a positive parameter. The regularisation term in equation (\ref{eqn:Flm}) is omitted in equation (\ref{eqn:Fsp}).  Our experiences to date have not indicated that regularisation is required but it could be included should the need arise.  The rationale leading to $\chi _{\rm{sp}} ^2$ is the same as for $\chi _{\rm{lm}} ^2$ in the previous section. Similarly, the function $S_{\rm{sp}}$ has the same form as $S_{\rm{lm}}$ but is defined replacing the luminosity weight terms by the spectral line weight equivalents.

Spectral line weight adaption comes from maximising $F_{\rm{sp}}(\bm{a})$ with respect to the particle line strength weights ($\partial F_{\rm{sp}} / \partial a_i = 0,\;\; \forall i$) leading to equations of the form
\begin{equation}
	\frac{d}{dt} a_i = - \epsilon_{\rm{sp}} a_i \frac{\partial}{\partial a_i}\left (\frac{1}{2} \chi_{\rm{sp}} ^2 \right),
	\label{eqn:spwtevoln}
\end{equation}
where the overall rate of adaption is controlled by $\epsilon_{\rm{sp}}$. There is one of these equations for every spectral absorption line being modelled.  Comparing equation (\ref{eqn:spwtevoln}) with equation (\ref{eqn:meanspec}), equation (\ref{eqn:spwtevoln}) is concerned with the adaption of individual particle weights whereas equation (\ref{eqn:meanspec}) shows how to calculate a model observable from the particle weights.

Taking equations (\ref{eqn:wtevoln}) and (\ref{eqn:spwtevoln}) together, we have two sets of adaption equations with one set handling luminosity weights and the other, spectral line weights.  The two sets are linked via the luminosity weights.  We have chosen to keep $F_{\rm{lm}}$ and the $F_{\rm{sp}}$ separate.  We could have decided to include $\chi ^2 _{\rm{sp}}$ in $F_{\rm{lm}}$ (more properly, $\chi ^2 _{\rm{lm}}$).  Also, we do not attempt to take into account empirical relationships between spectral lines.  We will examine these options in a later investigation.

\section{Application to external galaxies}\label{sec:external}

A spherical toy Plummer model was used during testing to confirm that the joint weight adaption equations (equations \ref{eqn:wtevoln} and \ref{eqn:spwtevoln}) performed together correctly, with observables being reproduced and both the luminosity and spectral line weights having converged. Having achieved this, the toy model was discarded in favour of using real galaxies.

\subsection{Observables}\label{sec:observables}
Our rationale for using ATLAS$^{\rm{3D}}$ data is that all the kinematic and line strength data we require for our M2M models is publicly available from the ATLAS$^{\rm{3D}}$ website\footnote{http://purl.org/atlas3d} and is readily convertible for use in chemo-M2M models.  More formally we take data from four ATLAS$^{\rm{3D}}$ papers, \citet{AtlasI}, \citet{AtlasII}, \citet{AtlasXV}, and \cite{AtlasXX}.  In conjunction with the ATLAS$^{\rm{3D}}$ value of galaxy inclination, we use their surface brightness multi-Gaussian expansions (MGEs) to provide us with luminosity constraints and the gravitational potential for orbiting M2M particles.  For M2M purposes, a nominal $10$ percent relative error is used with the luminosity constraints calculated from the MGEs.  For our kinematic constraints we take the ATLAS$^{\rm{3D}}$ line-of-sight velocity and velocity dispersion data and error values. Similarly, we utilise the ATLAS$^{\rm{3D}}$ H$\beta$, Fe5015 and Mg$\,b$ line strength data (\citealt{AtlasXXX}).  In common with previous projects (for example, \citealt{Cappellari2006}, \citealt{Long2012}, \citealt{Morganti2012}), prior to M2M modelling, all data are symmetrised as appropriate to axisymmetric modelling. In our case, we maintain the Voronoi cells the data are associated with and use the cells in M2M modelling to bin particle data in the creation of model observables.  

We apply chemo-M2M modelling to four ATLAS$^{\rm{3D}}$ galaxies chosen for their simple kinematic properties so that we may focus on model behaviour when absorption line strength data are employed.  In more detail, we choose elliptical and S0 galaxies which have high data quality, have zero dark matter fraction, have low separation of kinematic and photometric axes (less than $5^{\circ}$), and are not believed to have a bar.  The S0 galaxies and their inclinations to the line-of-sight are NGC 1248 ($42 ^{\circ}$ inclination), NGC 3838 ($79 ^{\circ}$), and NGC 4452 ($88^{\circ}$), while the elliptical galaxy is NGC 4551 ($63 ^{\circ}$).  Inclinations have been taken from \citet{AtlasXV}.  Note that there are no face-on ATLAS$^{\rm{3D}}$ galaxies meeting our requirements.

\subsection{M2M Miscellany}\label{sec:misc}
We create axisymmetric M2M models of our selected galaxies using the multi-Gaussian surface brightness expansions (MGEs) described in the ATLAS$^{\rm{3D}}$ paper \citet{AtlasXXI}, and techniques from \citet{Emsellem1994}.  For consistency with the self-consistent JAM modelling in \citet{AtlasXV} we do not use a dark matter potential.

The initial conditions on the particles are created using a scheme similar to that described in \citet{Long2012} (a three isolating integral scheme without the model line-of-sight velocity being adjusted to match the observed data). For modelling S0 galaxies, we modify the sampling of the angular momentum integral to create more circular than radial orbits.   We achieve this by defining a `circularity measure' $C$ to give an approximate indication of how circular our individual particle orbits are as
\begin{equation}
	C = \frac{|L_z|}{|L_z(E)|} - 1
\label{eqn:circ}
\end{equation}
where $L_z$ and $E$ are a particle's angular momentum and energy.  $L_z(E)$ is the angular momentum of the circular orbit in the galaxy's equatorial plane with the same energy as the particle.  A completely circular orbit has $C=0$ while a highly radial orbit has $C=-1$. We sample logarithmically on $|C|$ and then use the value obtained to calculate $L_z$.  The initial value of the luminosity weights is set equal to $1/N$ for all particles.  The initial values for the spectral line weights are created by sampling from Gaussians with mean the measured line strength value and dispersion, the associated measurement error.

In addition to the constraints already described, we employ two further constraints.  The first is a sum of luminosity weights constraint, as used in \citet{Long2012}, which constrains $\sum _i w_i = 1$.  The second is a sum of weights constraint for the spectral line weights.  In this case, for the spatial region that we have line strength data, we require that the model sum of weights equals the sum of the measured  line strengths.  Both constraints act to prevent models being free to increase or decrease the total surface luminosity or line strength from that measured.  Parametrisation takes place through the parameters $\lambda_{\rm{lm,sum}}$ and $\lambda_{\rm{sp,sum}}$.

The units we use within the M2M models in this paper are effective radii for length, $10^7$ years for time, and mass in units of the solar mass $M_{\odot}$.  Line strength data values are in Angstrom.

In Table \ref{tab:params} we list all the parameters identified in Section \ref{sec:m2m} and the typical values we use in our M2M models.  The values vary slightly for each galaxy.  The detailed rationale for the $\lambda _k$ parameters in equation (\ref{eqn:chilambda}) is recorded in \citet{Long2010, Long2012} (numerical balancing of the luminosity weight adaption equations).  The values for the $\lambda _k$ are calculated and applied automatically by our M2M implementation during a modelling run (there is no need to set these values manually). Initially, we used a small amount of regularisation ($\mu_{\rm{lm}} = 2$) to support the luminosity and surface brightness constraints. However with mean $\chi ^2$ per bin values $< 1$ being achieved for the luminosity constraints (which means that the model gravitational potential and mass density satisfy the usual Poisson equation), regularisation was not adding any benefit and was subsequently not used ($\mu_{\rm{lm}} = 0$).

\begin{table}
	\centering
	\caption{M2M parameters}
	\label{tab:params}
	\begin{tabular}{ccc}
		\hline
		Parameter & Value & Comments \\
		\hline
		\multicolumn{2}{l}{\textbf{General}} \\
		$N$  &  $5 \times 10^5$  &  number of particles \\
		\multicolumn{2}{l}{\textbf{Luminosity weights}} \\
		$\epsilon_{\rm{lm}}$ & $1.0 \times 10^{-4}$ & adaption rate \\
		$\alpha_{\rm{lm}}$   & $5.0 \times 10^{-2}$ & exp. smoothing\\
		$\mu_{\rm{lm}}$      & $0$                & regularisation\\
		$\lambda _{\rm{lm,sum}}$ & $10^3$    & sum of weights \\
		\multicolumn{2}{l}{\textbf{Spectral line weights}} \\
		$\epsilon_{\rm{sp}}$ & $1.0 \times 10^{-4}$ & adaption rate\\ 
		$\alpha_{\rm{sp}}$   & $5.0 \times 10^{-2}$ & exp. smoothing\\
		$\lambda _{\rm{sp,sum}}$ & $10^{-4}$    & sum of weights \\		
		\hline
	\end{tabular}
	
\medskip
The M2M parameters and their typical values.  Values are galaxy-specific, and are adjusted slightly as necessary to met the specific modelling needs of each galaxy.   
\end{table}

\begin{figure}
\centering
	\includegraphics[width=75mm]{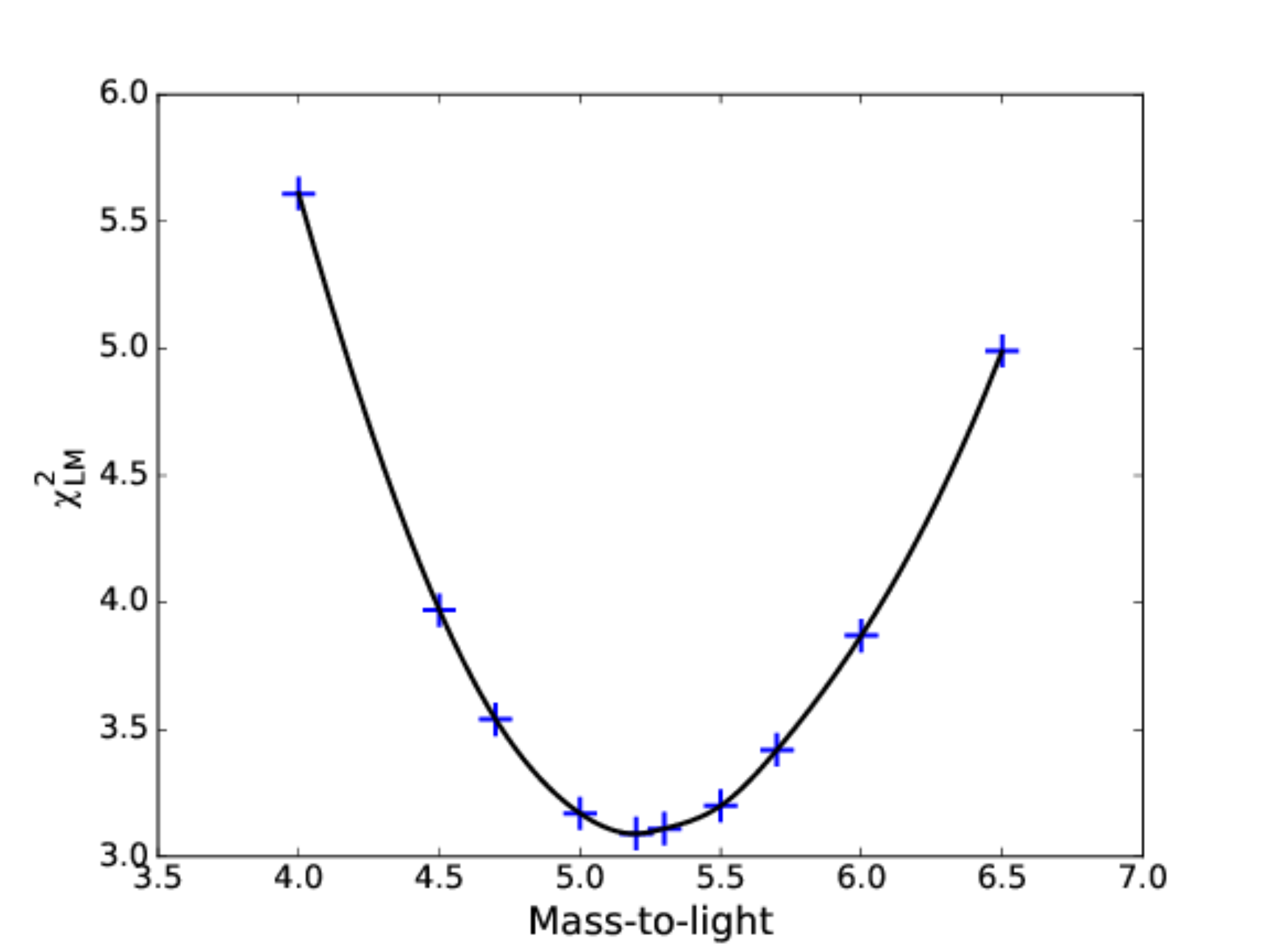}\\
	\caption{Mass-to-light ratio determination for NGC 4452.  The minimum in the $\chi_{\rm{lm}} ^2$ curve indicates the galaxy's ratio, $5.2$ in this case (see Section \ref{sec:misc}).}
\label{fig:mlratio4452}
\end{figure}

Whereas we have chosen to use the galaxy inclination angles from \citet{AtlasXV}, for the galaxy mass-to-light ratios we determine our own values using M2M modelling similar in approach to \citet{Long2012} - see Table \ref{tab:mlratios}. For a given galaxy, its mass-to-light ratio is determined by running several models at different ratios.  The galaxy's mass-to-light ratio is indicated by the minimum in the mass-to-light ratio to $\chi_{\rm{lm}} ^2$ curve.  An example for NGC 4452 is shown in  Figure \ref{fig:mlratio4452}.

\begin{table}
	\centering
	\caption{Galaxy inclinations and mass-to-light ratios}
	\label{tab:mlratios}
	\begin{tabular}{cccc}
		\hline
		Galaxy & Inclination & Mass-to-light ratio & Type \\
		\hline
		NGC 1248 & $42 ^{\circ}$ & $2.50$ & S0 \\
		NGC 3838 & $79 ^{\circ}$ & $4.00$ & S0 \\
		NGC 4452 & $88 ^{\circ}$ & $5.20$ & S0 \\
		NGC 4551 & $63 ^{\circ}$ & $4.89$ & E \\
		\hline
	\end{tabular}
	
\medskip
Galaxy inclinations and mass-to-light ratios.  The inclinations are taken from \citet{AtlasXV}, while the mass-to-light ratios are determined by M2M modelling.
\end{table}

\subsection{Results}\label{sec:results}

\begin{figure*}
\centering
\begin{tabular}{lcr}

\includegraphics[width=50mm]{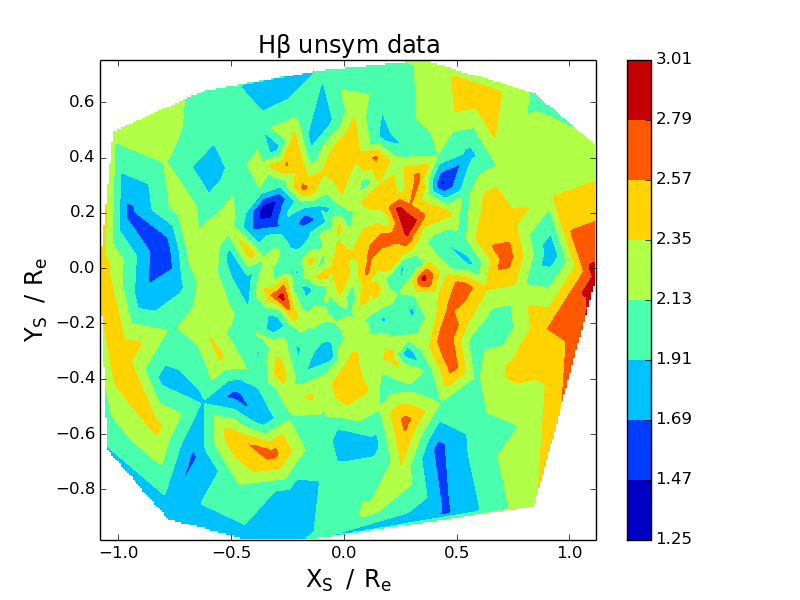} & \includegraphics[width=50mm]{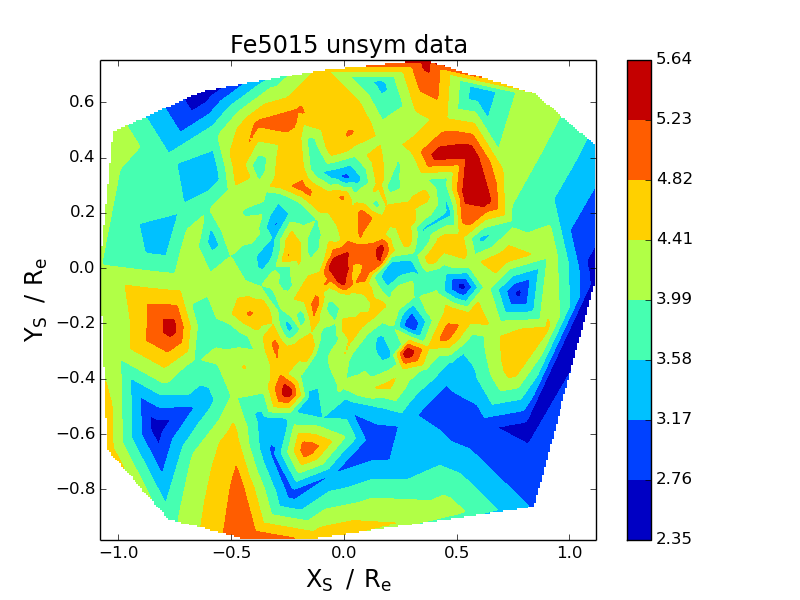}  & \includegraphics[width=50mm]{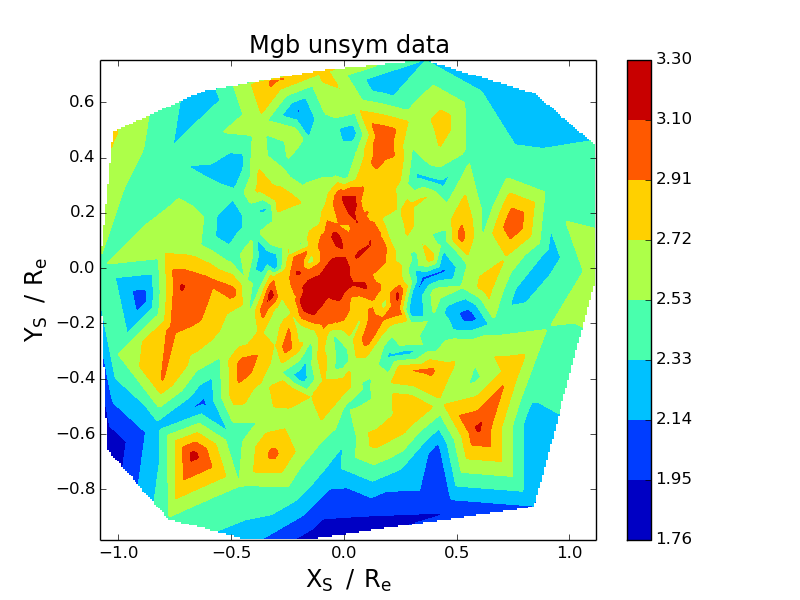}\\

\includegraphics[width=50mm]{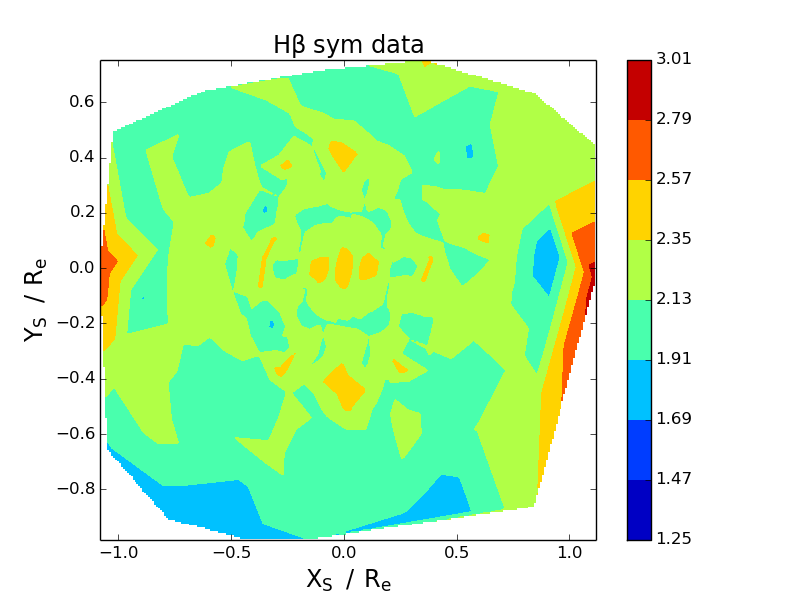} & \includegraphics[width=50mm]{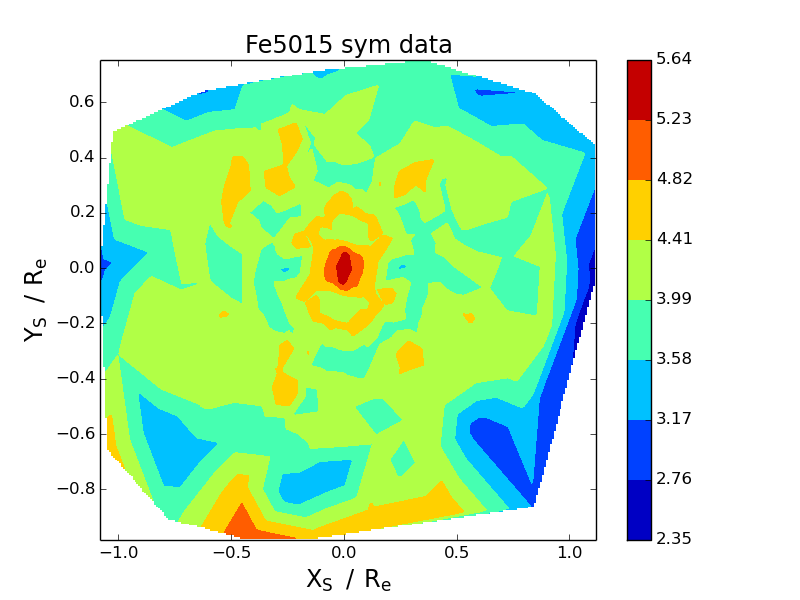}  & \includegraphics[width=50mm]{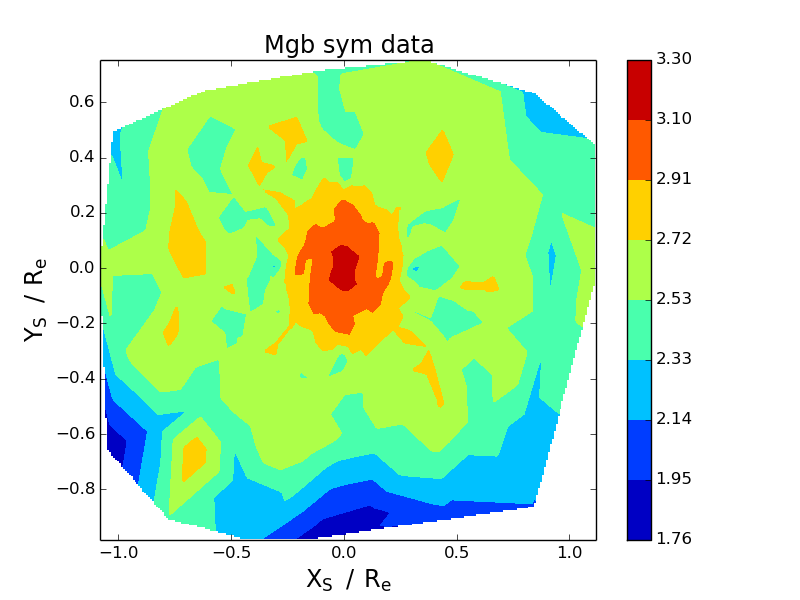}\\

\includegraphics[width=50mm]{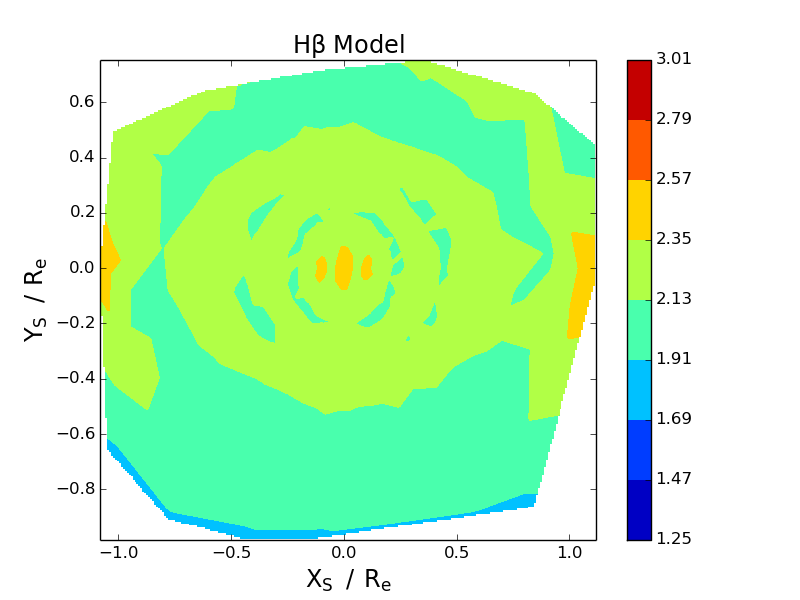} & \includegraphics[width=50mm]{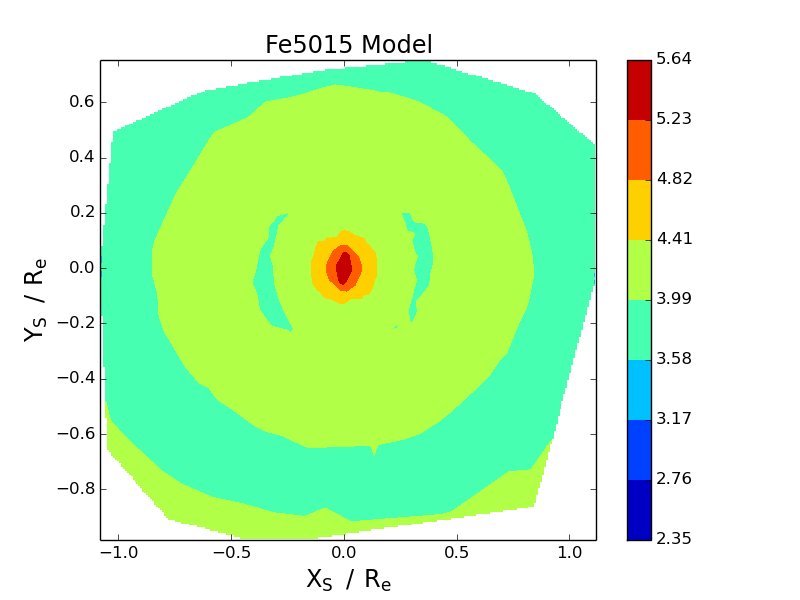}  & \includegraphics[width=50mm]{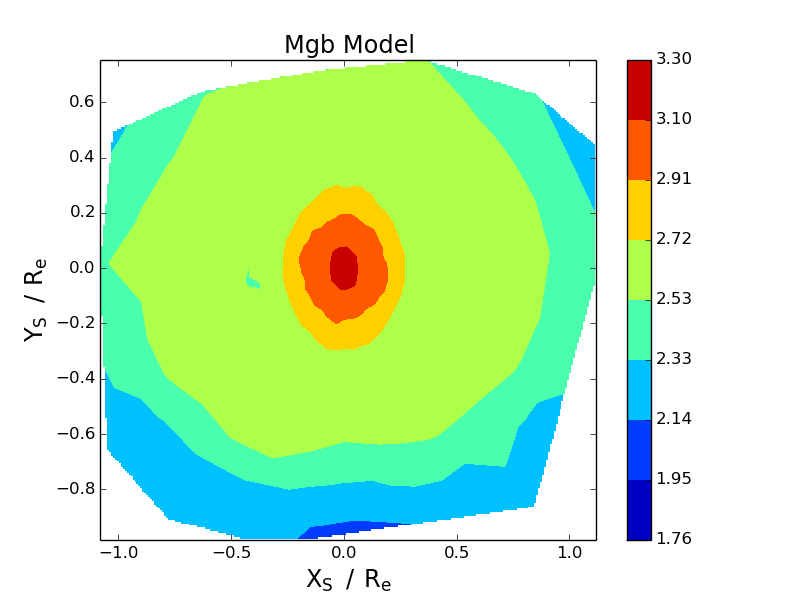}\\

\includegraphics[width=50mm]{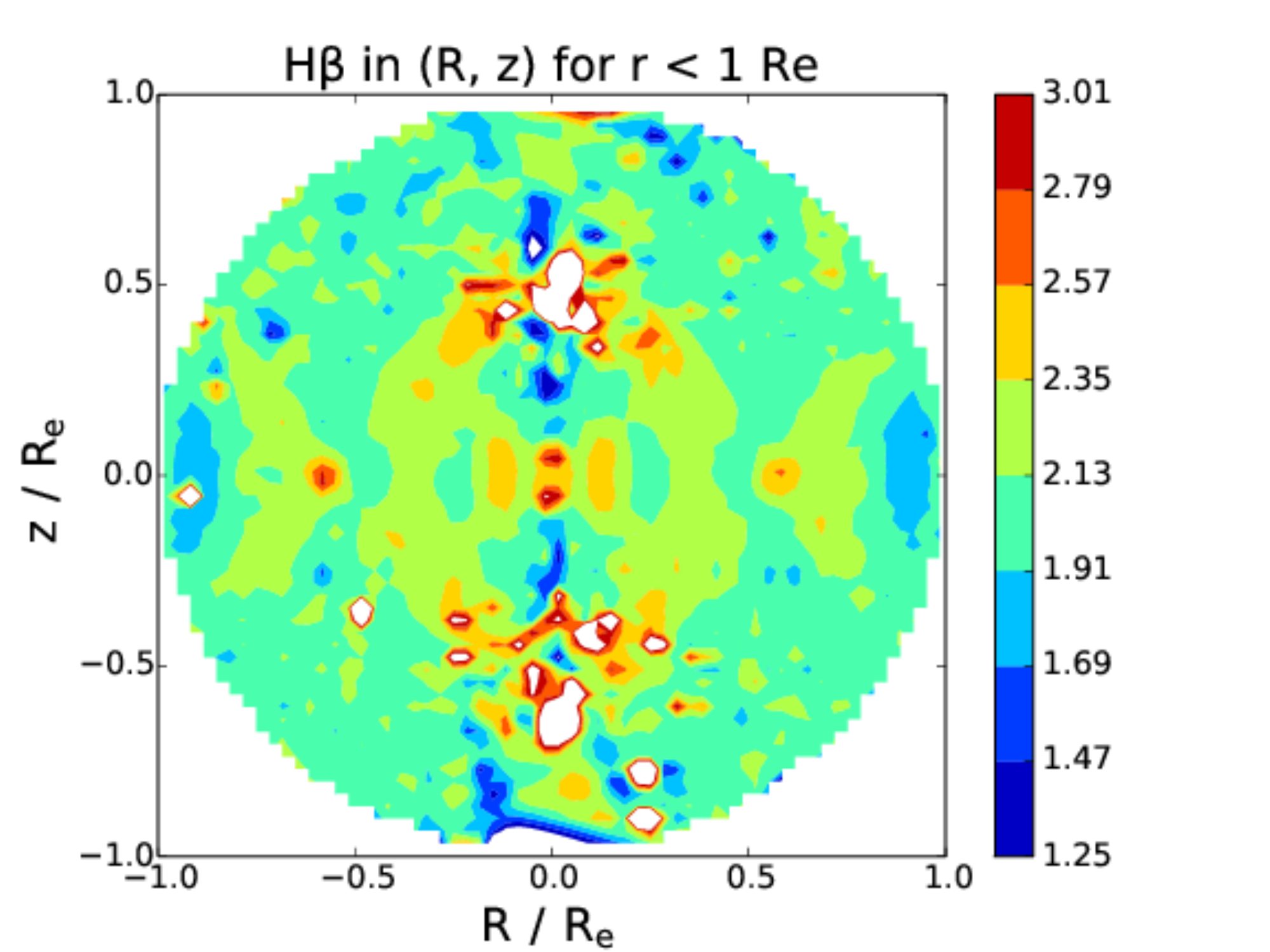} & \includegraphics[width=50mm]{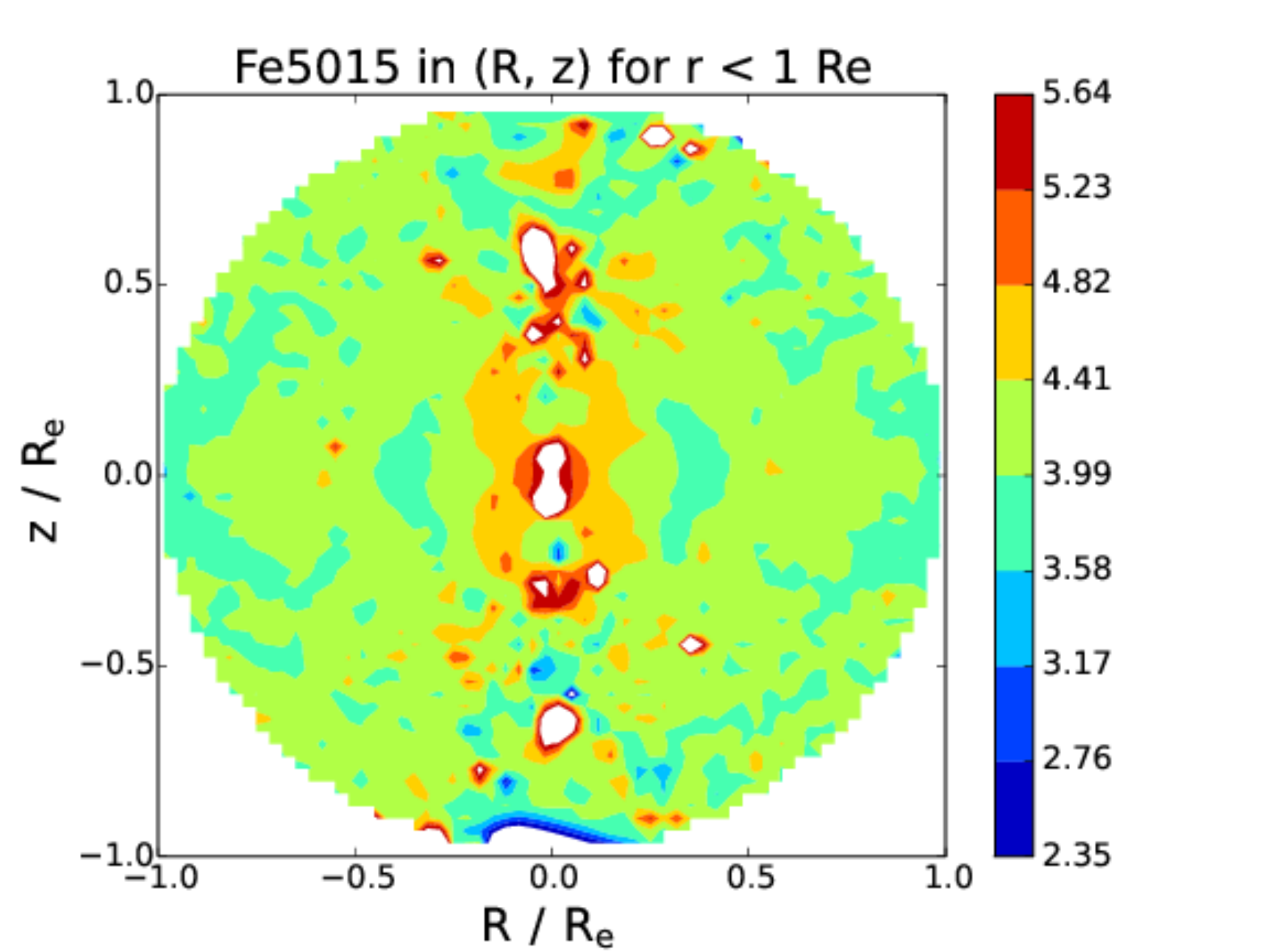}  & \includegraphics[width=50mm]{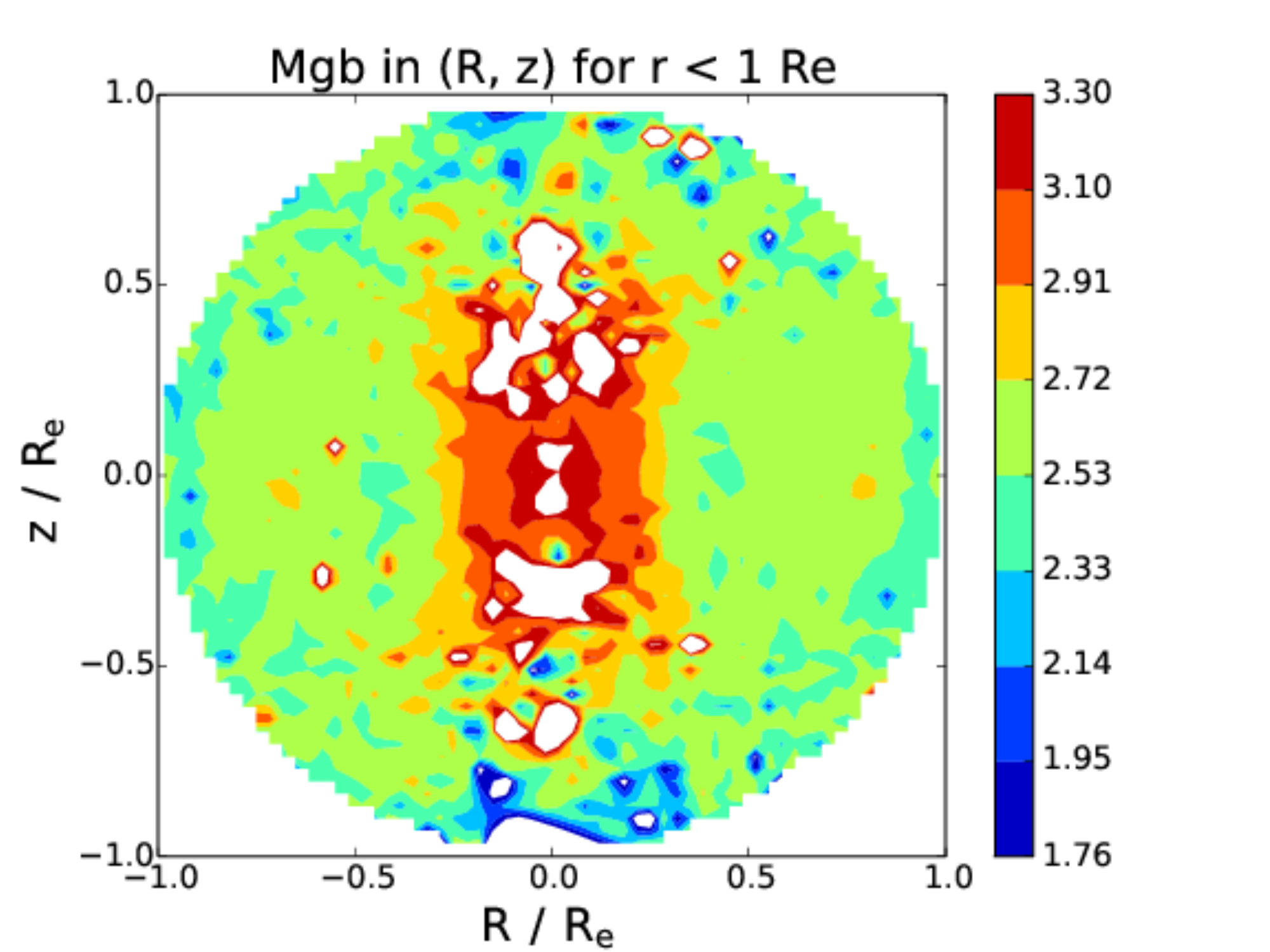}\\

\includegraphics[width=50mm]{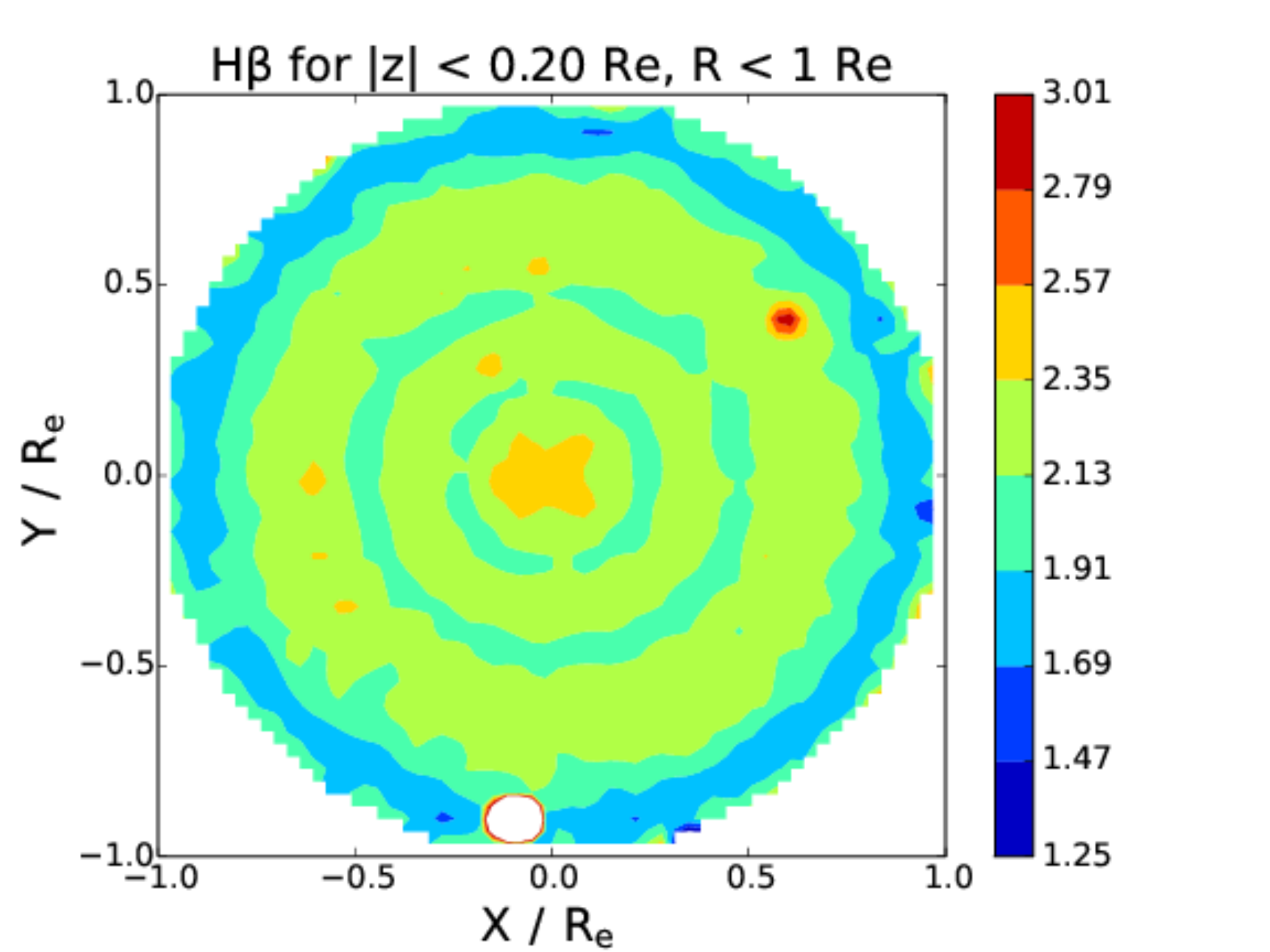} & \includegraphics[width=50mm]{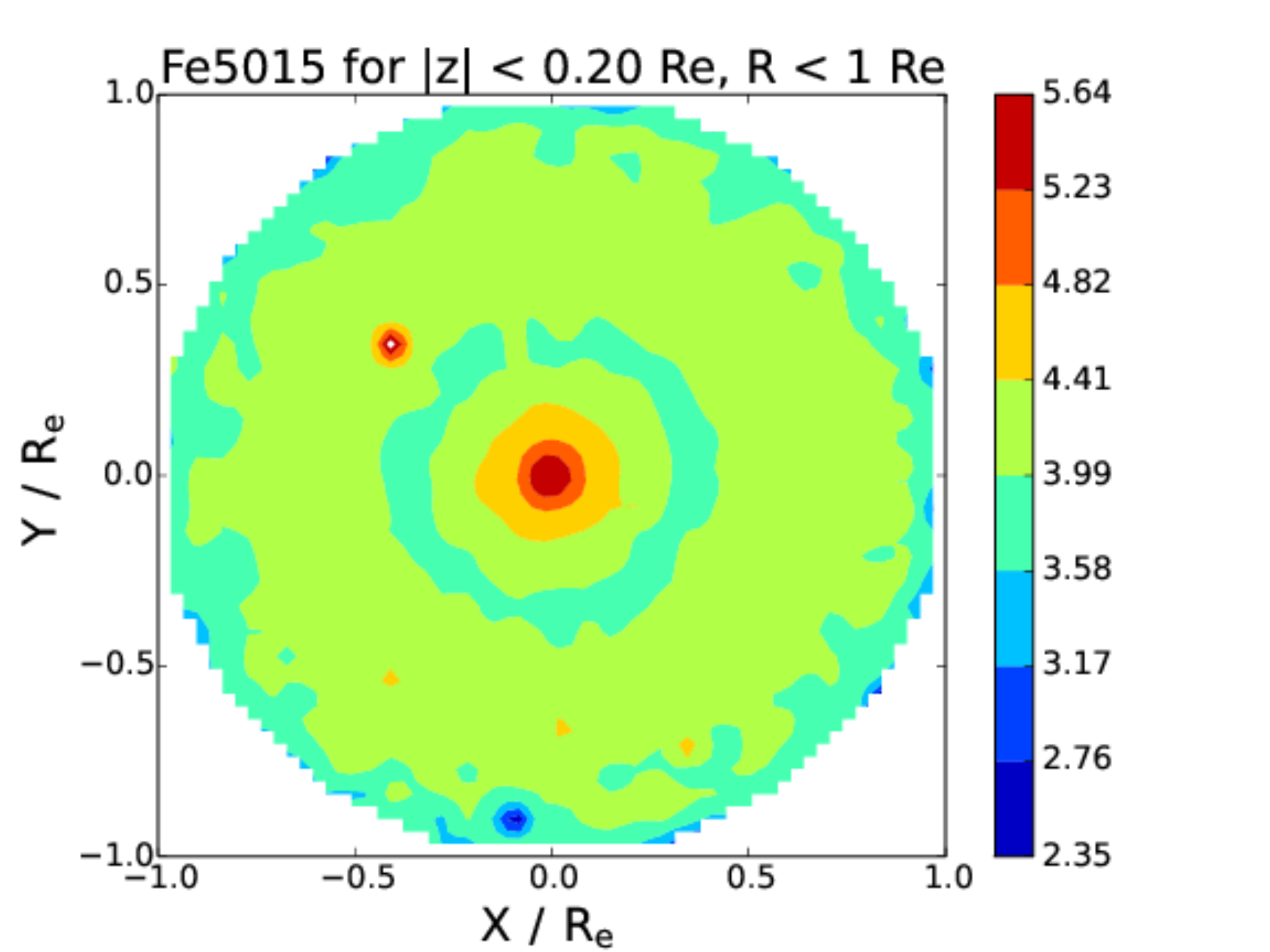}  & \includegraphics[width=50mm]{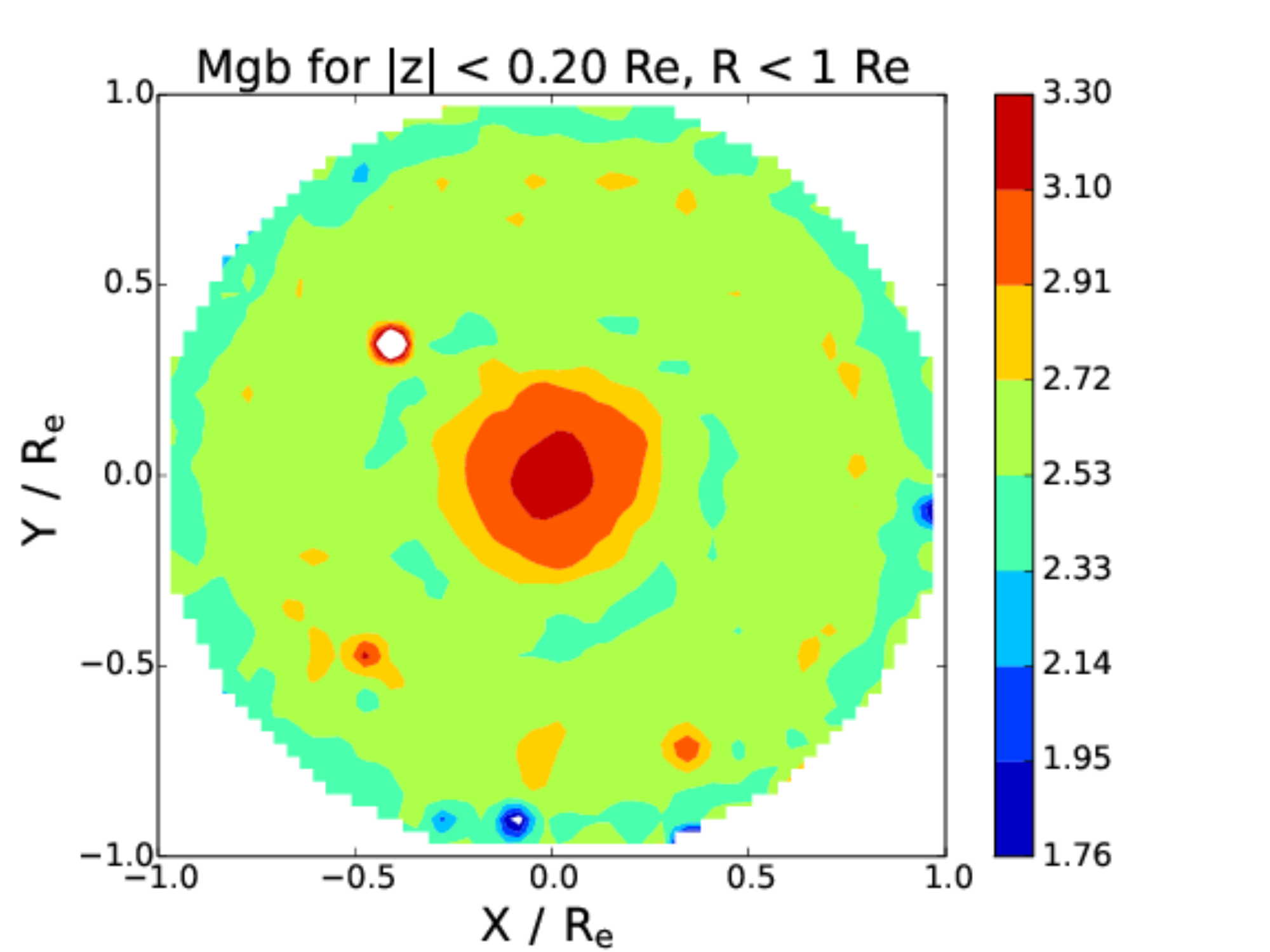}\\

\end{tabular}

\medskip
\caption{NGC 1248 line strength plots from chemo-M2M modelling.  The columns represent the absorption lines being modelled (H$\beta$, Fe5015 and Mg$\,b$).  The first row shows the released ATLAS$^{\rm{3D}}$ data, and the second and third rows show respectively the symmetrised data input to the M2M modelling process and the M2M reproduction of that data.  The fourth and fifth rows show an $(R, z)$ plot of the particle data for $r < 1 \; \rm{R_e}$ and an equatorial plot for $|z| < 0.2 \; \rm{R_e}$. Line strength units (see colour bars) are Angstrom (see Section \ref{sec:misc}).  An intensity plot of the galaxy may be found in \citet{AtlasI} (see figs. 5 and 6).}
\label{tab:NGC1248}
\end{figure*}

\begin{figure*}
\centering
\begin{tabular}{lcr}

\includegraphics[width=50mm]{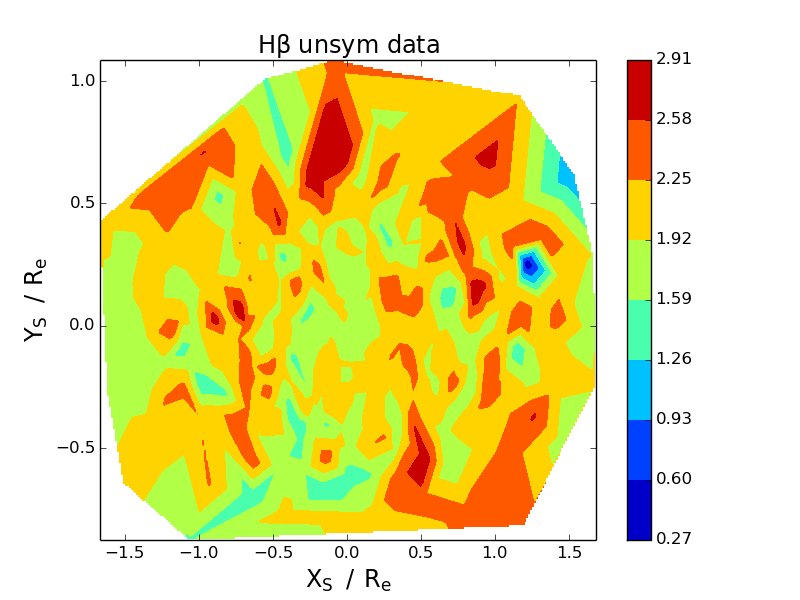} & \includegraphics[width=50mm]{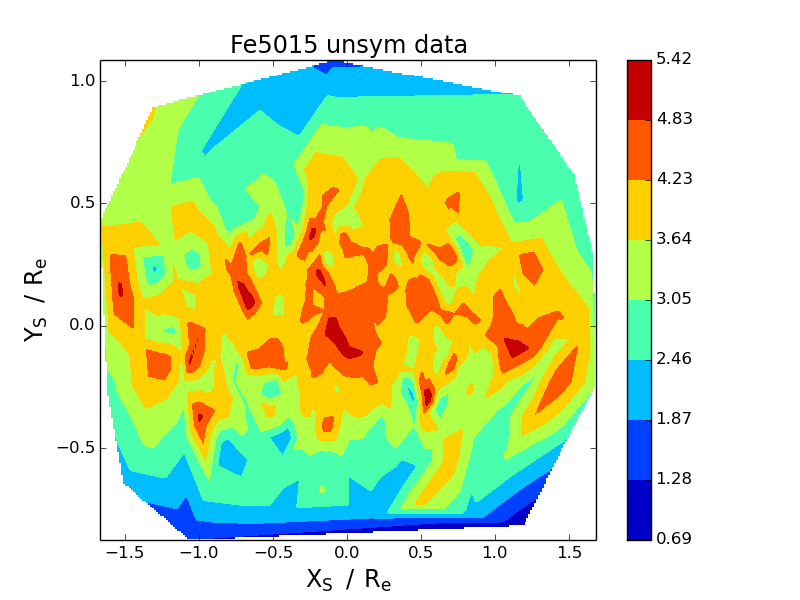}  & \includegraphics[width=50mm]{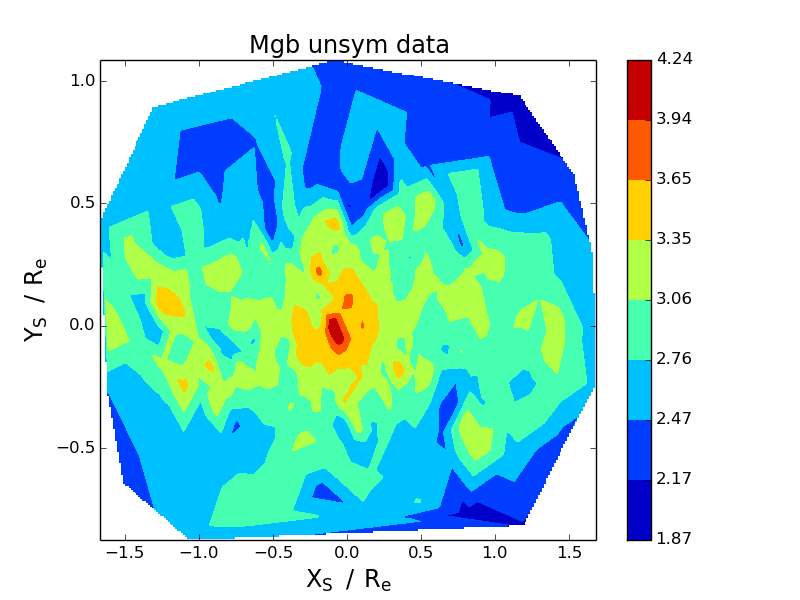}\\

\includegraphics[width=50mm]{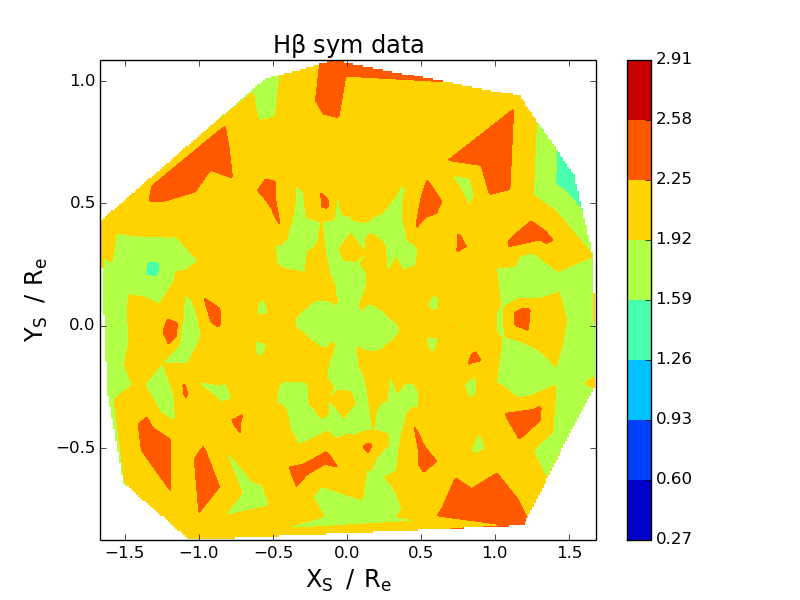} & \includegraphics[width=50mm]{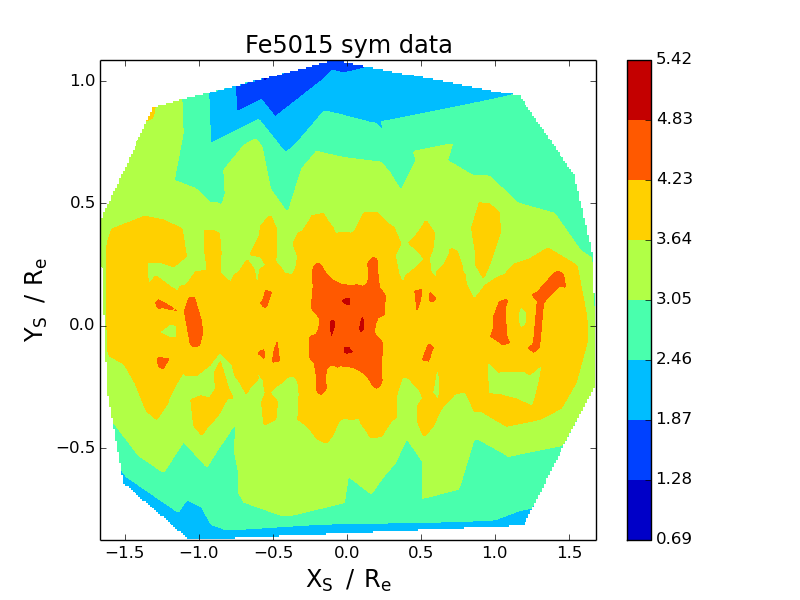}  & \includegraphics[width=50mm]{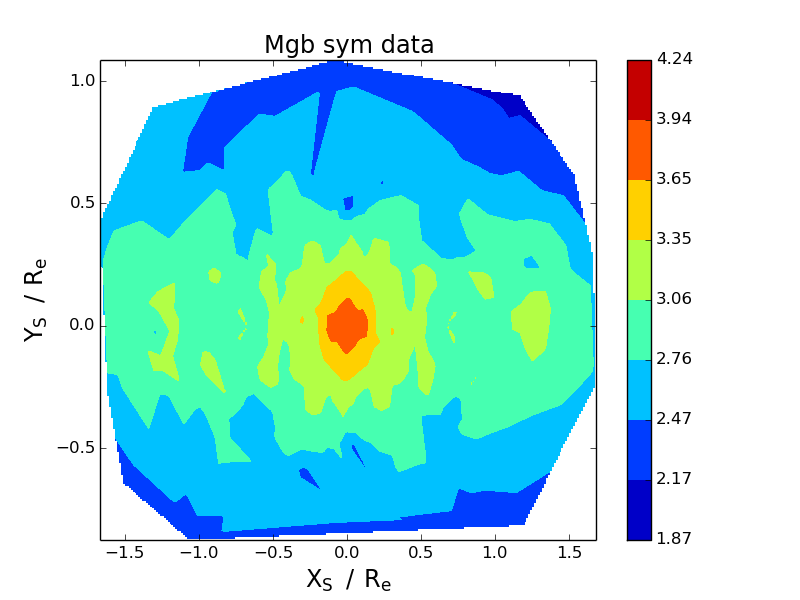}\\

\includegraphics[width=50mm]{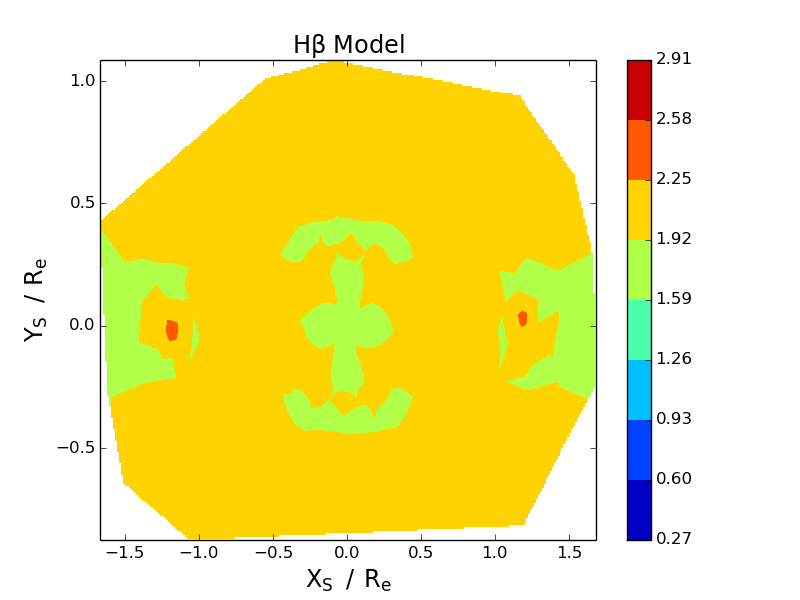} & \includegraphics[width=50mm]{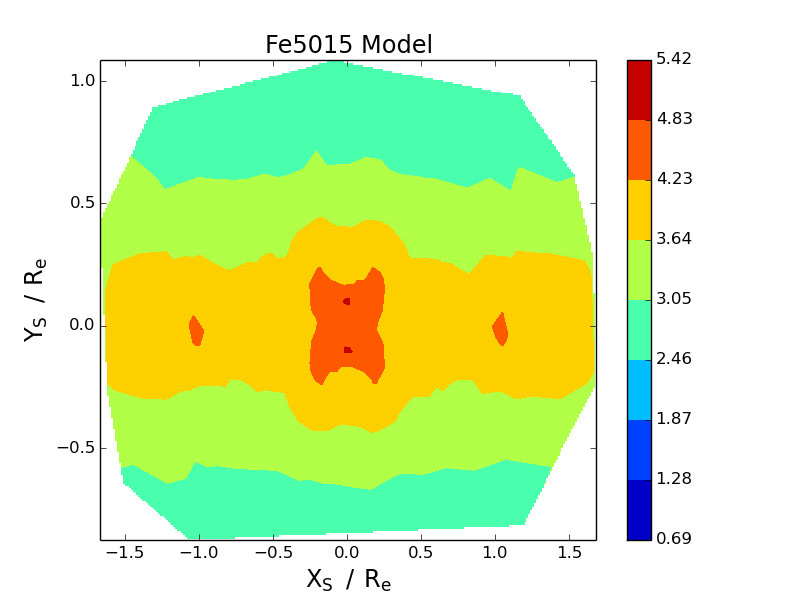}  & \includegraphics[width=50mm]{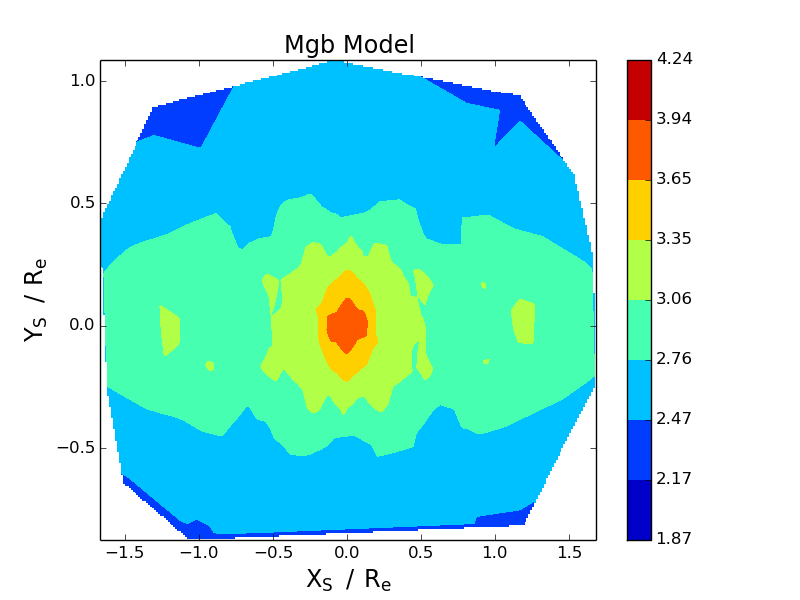}\\

\includegraphics[width=50mm]{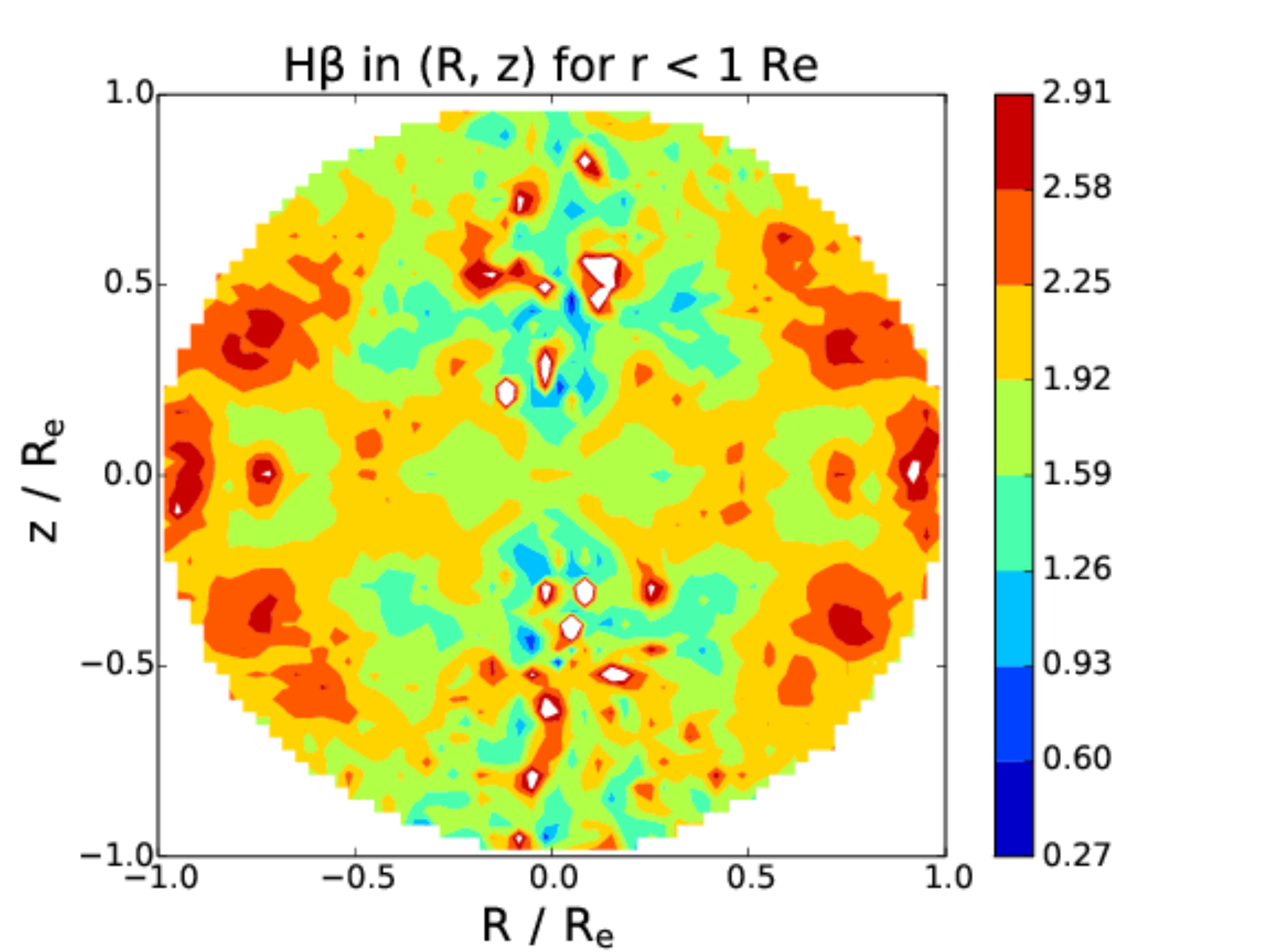} & \includegraphics[width=50mm]{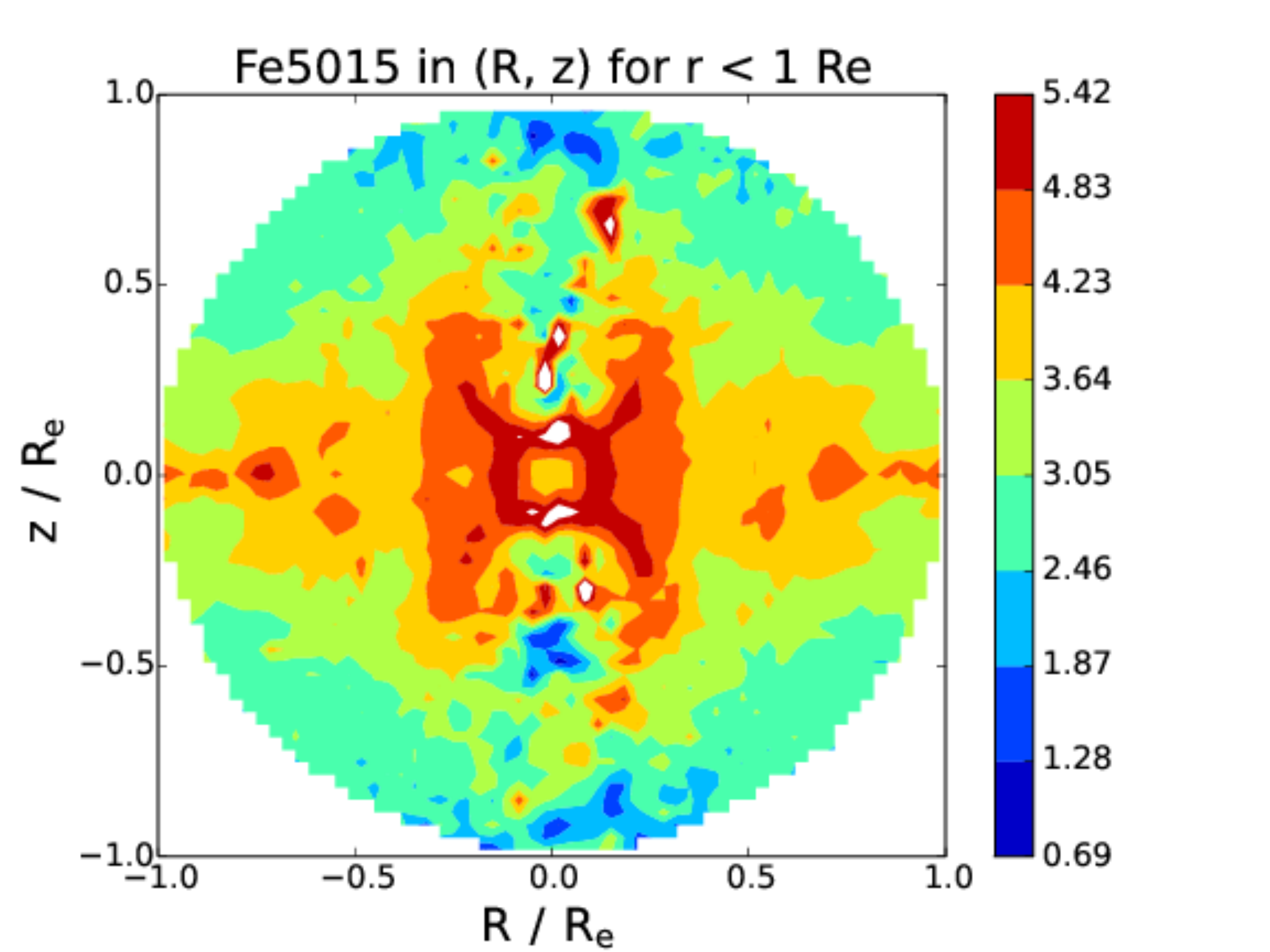}  & \includegraphics[width=50mm]{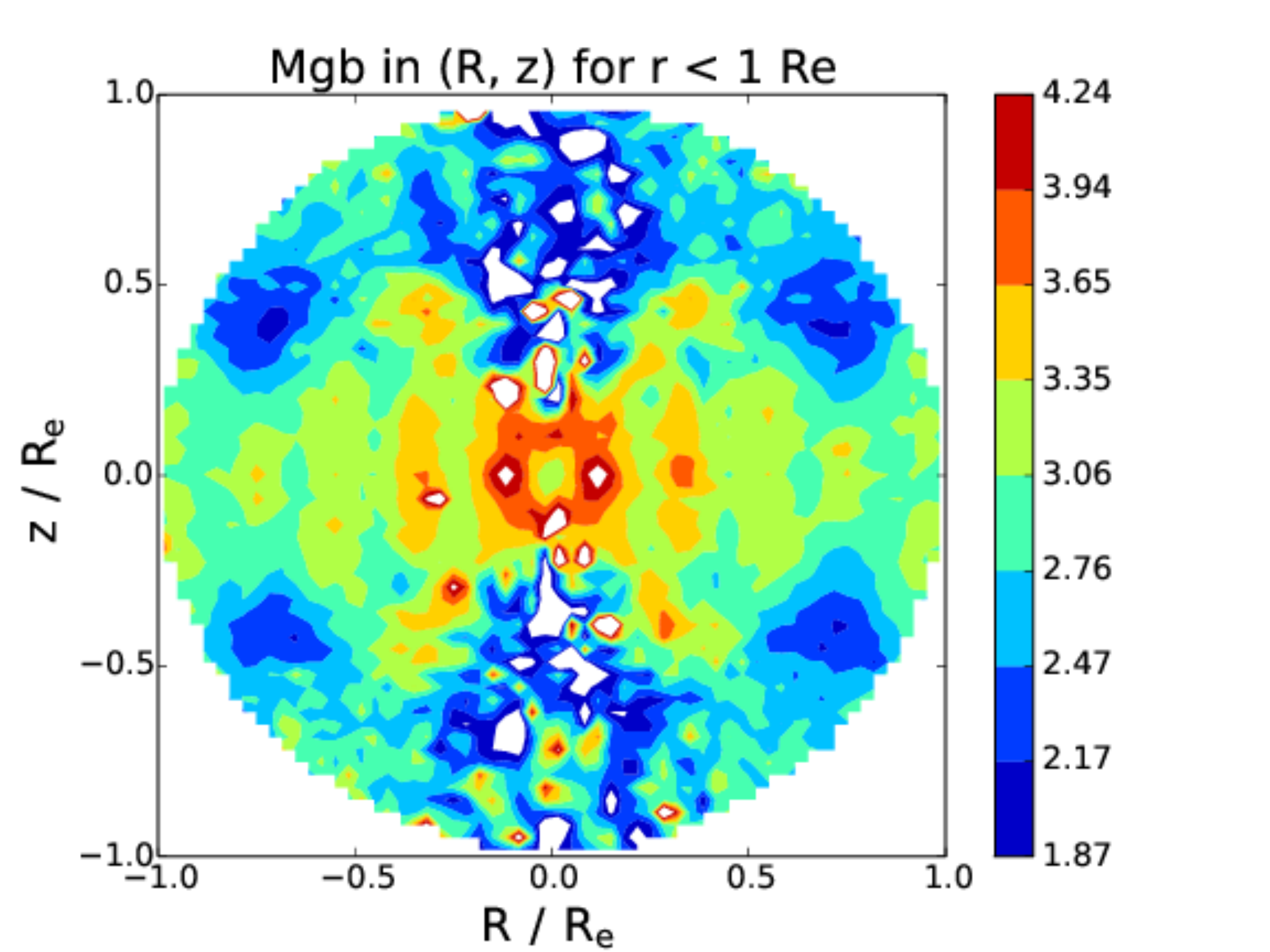}\\

\includegraphics[width=50mm]{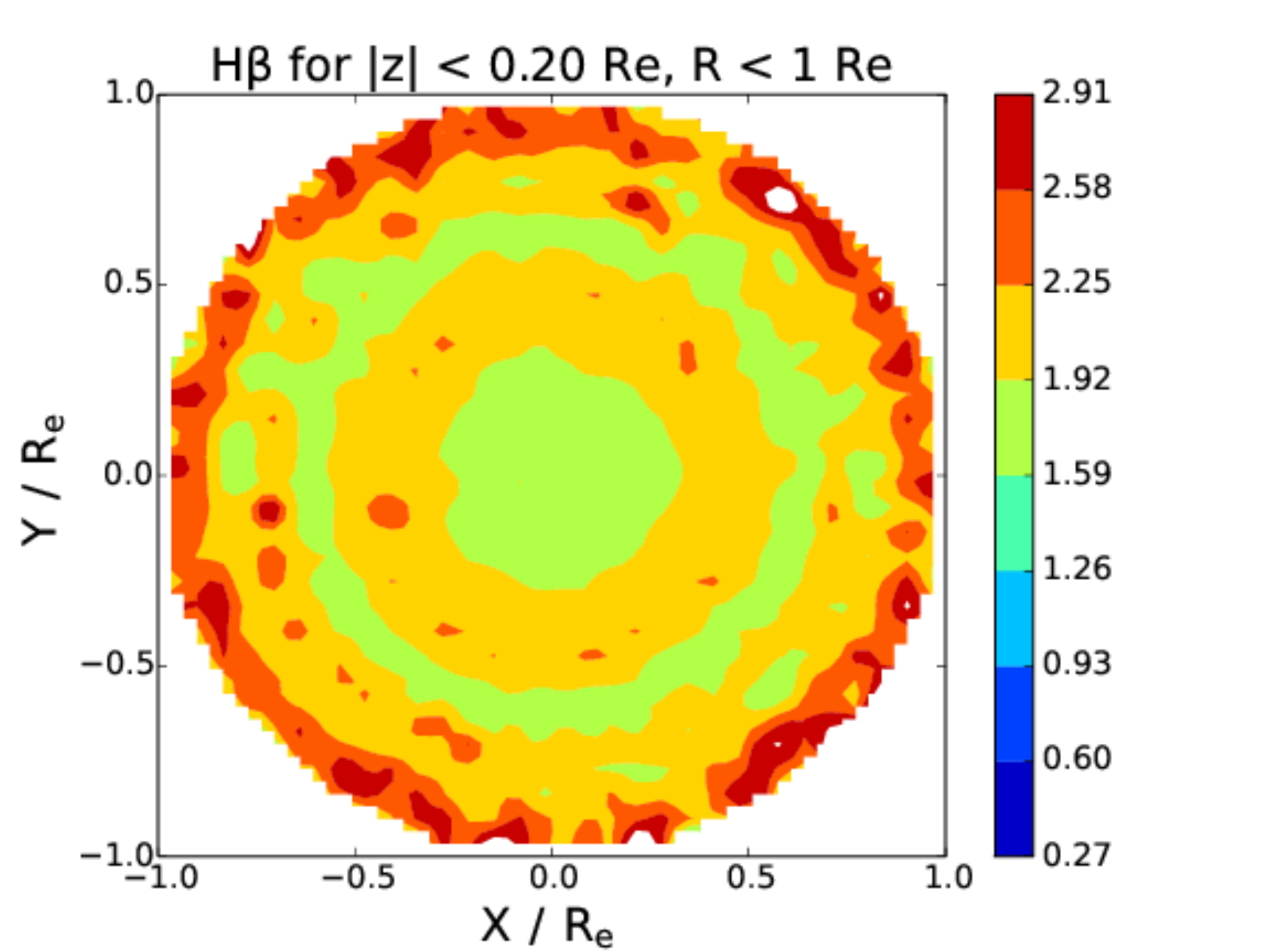} & \includegraphics[width=50mm]{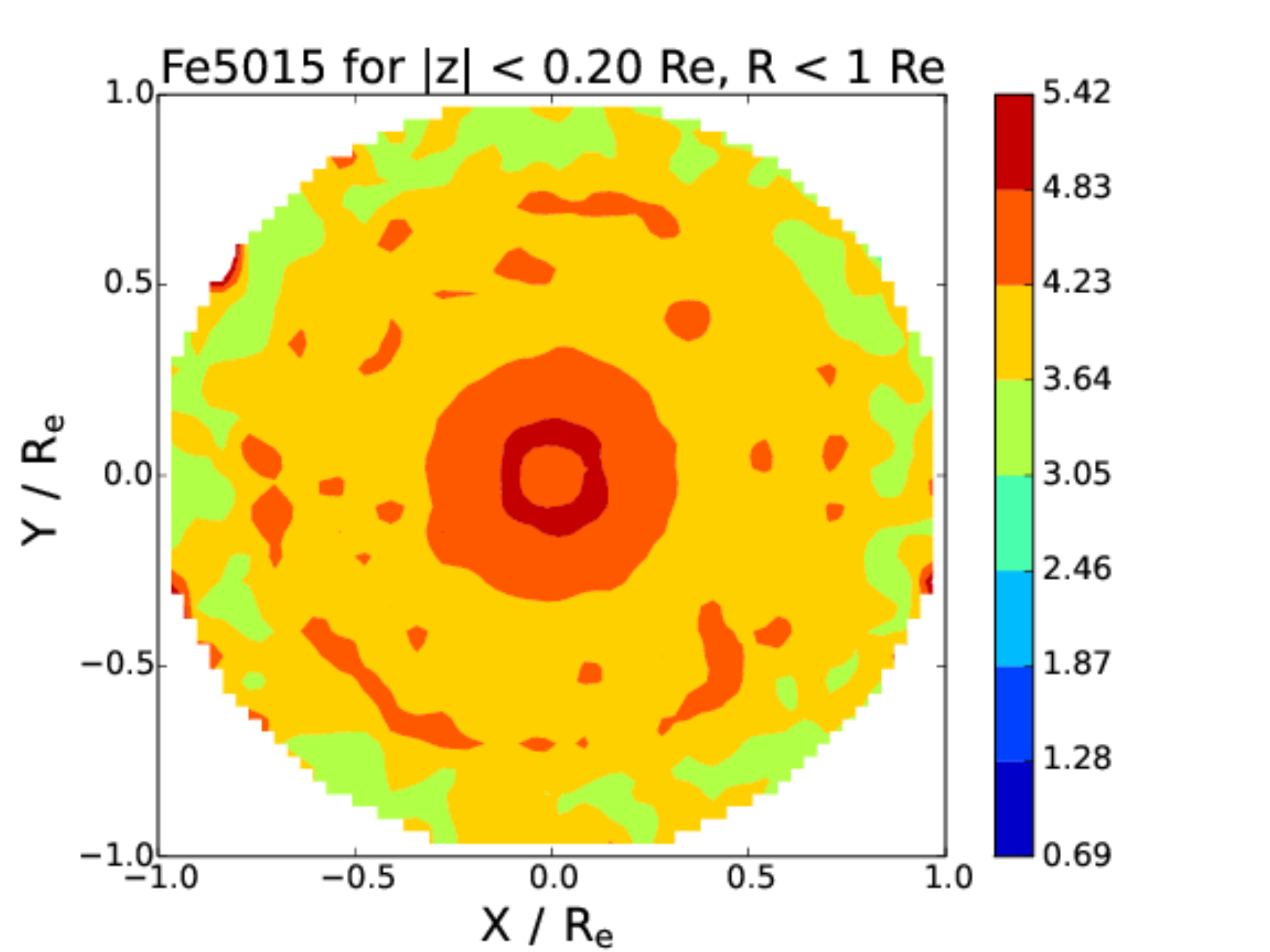}  & \includegraphics[width=50mm]{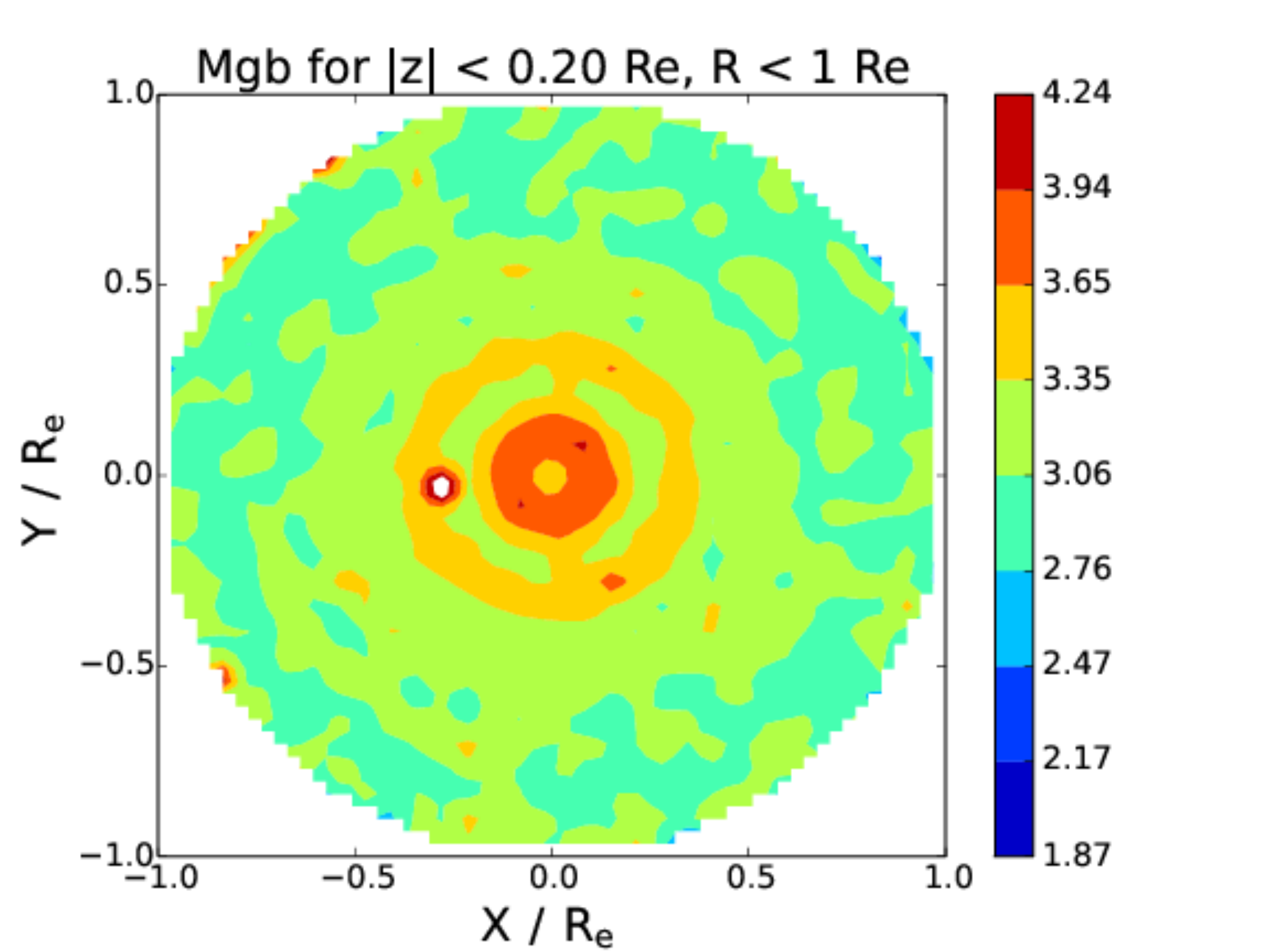}\\

\end{tabular}

\medskip
\caption{NGC 3838 line strength plots from chemo-M2M modelling.  See Figure \ref{tab:NGC1248} for a description of the rows and columns.  The difference in the H$\beta$ plotted area, by comparison with the other two spectral lines in the first three rows, is due to the removal of a single anomalously high data point.  White regions within plots indicate data values outside of the colour bar range.  For comparison purposes, the data ranges for each column have been standardised on the range for the released ATLAS$^{\rm{3D}}$ data. }
\label{tab:NGC3838}
\end{figure*}

\begin{figure*}
\centering
\begin{tabular}{lcr}

\includegraphics[width=50mm]{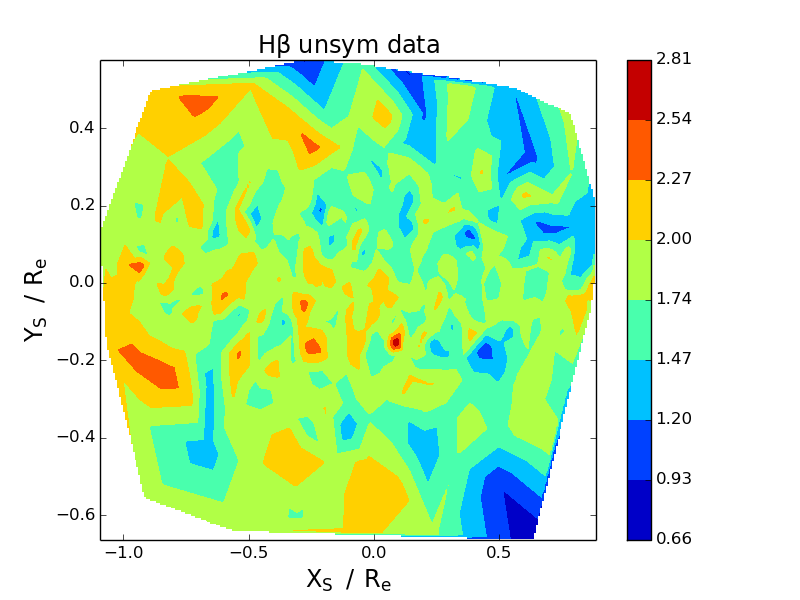} & \includegraphics[width=50mm]{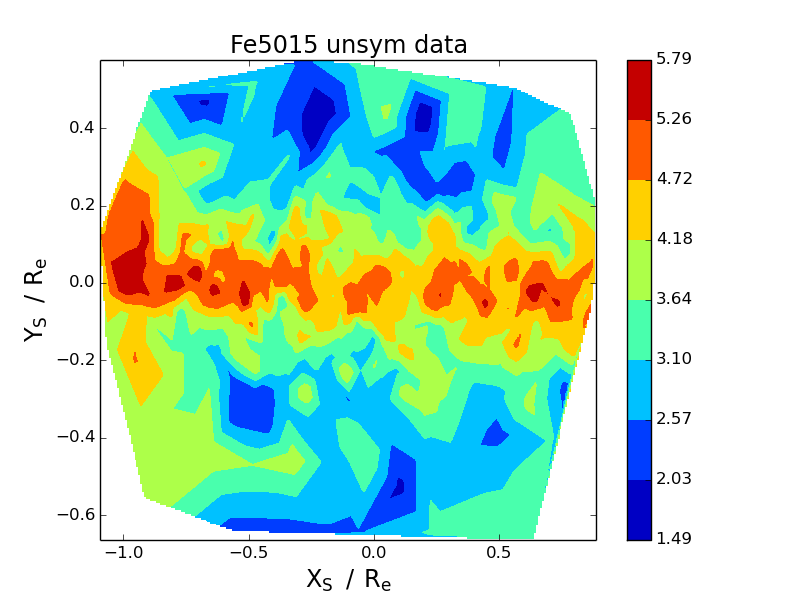}  & \includegraphics[width=50mm]{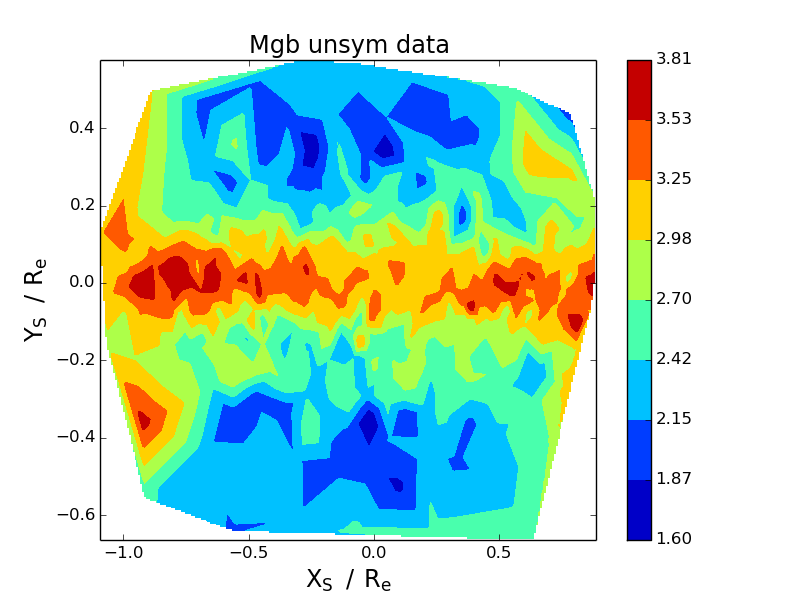}\\

\includegraphics[width=50mm]{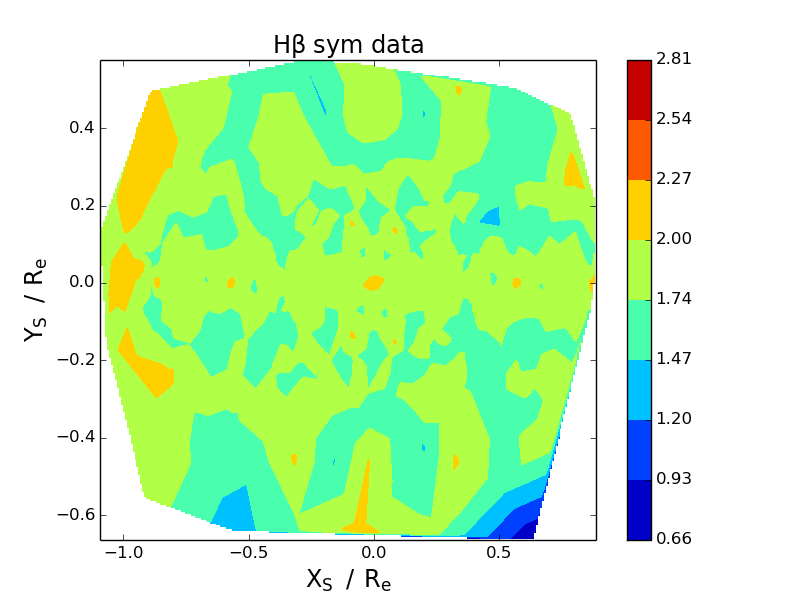} & \includegraphics[width=50mm]{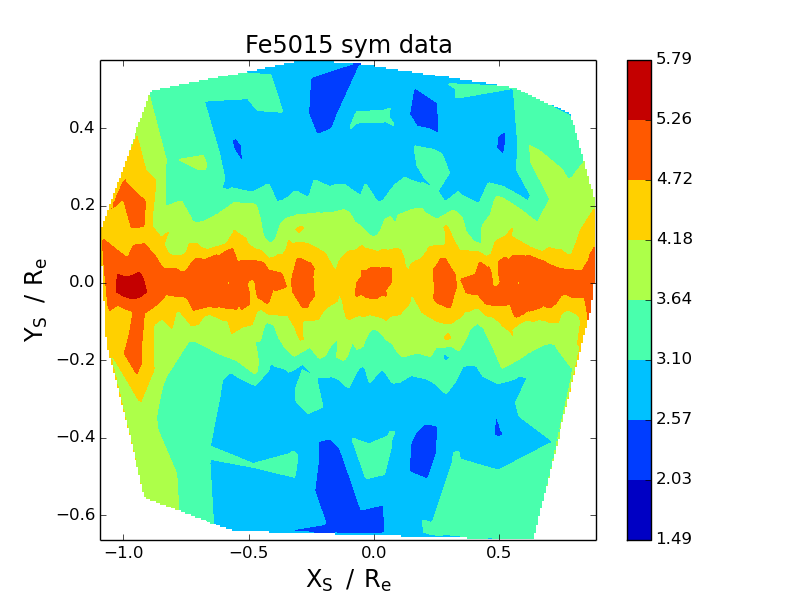}  & \includegraphics[width=50mm]{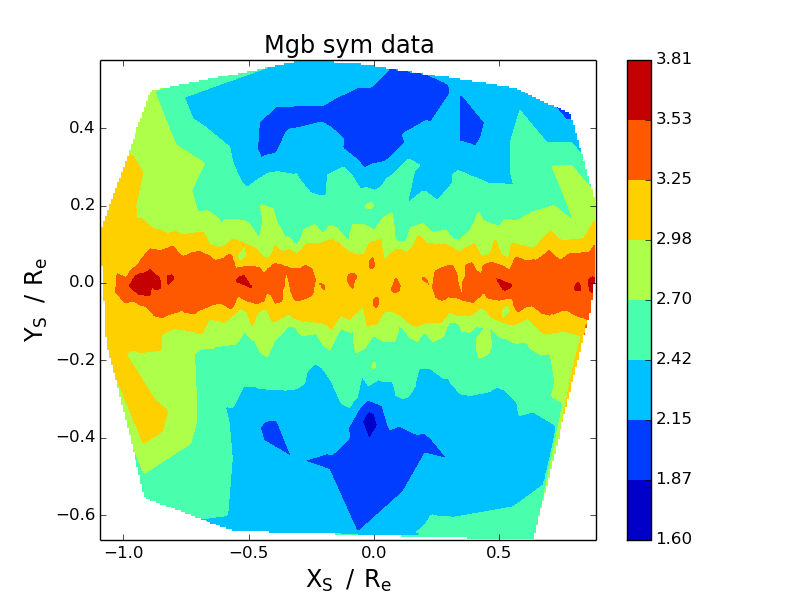}\\

\includegraphics[width=50mm]{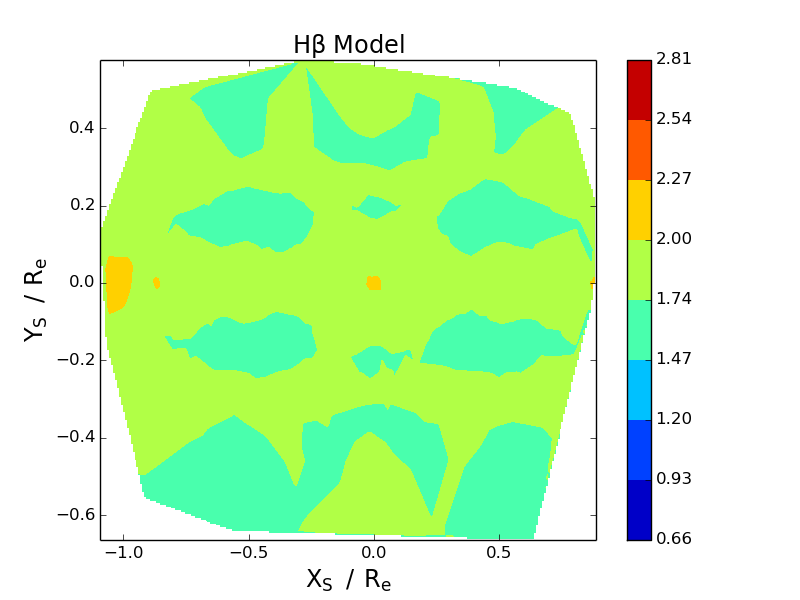} & \includegraphics[width=50mm]{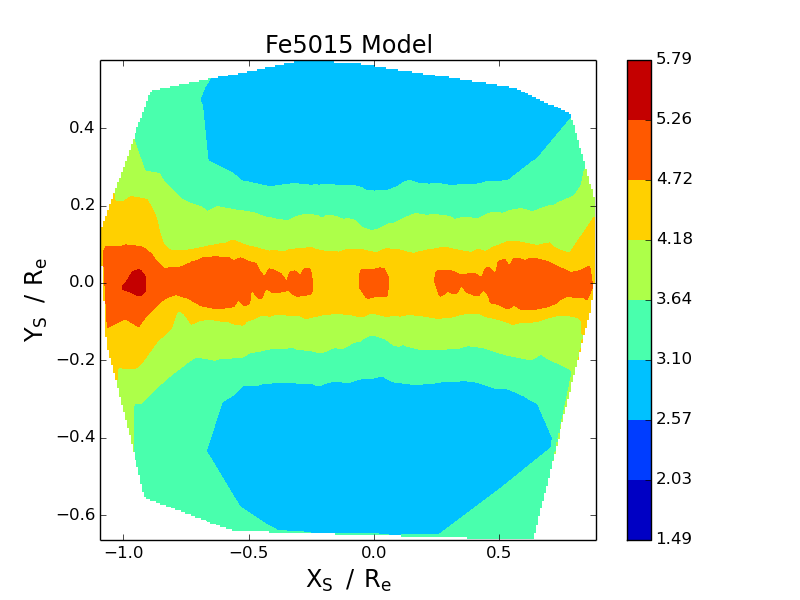}  & \includegraphics[width=50mm]{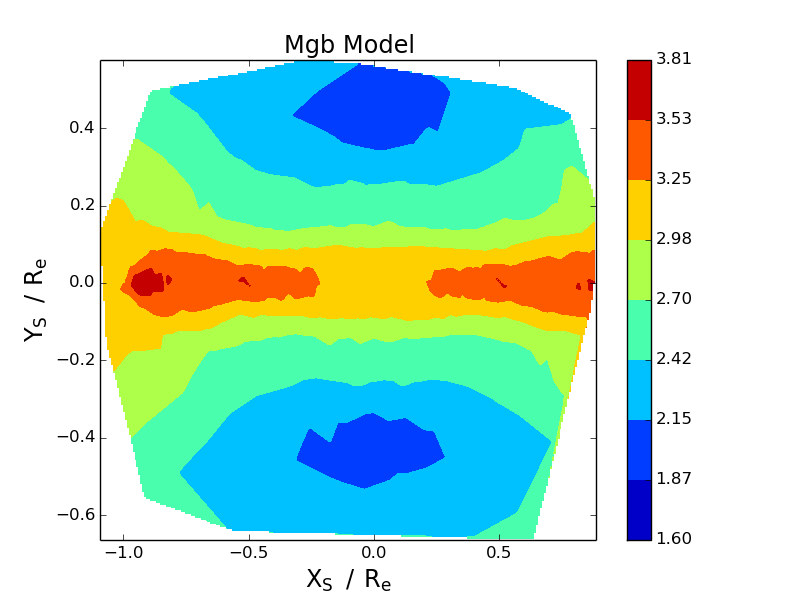}\\

\includegraphics[width=50mm]{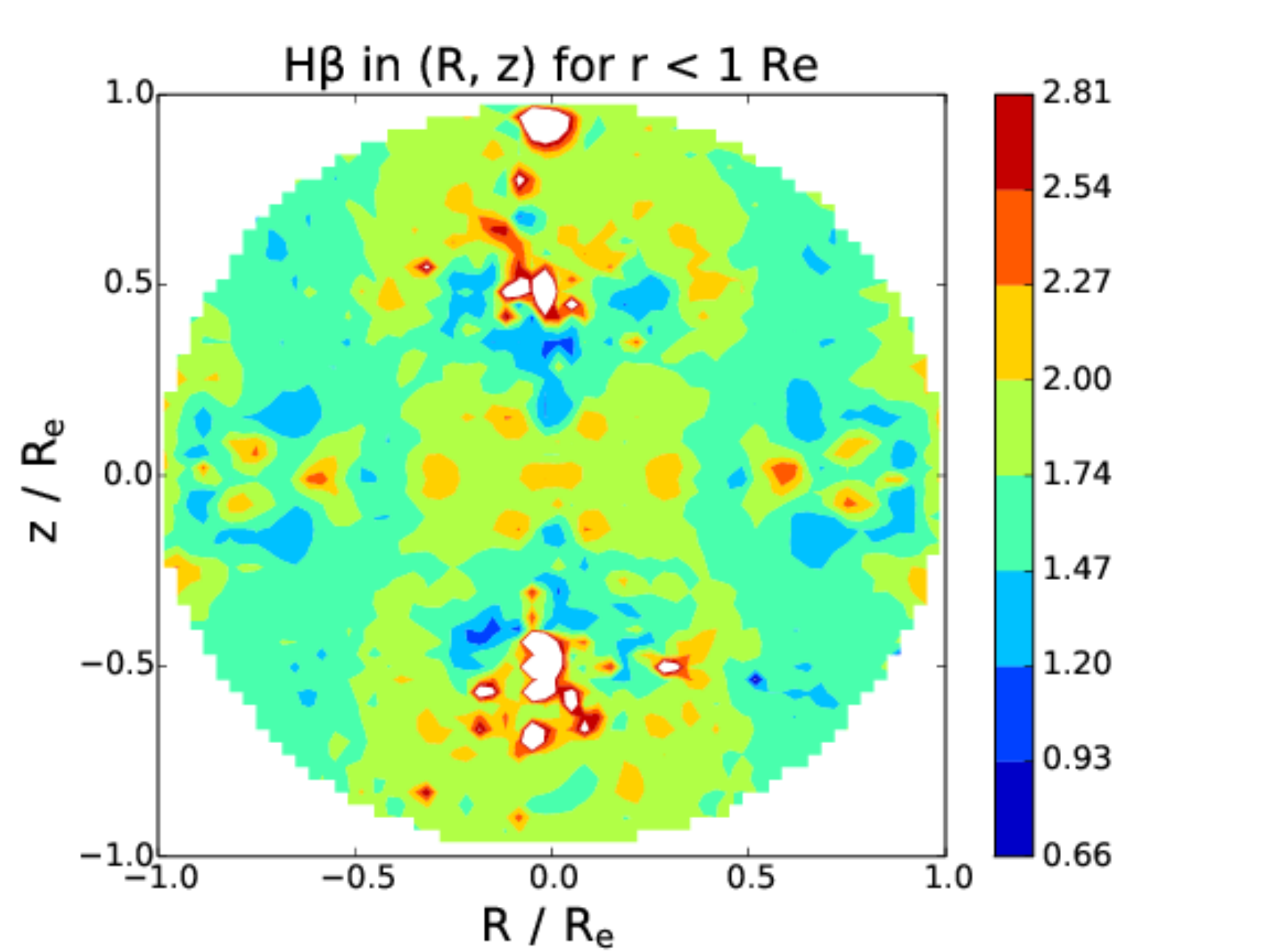} & \includegraphics[width=50mm]{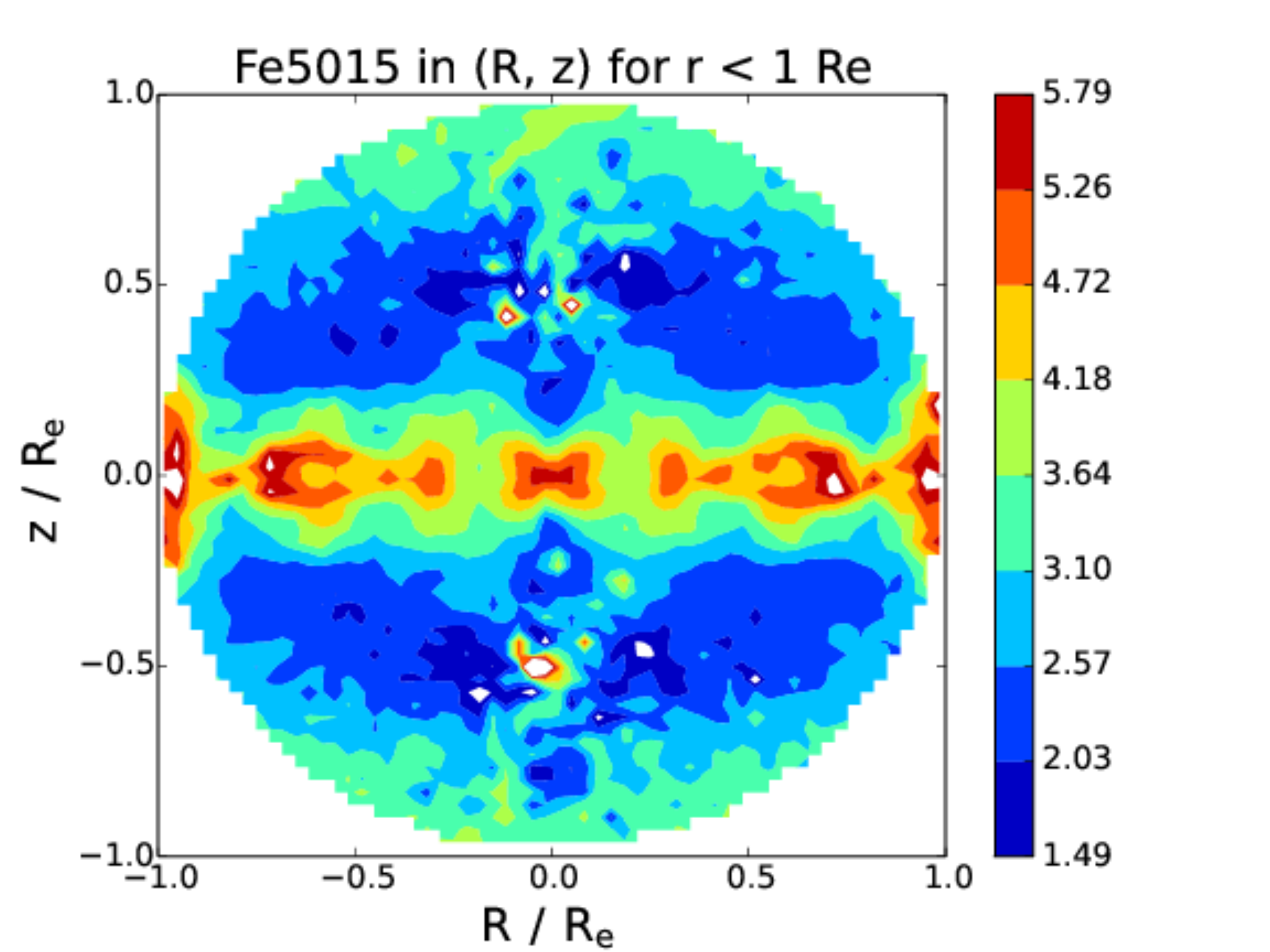}  & \includegraphics[width=50mm]{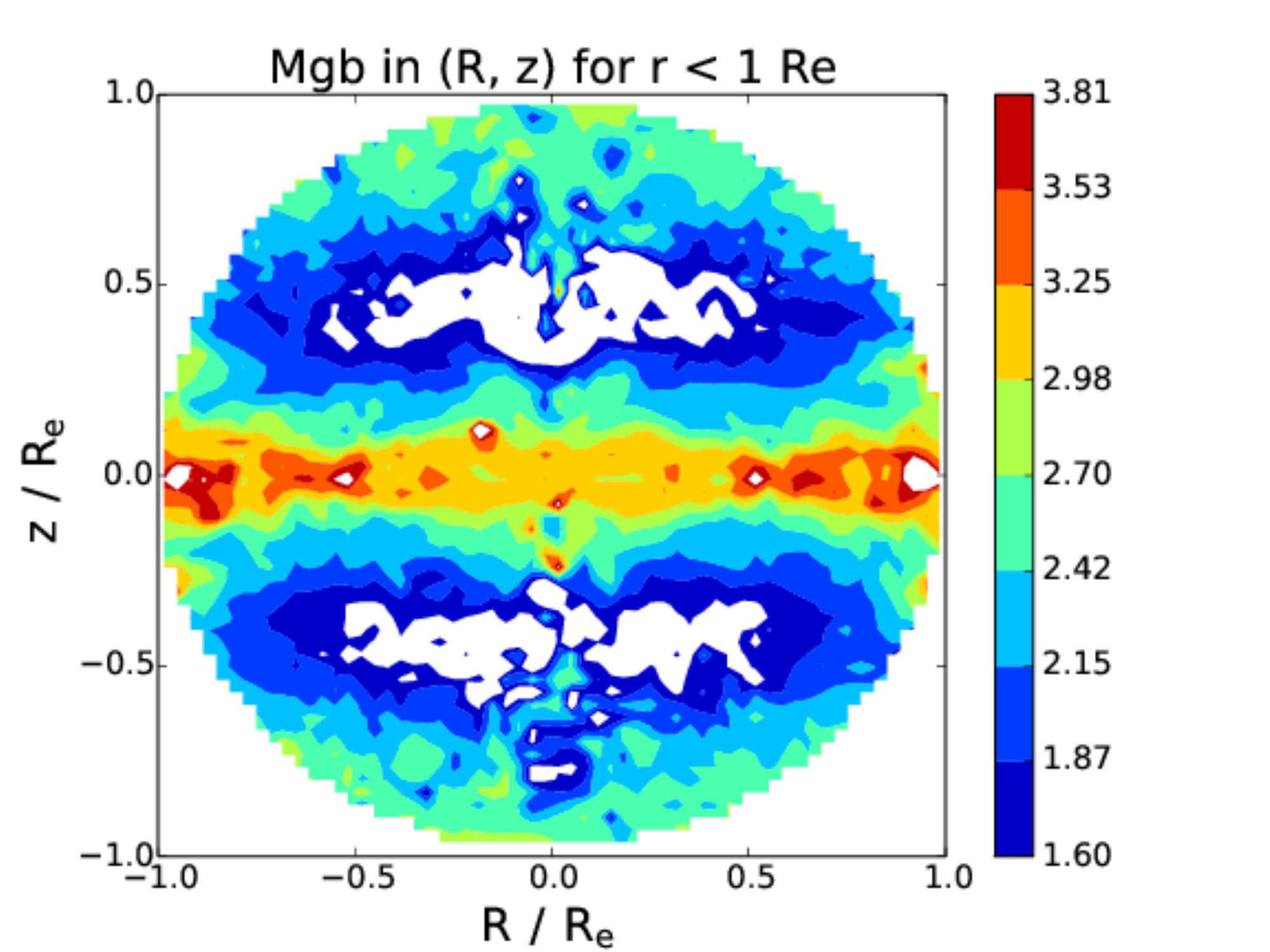}\\

\includegraphics[width=50mm]{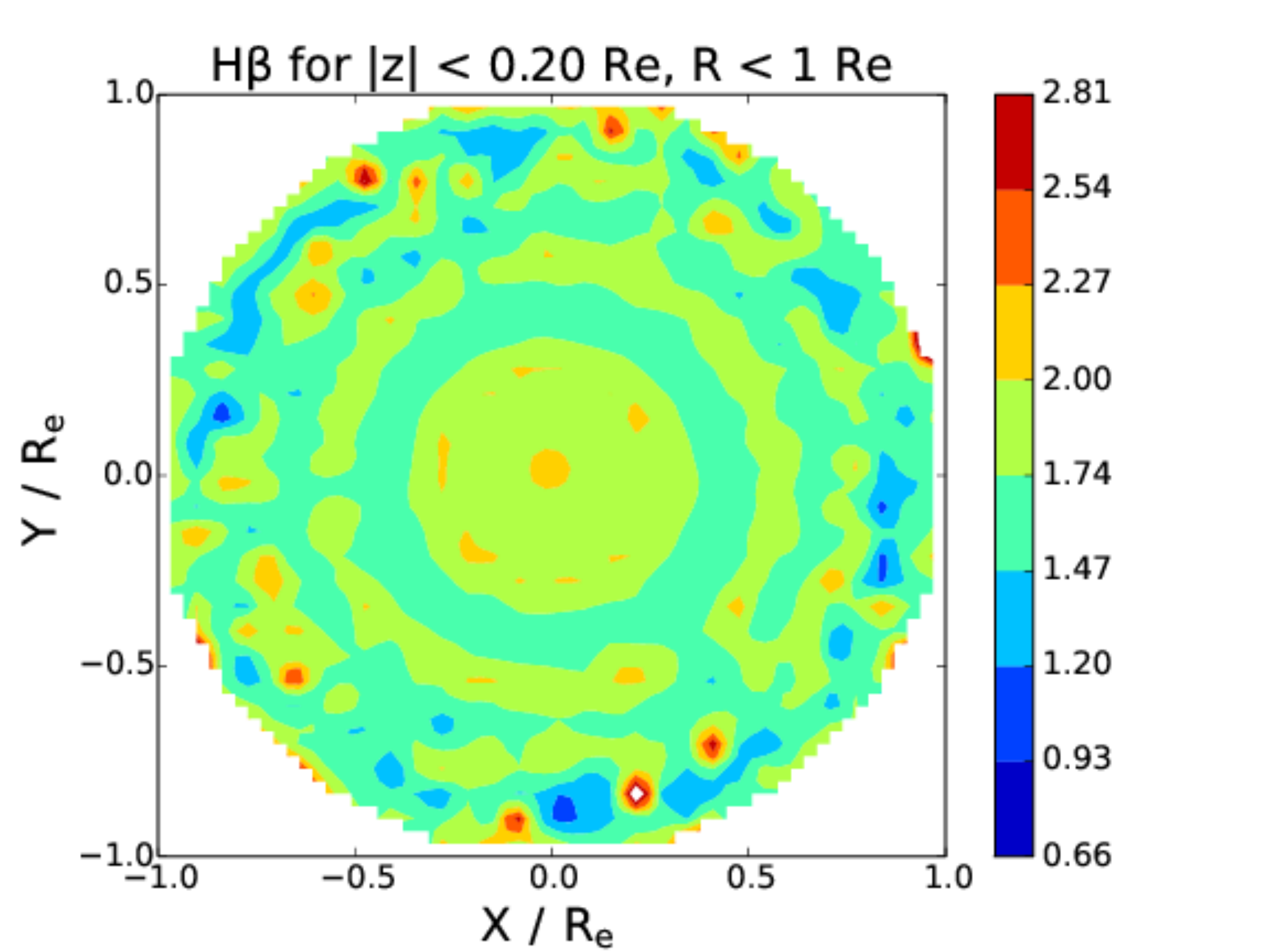} & \includegraphics[width=50mm]{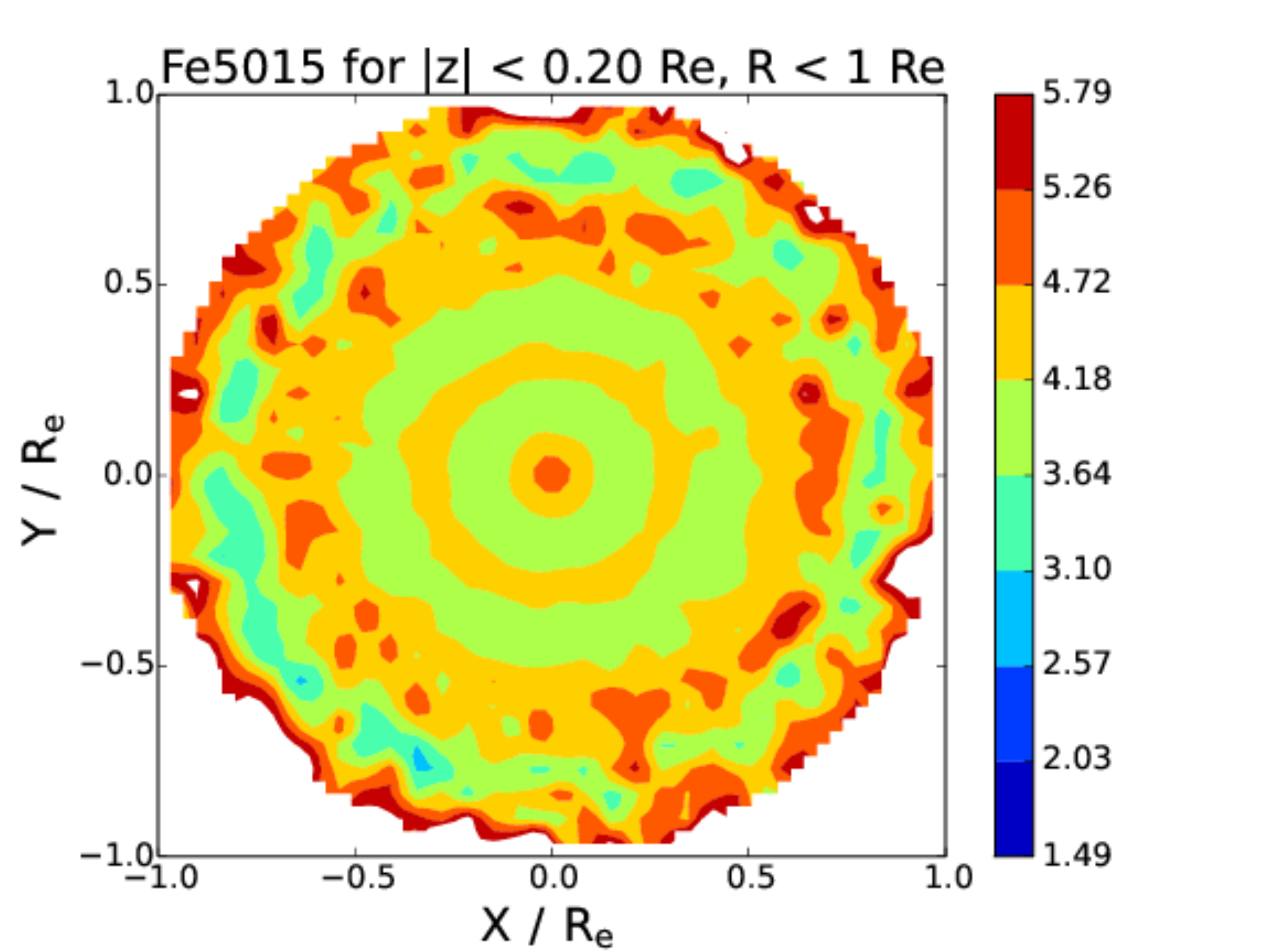}  & \includegraphics[width=50mm]{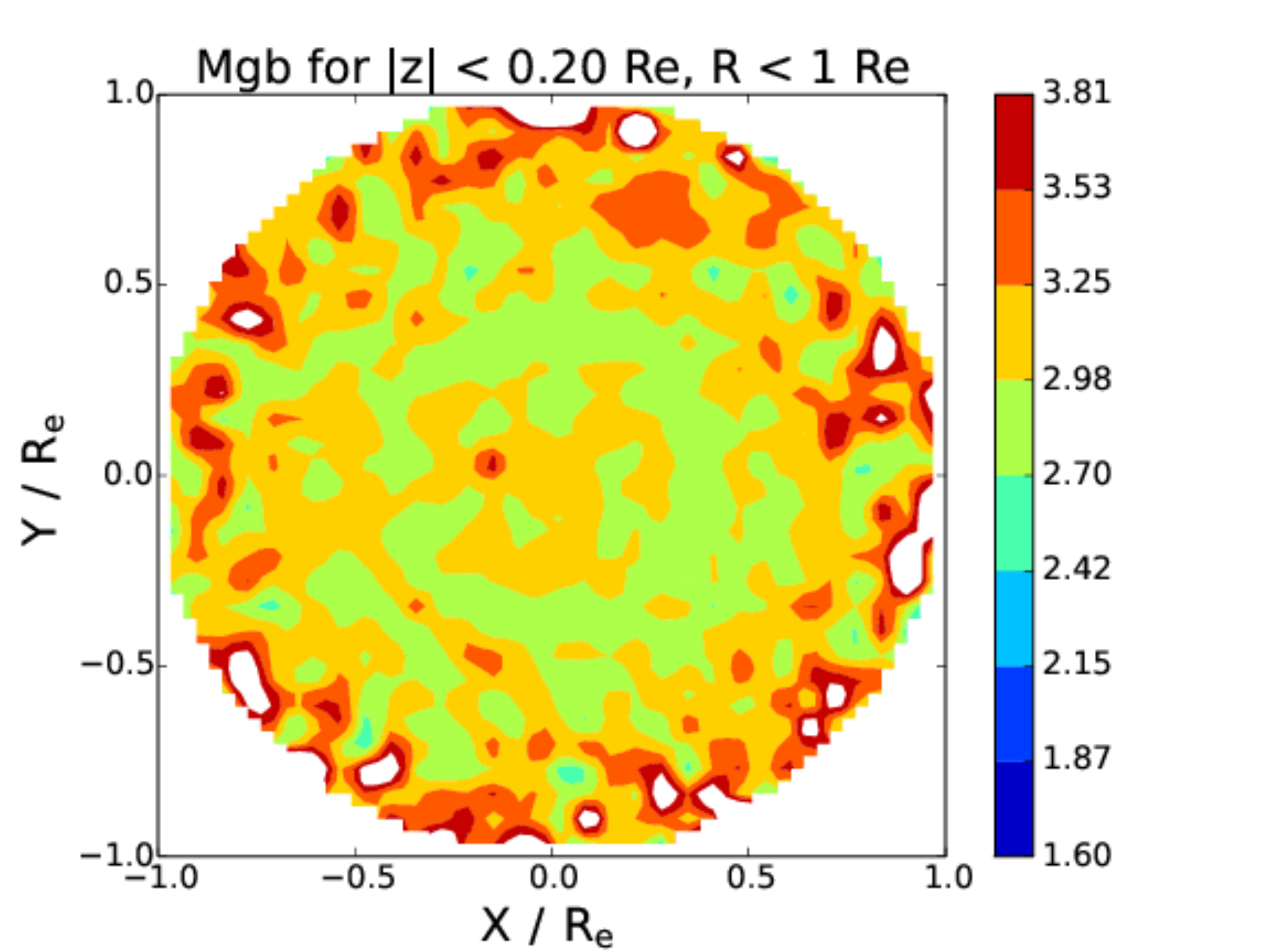}\\

\end{tabular}

\medskip
\caption{NGC 4452 line strength plots from chemo-M2M modelling.  See Figure \ref{tab:NGC1248} for a description of the rows and columns.}
\label{tab:NGC4452}
\end{figure*}

\begin{figure*}
\centering
\begin{tabular}{lcr}

\includegraphics[width=50mm]{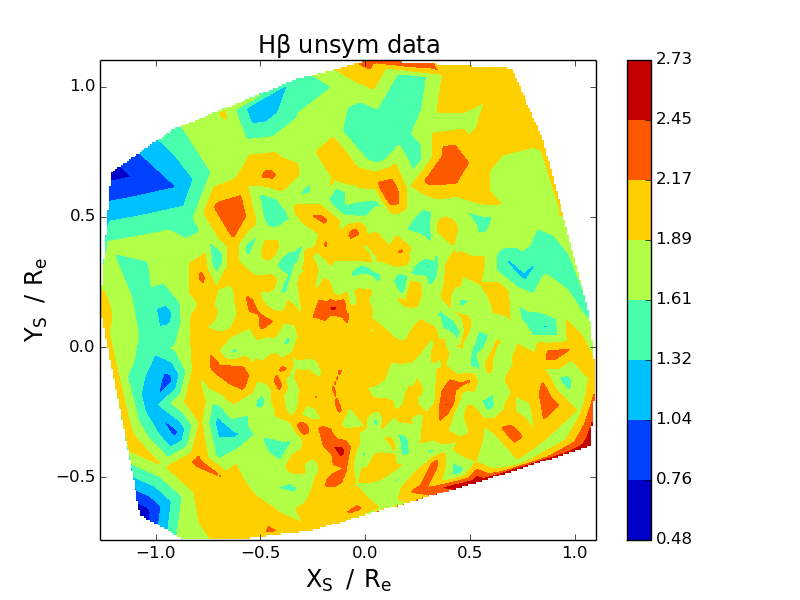} & \includegraphics[width=50mm]{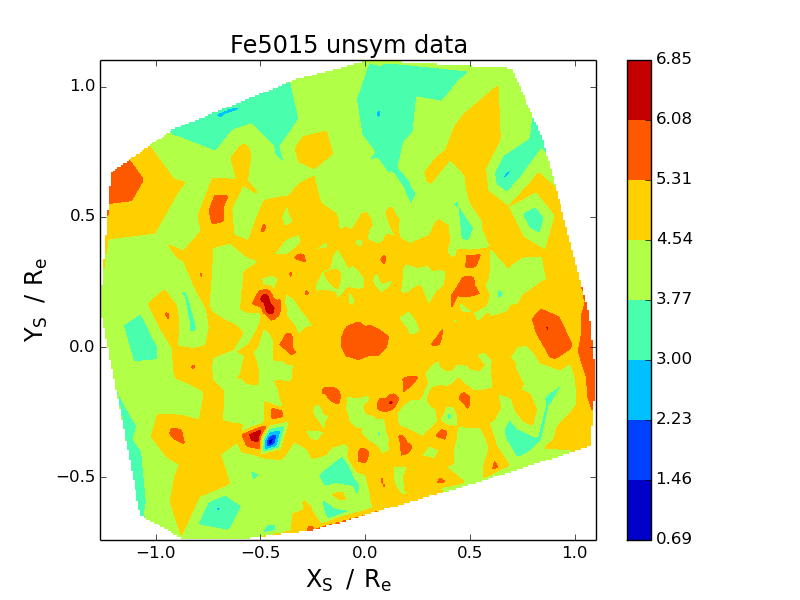}  & \includegraphics[width=50mm]{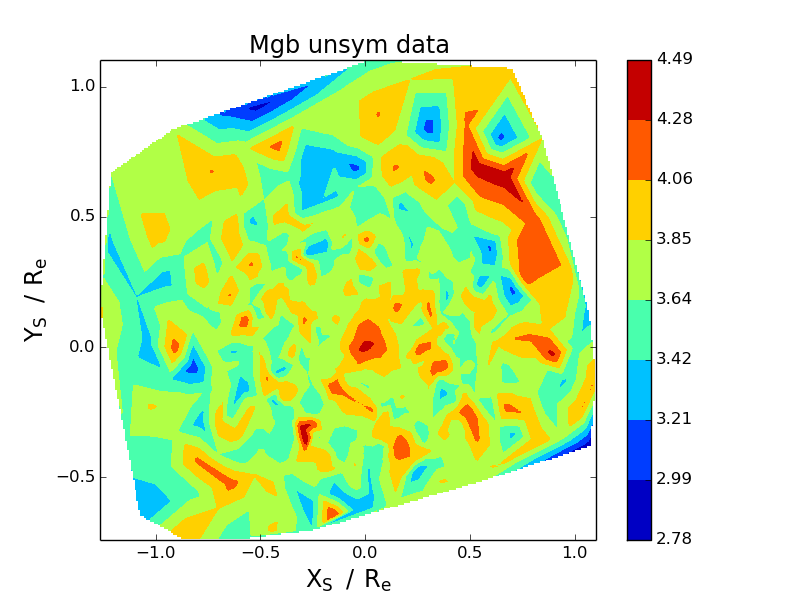}\\

\includegraphics[width=50mm]{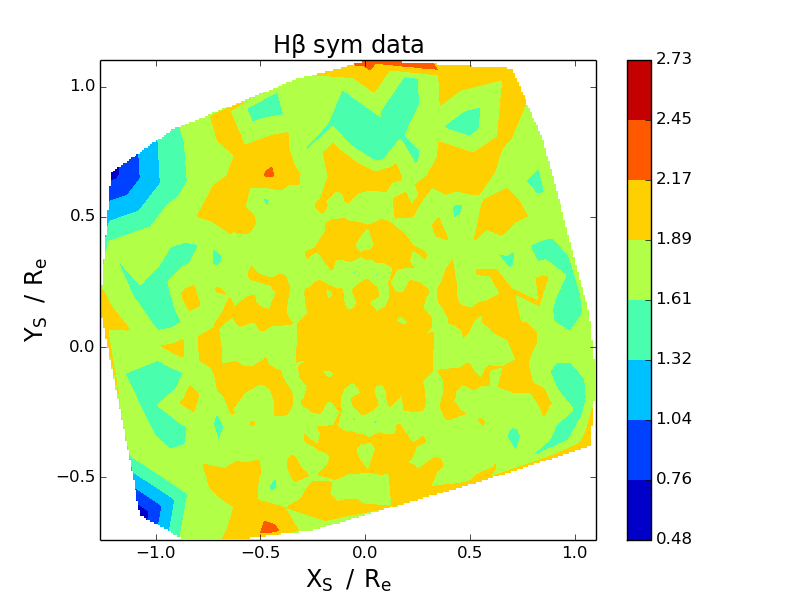} & \includegraphics[width=50mm]{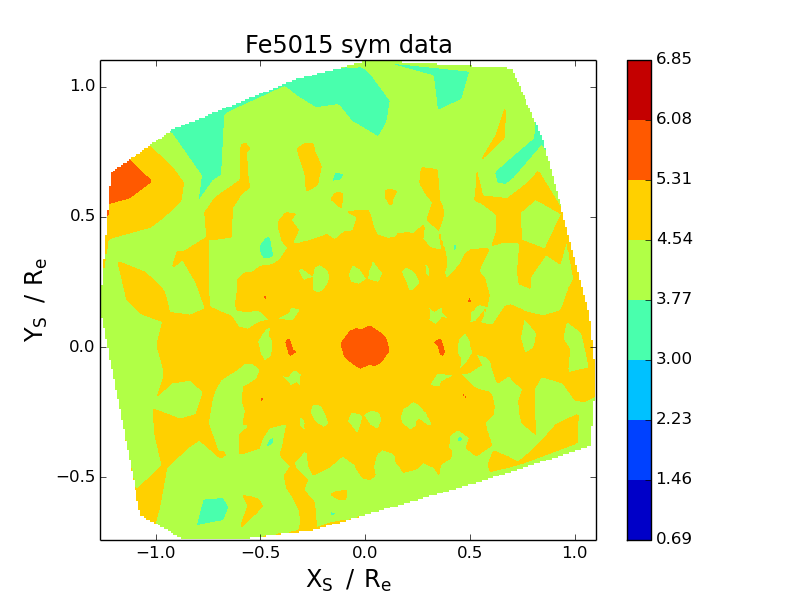}  & \includegraphics[width=50mm]{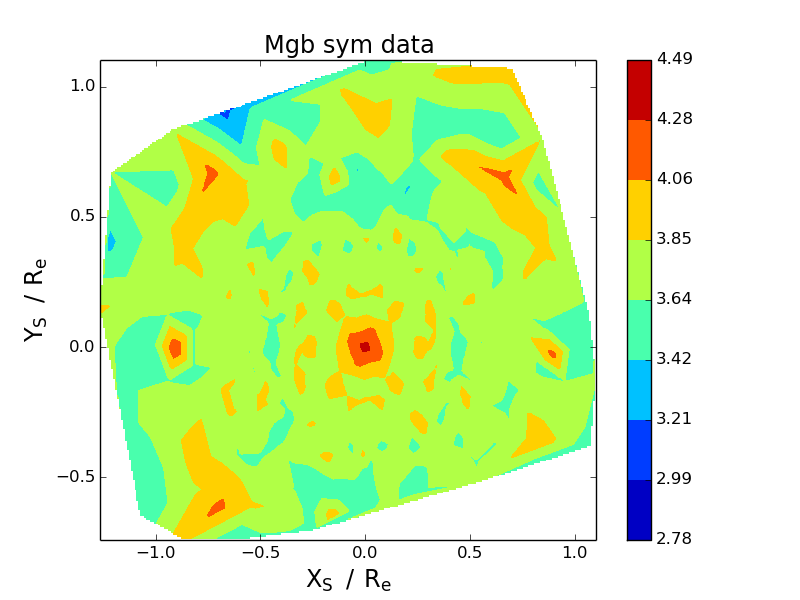}\\

\includegraphics[width=50mm]{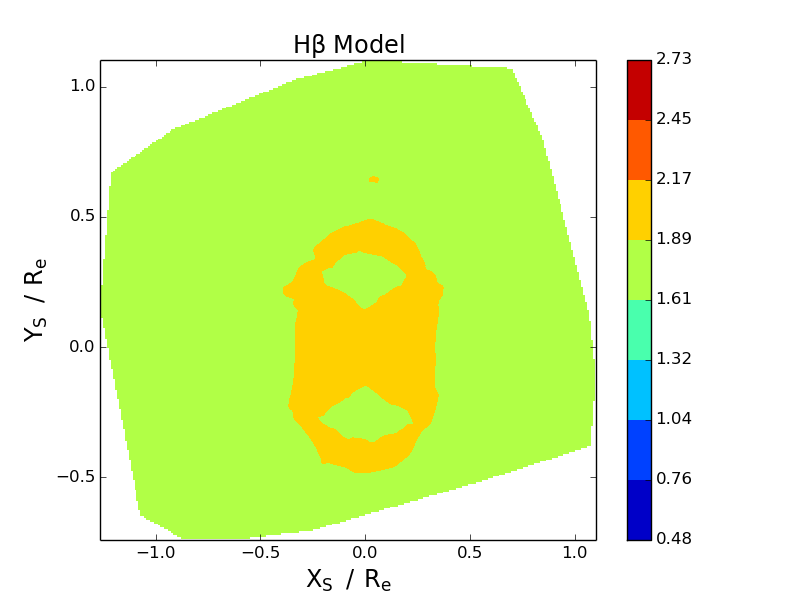} & \includegraphics[width=50mm]{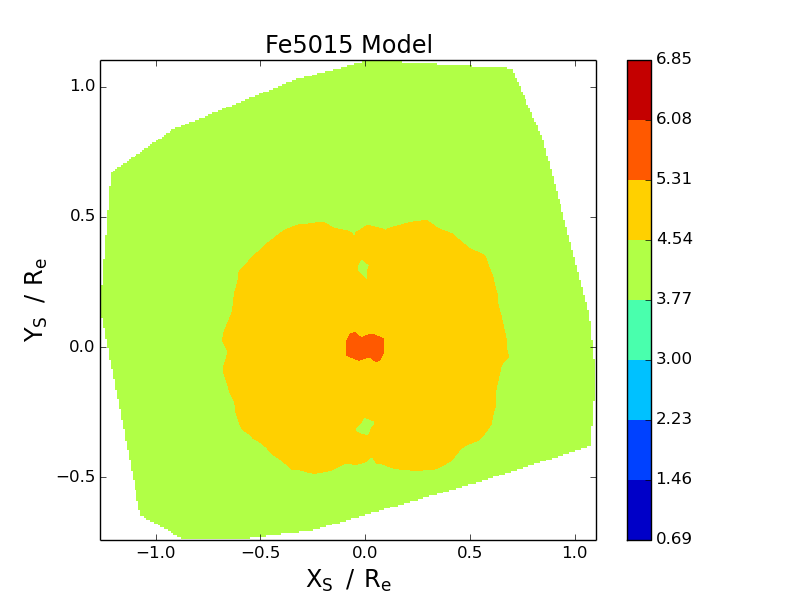}  & \includegraphics[width=50mm]{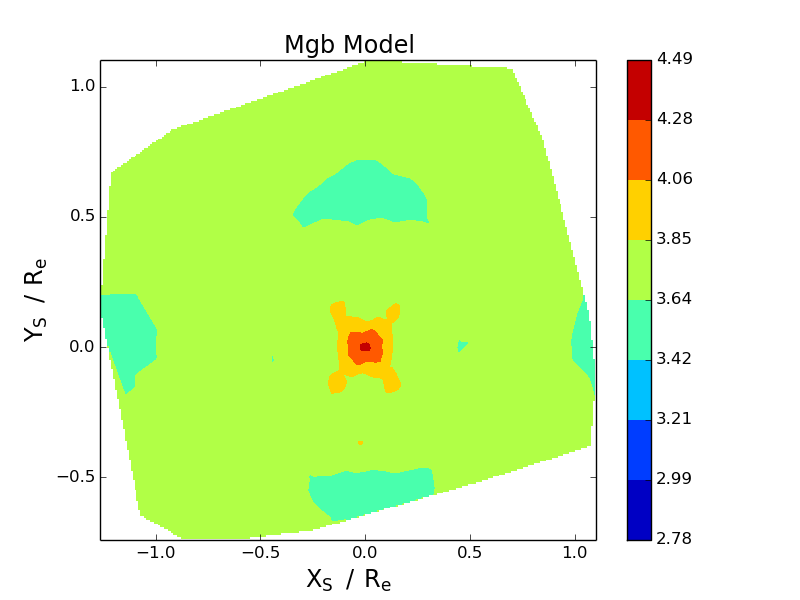}\\

\includegraphics[width=50mm]{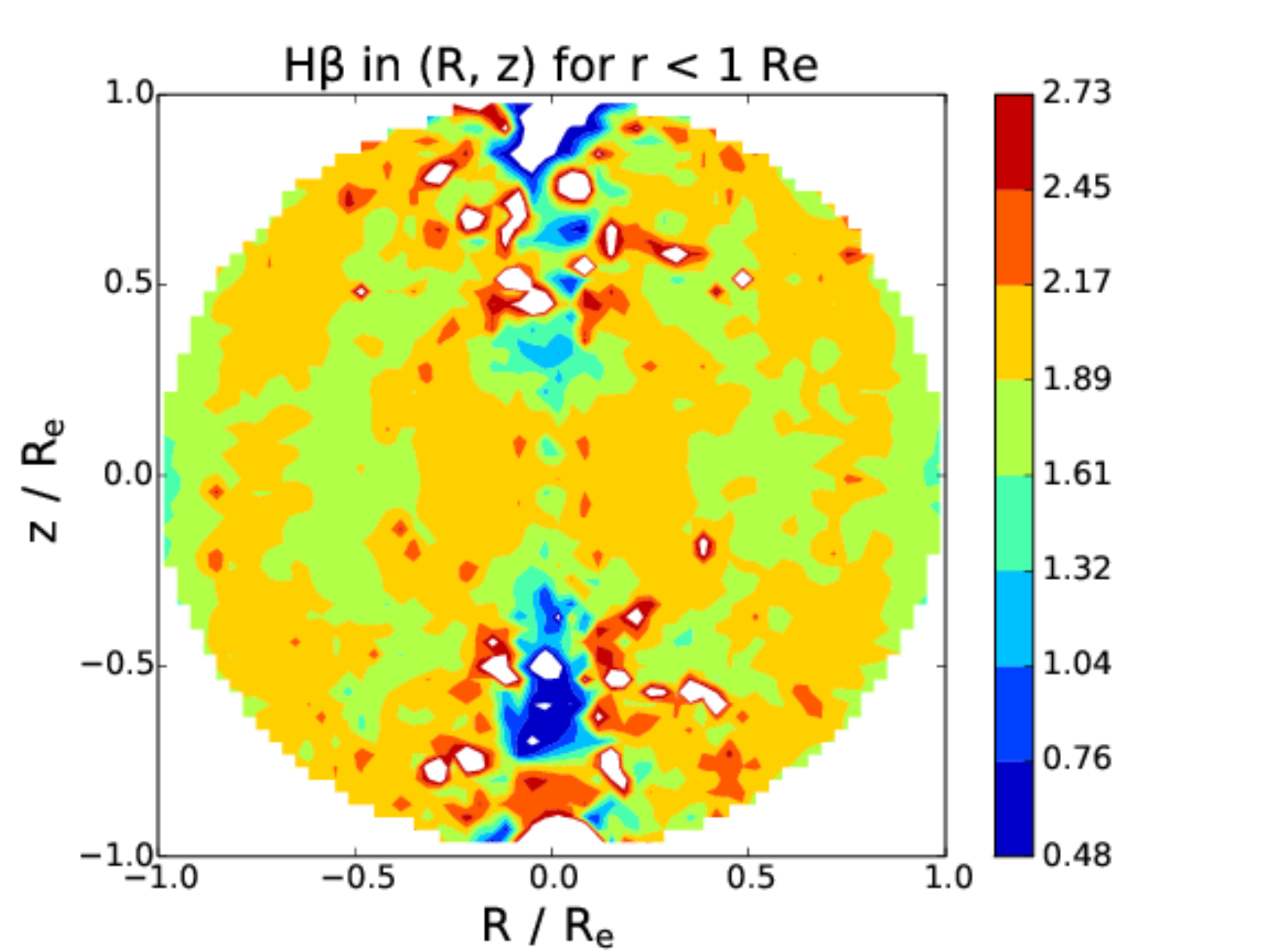} & \includegraphics[width=50mm]{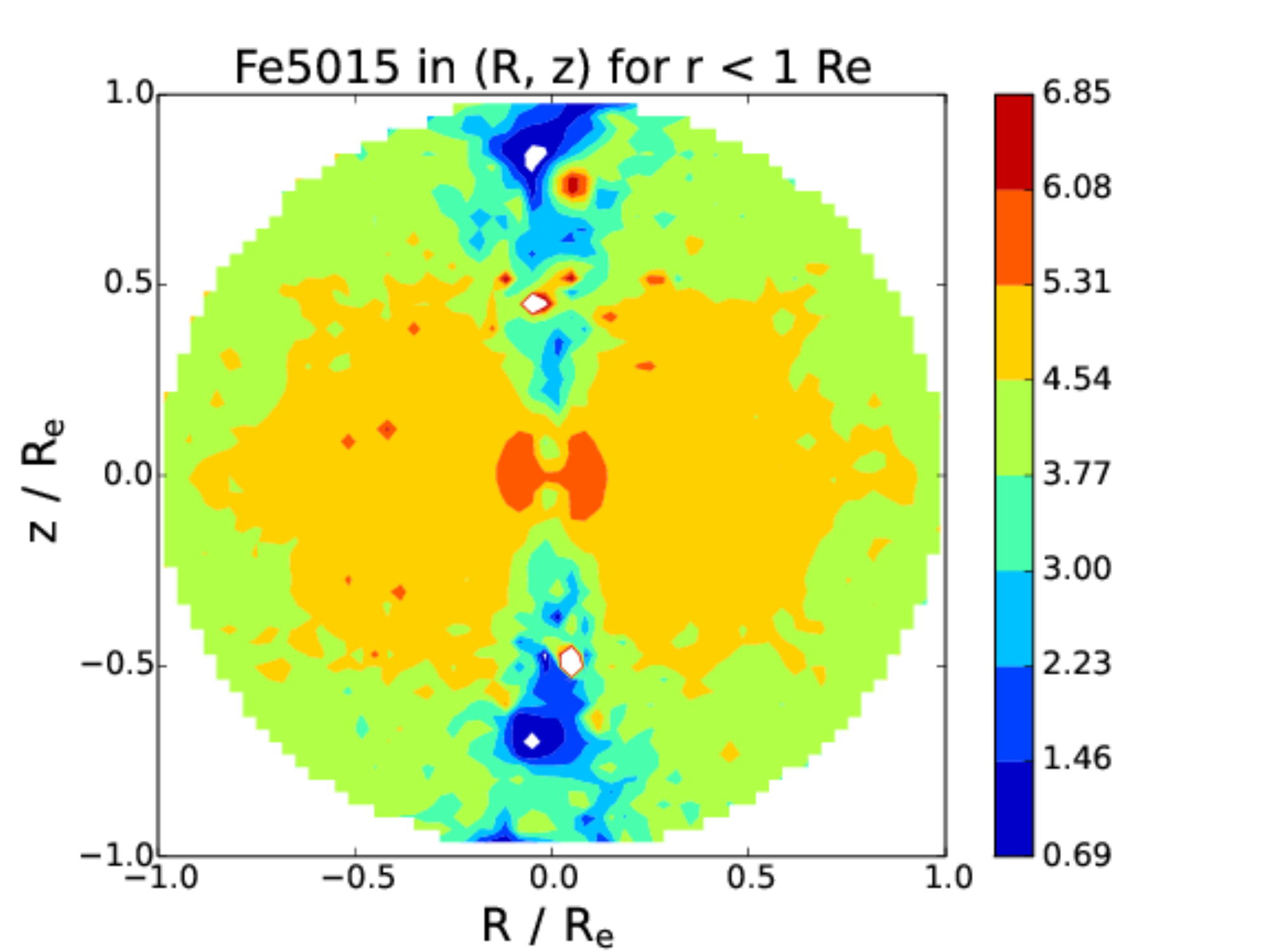}  & \includegraphics[width=50mm]{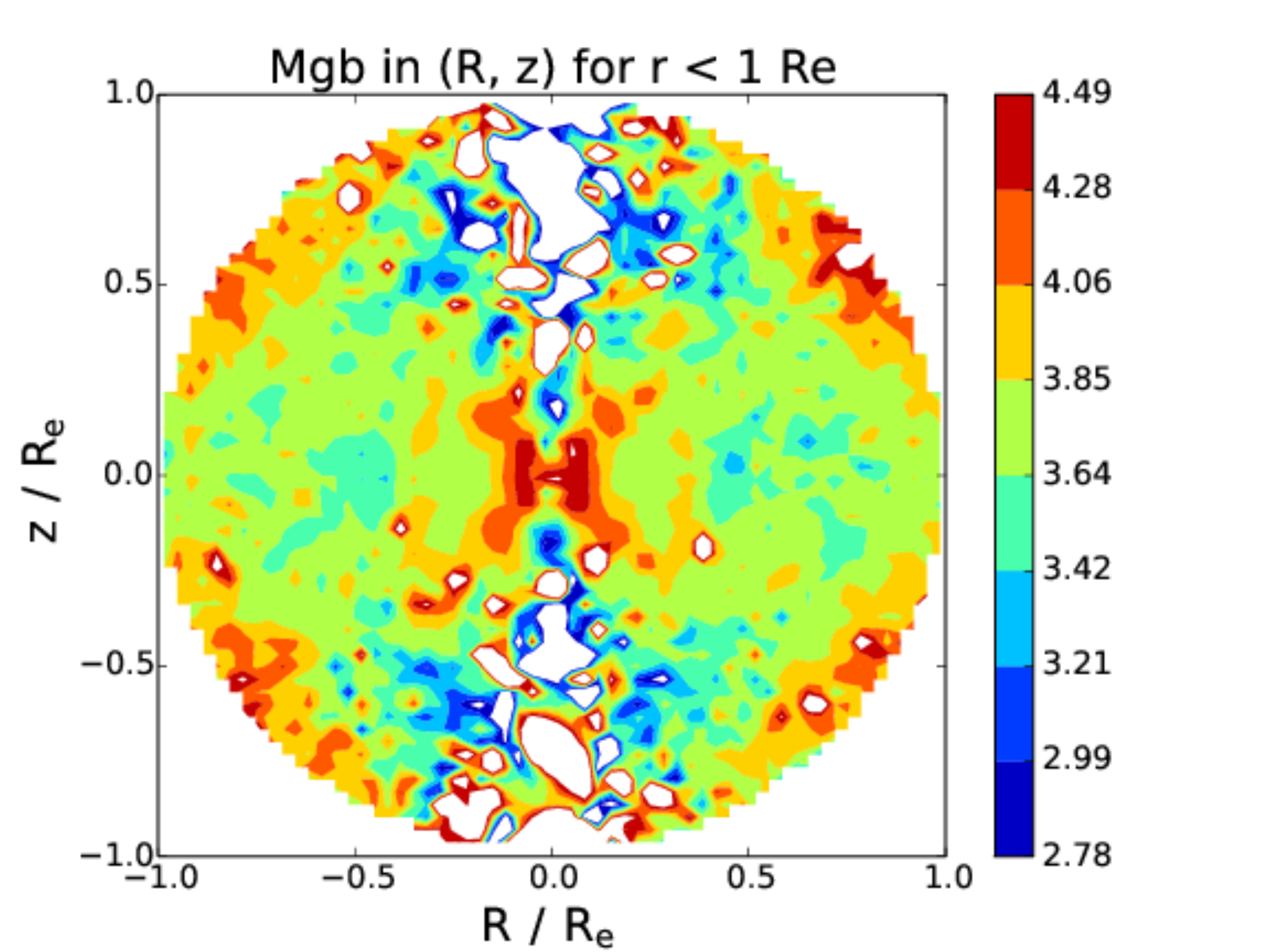}\\

\includegraphics[width=50mm]{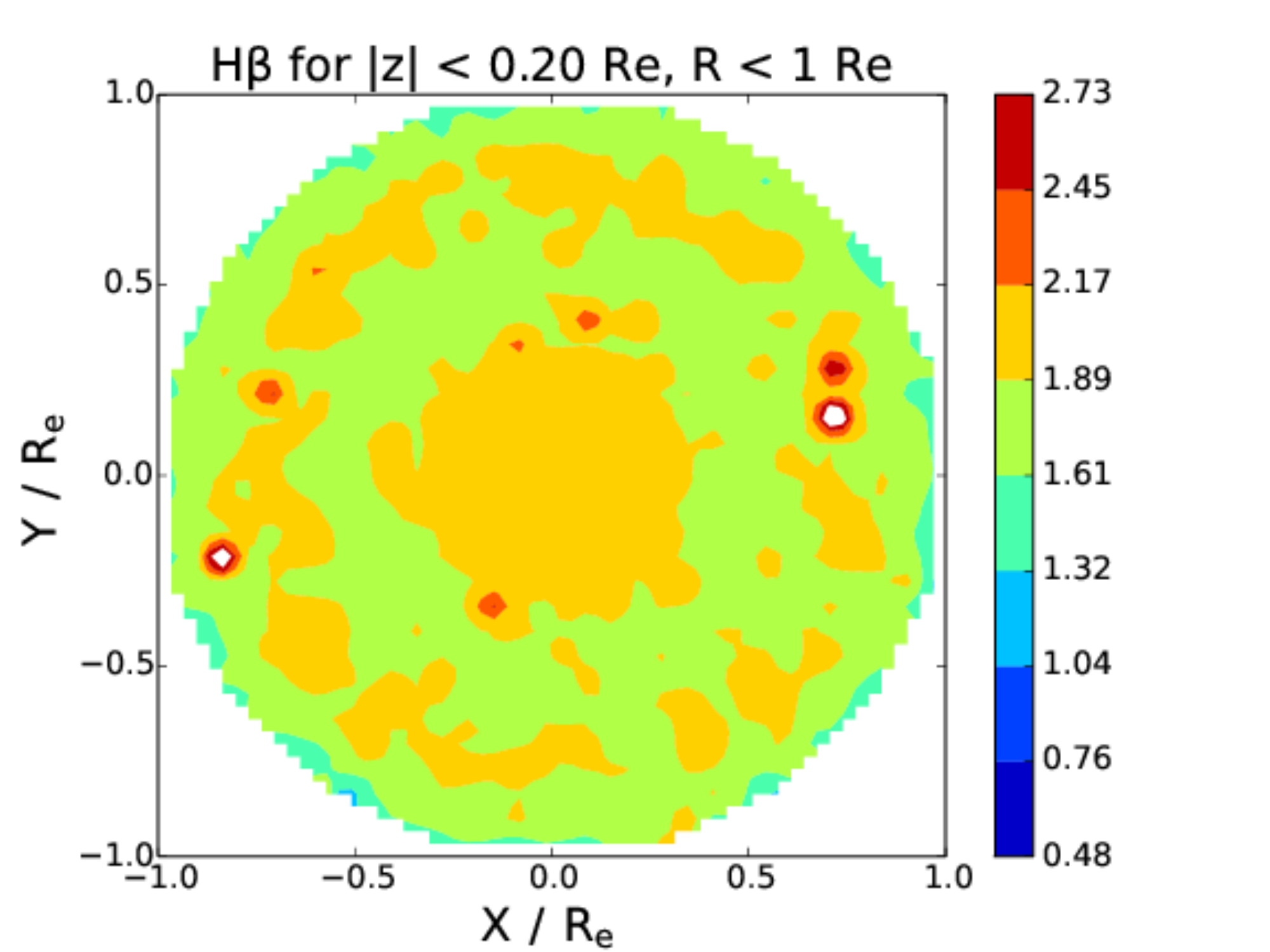} & \includegraphics[width=50mm]{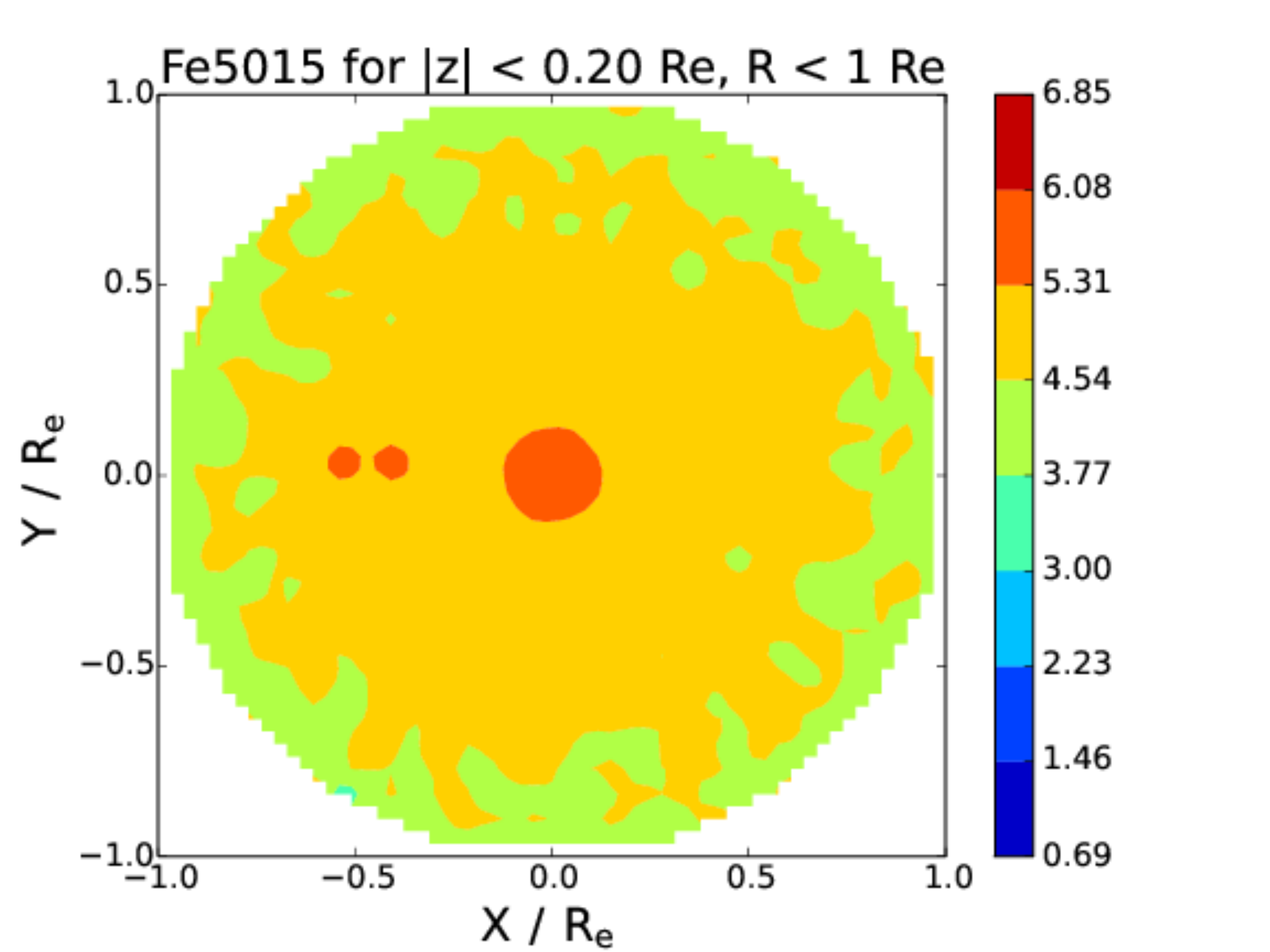}  & \includegraphics[width=50mm]{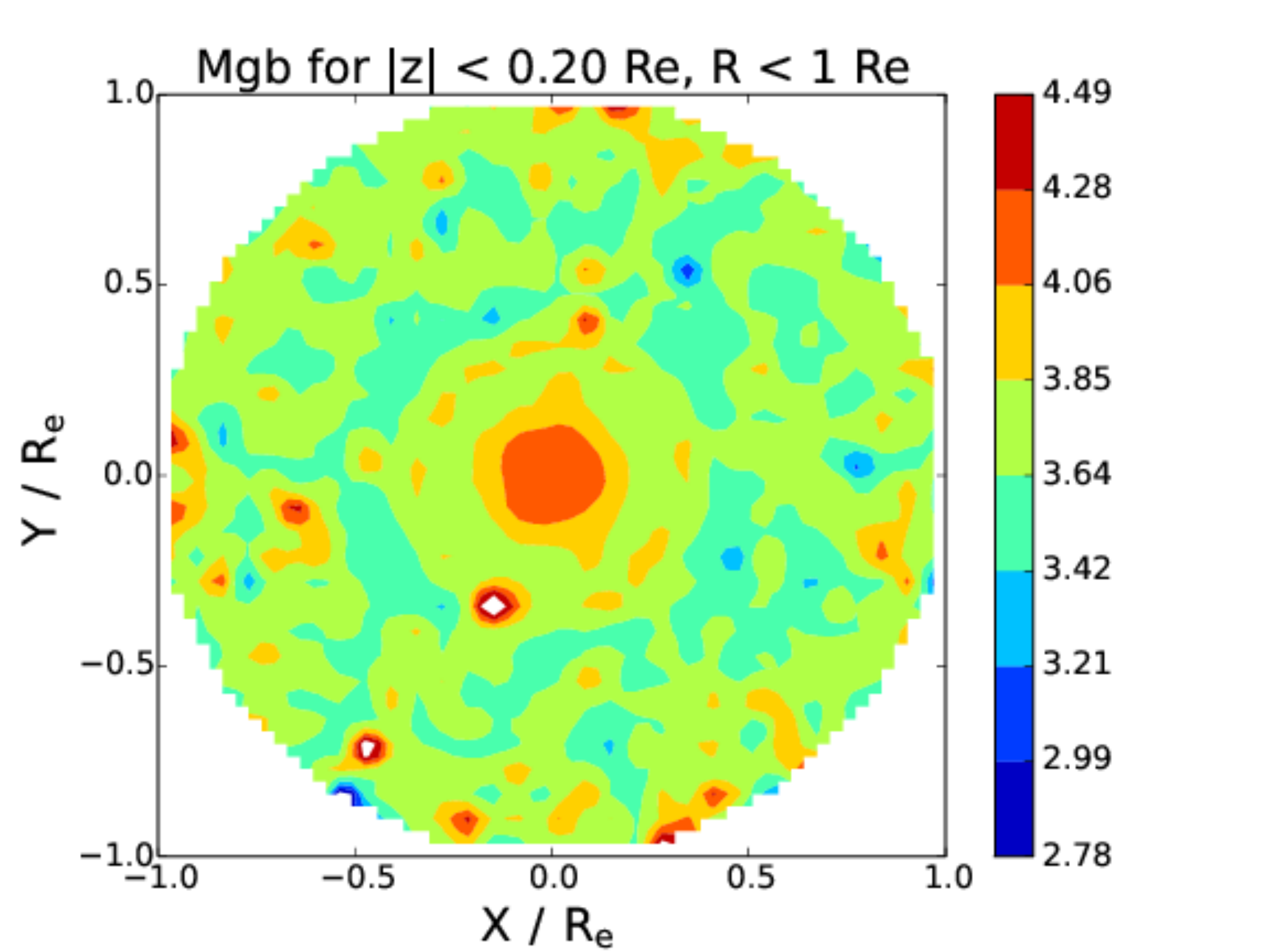}\\

\end{tabular}

\medskip
\caption{NGC 4551 line strength plots from chemo-M2M modelling.  See Figure \ref{tab:NGC1248} for a description of the rows and columns.} 
\label{tab:NGC4551}
\end{figure*}

For our four galaxies, the results of the M2M modelling runs are shown in Figs. \ref{tab:NGC1248}, \ref{tab:NGC3838}, \ref{tab:NGC4452} and \ref{tab:NGC4551}. The columns represent the absorption lines being modelled.  The first row in the figures shows the released ATLAS$^{\rm{3D}}$ data, and the second and third rows show respectively the symmetrised data input to the M2M modelling process and the M2M reproduction of that data.  The fourth and fifth rows examine the 3D distribution of the end of run particle line strength data.  The fourth row shows an $(R, z)$ plot restricted to a spherical radius of $r < 1 \; \rm{R_e}$, while the fifth row, an equatorial plot for $|z| < 0.2 \; \rm{R_e}$. For the kinematic data, the data and model $v_{\rm{rms}}$ maps are contained in the appendices.  We also leave until the appendices figures showing the end of modelling run particle weight distributions.  For all observables in all galaxies, the individual $\chi^2$ per bin values are $\approx1$. Weight convergence is $>95\%$ for both the luminosity and spectral weights.  Overall, we have successfully modelled the symmetrised 2D line strength data at the same time as modelling the observed kinematics.

\begin{figure}
\centering
	\includegraphics[width=75mm]{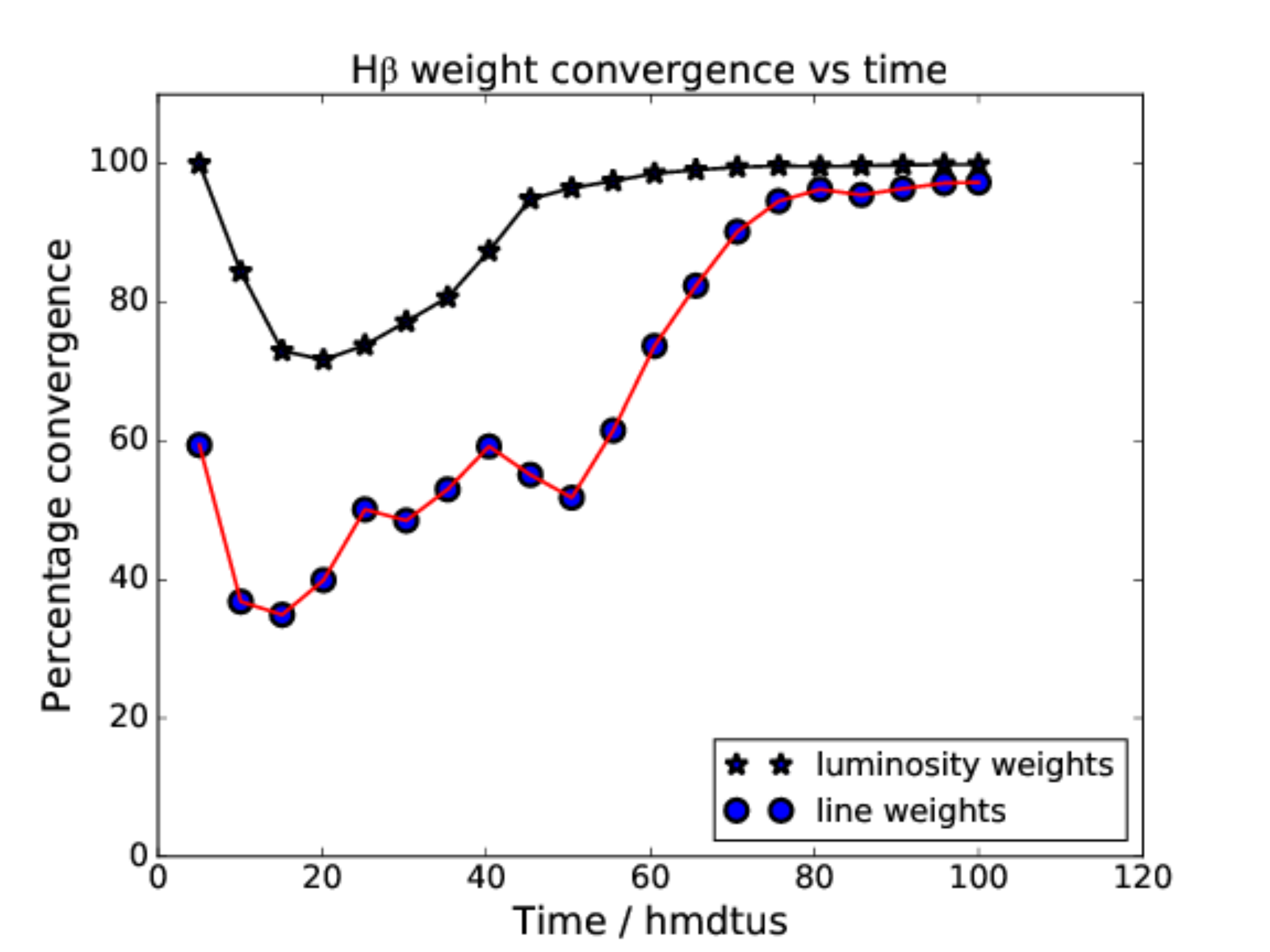}\\
	\caption{Convergence of the luminosity and H$\beta$ spectral weights over time taken from the M2M model for NGC 4551 (see Section \ref{sec:results}).  Given that spectral weight calculations are linked to the luminosity weights, once the luminosity weights approach convergence so the fluctations in the overall spectral weight percentages disappear and the spectral weights converge.  Time in this figure is measured in 'hmdtus` - half mass dynamical time units.}
\label{fig:wtevoln}
\end{figure}

\begin{figure}
\centering
	\includegraphics[width=75mm]{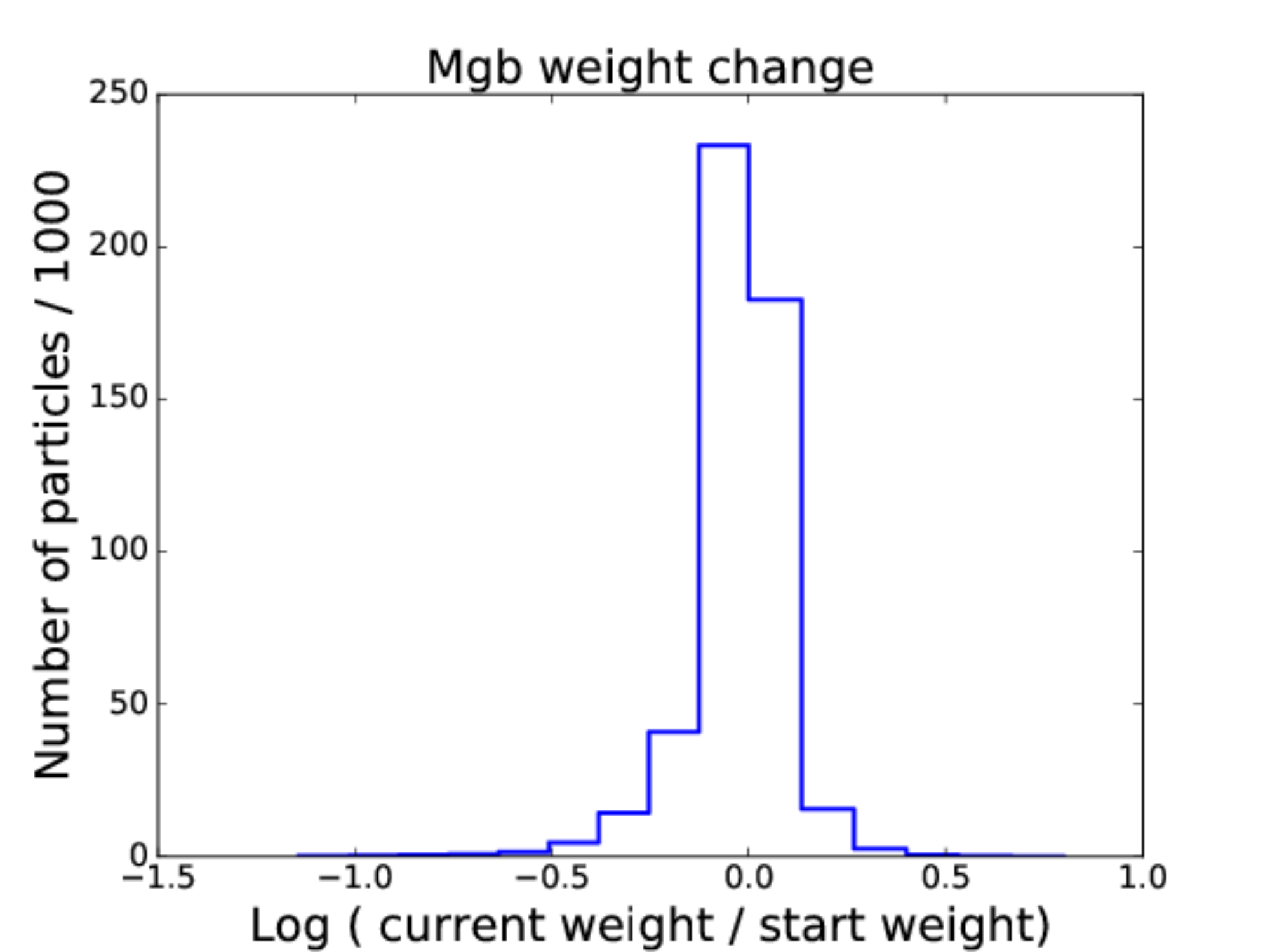}\\
	\caption{Histogram showing the number of particles in bins of the log of the ratio of the final to initial Mg$\,b$ weights.  Taken from the M2M model for NGC 4452, the histogram shows the extent to which the initial weights have been adjusted (see Section \ref{sec:results}).}
\label{fig:wtLz}
\end{figure}

We show two additional plots, taken from the various galaxy models, to provide insight into the modelling process.  Figure \ref{fig:wtevoln} shows the particle weight convergence over time for both the luminosity and spectral weights.  Given that spectral weight calculations are linked to the luminosity weights, once the luminosity weights approach convergence so the fluctations in the overall spectral weight calculations disappear and the spectral weights proceed smoothly to convergence.  In Figure \ref{fig:wtLz} we show the typical extent to which the initial spectral weights have been adjusted by the adaption processes.  The figure is as expected, showing both increases and decreases in weights.  Finally, given the theory in Section \ref{sec:m2m}, no correlations are anticipated between the luminosity and spectral weights and this is borne out in practice.

Modelling the 2D line strength data has performed as expected.  Given we are using 3D particle models, we also obtain a 3D particle chemical representation of the galaxies.  It must not be forgotten that the observed data has been symmetrised to facilitate modelling, and as a consequence the model is an `averaged' representation of the galaxy.  In light of this, as is already the case with 3D kinematic data, the 3D representation must be interpreted carefully.  For example, the equatorial plane plots (row 5) in Figs. \ref{tab:NGC1248}, \ref{tab:NGC3838}, \ref{tab:NGC4452} and \ref{tab:NGC4551} show more ring / annulus artefacts than might be expected from an examination of the unsymmetrised raw data (row 1).  Overall the Fe5015 and Mg$\,b$ lines appear to be better modelled than the H$\beta$ absorption line.  This might be due to the small H$\beta$ data range, by comparison with Fe5015 and Mg$\,b$. That the artefacts occur is inevitable given the symmetrisation of the data and can not be resolved by just using unsymmetrised data.  From the tests we have run, using unsymmetrised line strength data in our M2M models results in weight convergence of only $\approx 65\%$ and mean $\chi^2$ per bin $>>1$.  We will return to this matter in Section \ref{sec:discuss}.  Considering the elliptical galaxy vs. the S0 galaxies, the elliptical galaxy NGC 4552 appears to be more plausibly modelled than the S0 galaxies.  Considering the S0 galaxies only, the higher inclination pair NGC 3838 and NGC 4452 (see Table \ref{tab:mlratios}) appear better modelled than the lower inclination NGC 1248.

\section{Discussion}\label{sec:discuss}
From our results in Section \ref{sec:results}, it is clear that we are currently able to model successfully a symmetrised 2D line strength projection.  Though initial signs are positive, it is also clear that additional work is required to understand the extent to which the 3D maps from the M2M particles are a plausible representation of the real galaxy.  It should be possible to understand how well statistically the chemo-M2M method is able to recover 3D distribution by using data with known distributions from either population synthesis models or chemo-hydrodynamical cosmological simulations. More sophisticated M2M modelling will also help, for example, moving to triaxial potentials, or introducing perhaps a 3D line strength constraint, or introducing additional observational results to relate kinematics to line strength. Whether using population synthesis models to provide line strength profiles when modelling real galaxies is an appropriate approach is an open question.  Much will depend on the additional assumptions that will be needed.  The same is true for cosmological simulations.

Data symmetrisation is frequently used in a number of different methods for the stellar dynamical modelling of galaxies, for example Jeans equation modelling \citep{AtlasXV} and Schwarzschild modelling \citep{Remco2008}, as well as in M2M modelling \citep{Morganti2013}.  It is not obvious that any of these methods is well-suited to modelling, or indeed capable of modelling, asymmetric data.  Perhaps action-based distribution function modelling (for example, \citealt{Sanders2015}, \citealt{Posti2015}) will offer a solution.  An alternative approach might be to decompose the observed data into a symmetric part and an asymmetric part, modelling the symmetric part with M2M and the asymmetric part with a different technique perhaps.

We have not so far attempted to model using absorption line data from individual stars.  Modelling with such data is indeed possible using a mechanism similar to that used in M2M models for modelling radial velocities (see for example  \citealt{Long2010} or \citealt{Zhu2014}).  The likelihood terms come from considering the line-of-sight line strength distribution convolved with a Gaussian constructed from the observed data.  Overall this approach will act to improve the accuracy and credibility of the 3D line strength maps.

We have used axisymmetric models for simplicity.  However, if we wish to understand the chemo-orbital structure of a galaxy then models which allow a greater variety of orbit types are required, for example, triaxial or self-gravitating models.  It will be interesting to understand whether there are any correlations between kinematics and spectral line data, in particular whether any relationships can be found between orbital families and absorption line strengths.  We have used galaxies at different inclinations to the line of sight but have found nothing of note in the the current investigation, apart from a possible concern regarding modelling lower inclination galaxies.  As expected, line strength modelling behaves no differently from luminosity or kinematic modelling at different inclinations.

As already indicated, the theory we have constructed in Section \ref{sec:enhtheory} is not the only approach.  For example, we could choose to merge $\chi_{\rm{sp}} ^2$ (equation \ref{eqn:Fsp}) into $F_{\rm{lm}}$ (equation \ref{eqn:Flm}) such that the kinematic aspects of the model are directly influenced by the line strength data.  Similarly, we might chose to include empirical relationships between different absorption lines. This and other variations require further investigation.

Within our modelling we are making an implicit assumption that stars exhibiting the spectral lines we are modelling with are following general motions represented by the kinematic data that has been extracted from the spectra.  Perhaps it might be more appropriate to extract the kinematic data associated with specific spectral lines and then to create spectral line specific models.  Alternatively, perhaps we should regard M2M particles with their spatial positions, velocities and weights as being the first step towards modelling a complete spectral data cube.

As a final comment, it may be possible to apply the approach we have taken to \citet{Schwarz1979} modelling by first solving for the luminosity weights and then taking those weights and solving for the spectral line weights.

\section{Conclusions}\label{sec:conclusions}
We have met the objectives we set out in the Introduction, Section \ref{sec:introduction}.  We have extended the M2M method into a chemo-dynamical method which is able to handle both kinematic and absorption line constraints simultaneously, and have applied the extended method to four ATLAS$^{\rm{3D}}$ galaxies.  We see signs that the extended method may enable us to start understanding the 3D line strength distribution.   However, as anticipated in the objectives, and as can be seen from the Discussion, Section \ref{sec:discuss}, much remains to be investigated to understand the limitations of the deprojected line strength maps, and to analyse the chemical footprint of orbits, before we are able to make robust predictions.  Clearly, a second paper is needed after this `investigation of concept' paper, and it is now in preparation (Li et al.).  Overall, we believe a promising first step has been taken in enhancing Syer \& Tremaine's made-to-measure method to perform chemo-dynamical modelling.

\begin{acknowledgements}
The reviewer is thanked for their prompt report.  The author acknowledges occasional communication and technical discussions with Shude Mao and Michael Merrifield during an early phase of the investigation.  Computer runs were performed on the \textit{Zen} high performance computer cluster of the National Astronomical Observatories, Chinese Academy of Sciences (NAOC). This work has been supported by the Strategic Priority Research Program ``The Emergence of Cosmological Structures'' of the Chinese Academy of Sciences Grant No. XDB09000000, and by the National Natural Science Foundation of China (NSFC) under grant numbers 11333003 and 11390372.
\end{acknowledgements}

\appendix

\section{Kinematic comparison}

\begin{figure*}
\centering
\label{tab:vrms}
\begin{tabular}{cc}

 \multicolumn{2}{c}{NGC 1248}  \\
 \includegraphics[width=50mm]{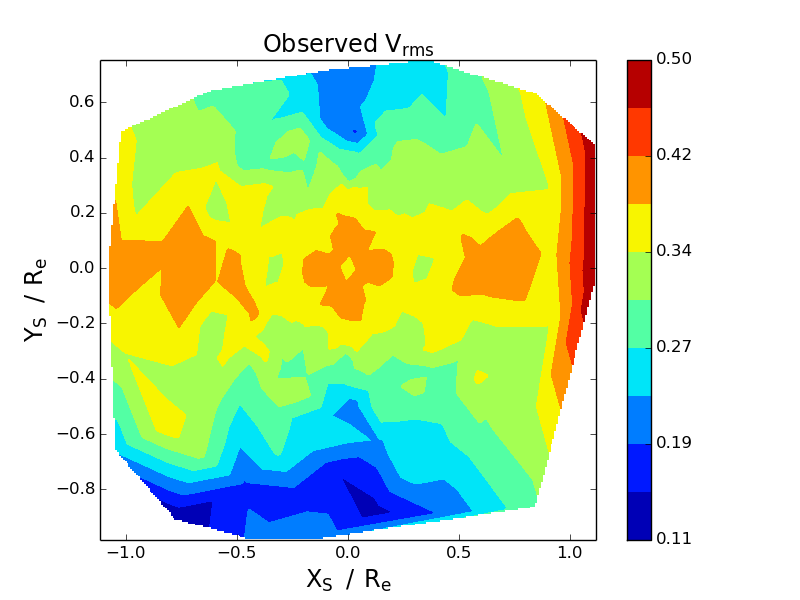} & \includegraphics[width=50mm]{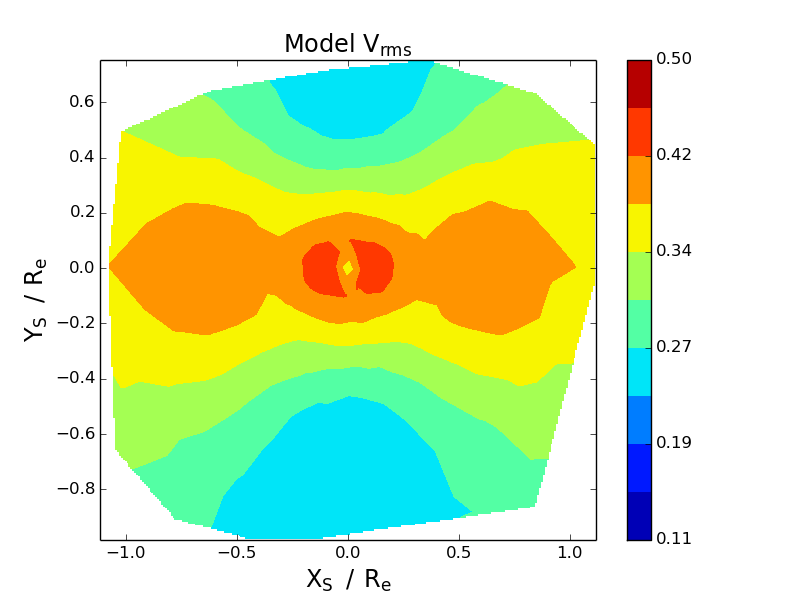}\\

 \multicolumn{2}{c}{NGC 3838} \\
 \includegraphics[width=50mm]{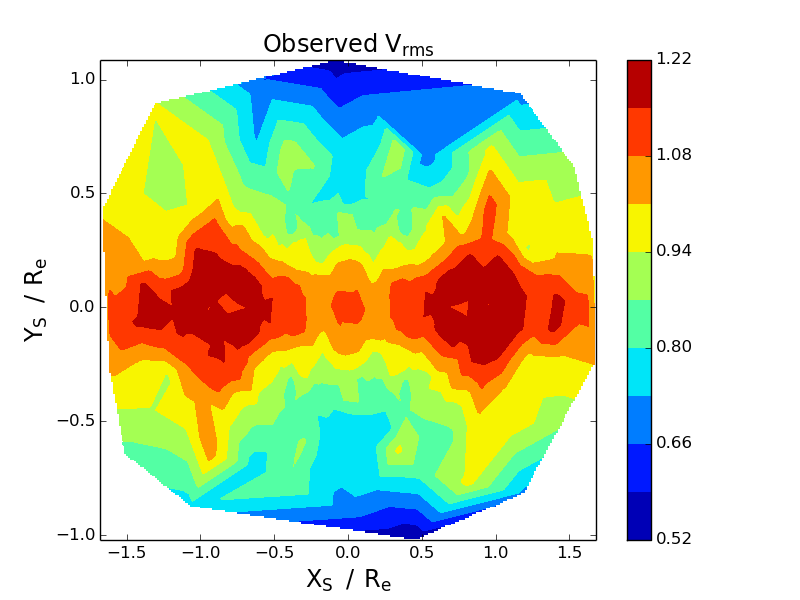} & \includegraphics[width=50mm]{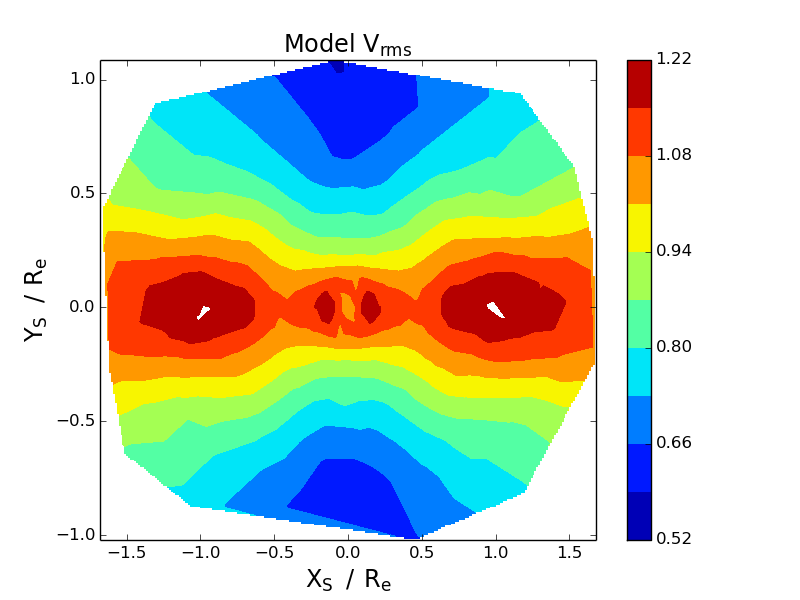}\\

 \multicolumn{2}{c}{NGC 4452} \\
 \includegraphics[width=50mm]{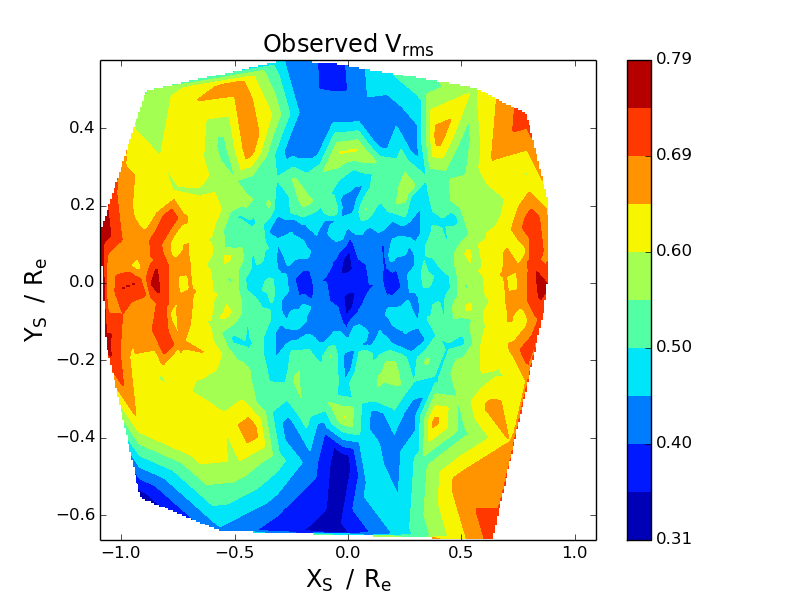} & \includegraphics[width=50mm]{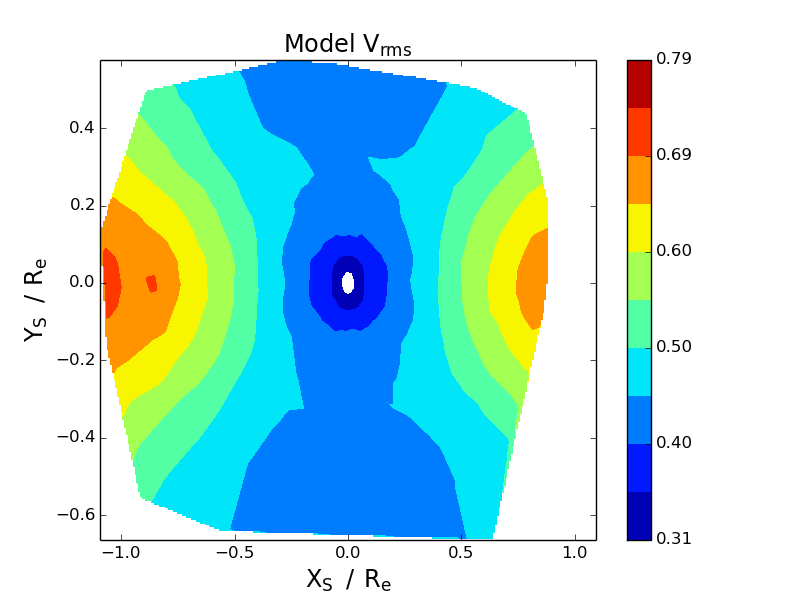}\\
 
 \multicolumn{2}{c}{NGC 4551} \\
 \includegraphics[width=50mm]{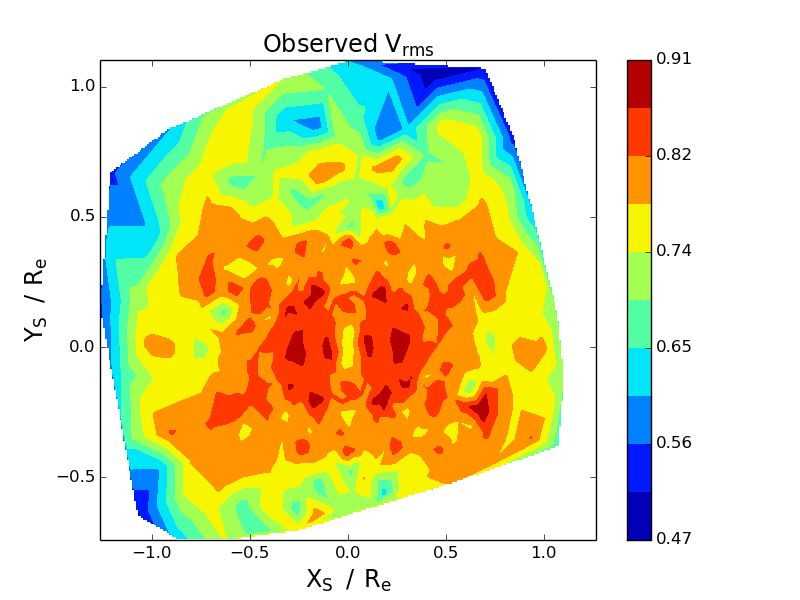} & \includegraphics[width=50mm]{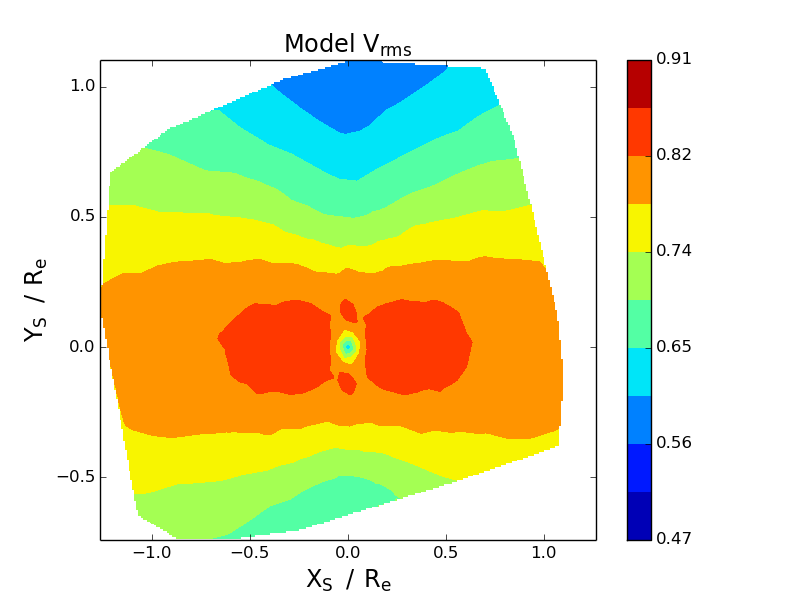}\\

\end{tabular}

\medskip
\caption{Galaxy $v_{\rm{rms}}$ comparison.  A comparison of the symmetrised data and M2M model $v_{\rm{rms}}$ maps.  For the model maps, the underlying mean $\chi^2$ per bin values are $\approx 1$.  Velocity units are as per Section \ref{sec:misc}.}
\end{figure*}

The main focus of the modelling exercise recorded here has been on the spectral line strength data not the kinematic data.  For completeness, in Figure \ref{tab:vrms}.1, we include a comparison of the data and model $v_{\rm{rms}}$ values.

\section{Particle weights}

\begin{figure*}
\centering
\label{tab:eor_lum}
\begin{tabular}{ccc}

 NGC 1248 & NGC3838 & NGC 4452 \\
 \includegraphics[width=50mm]{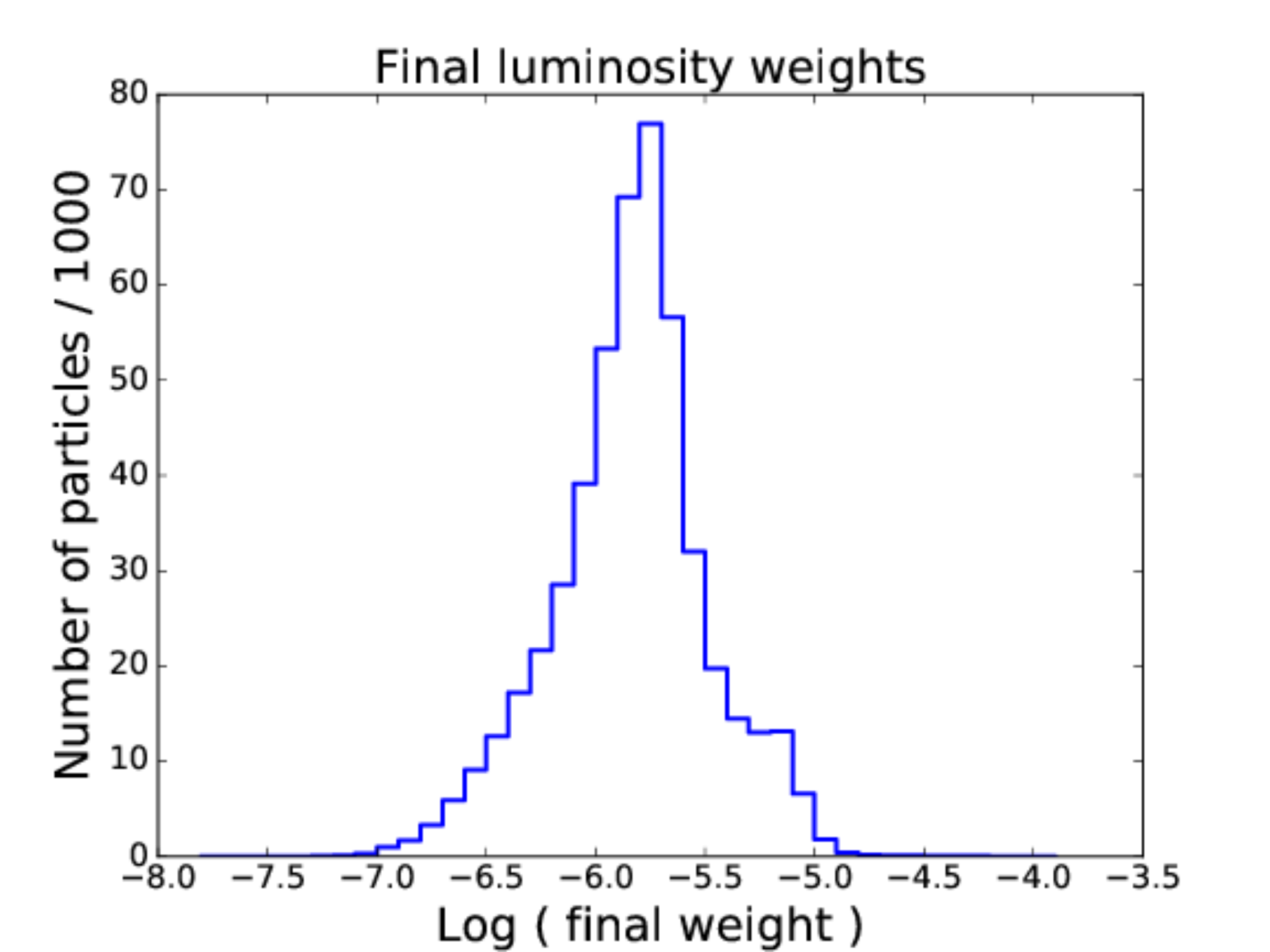} &  \includegraphics[width=50mm]{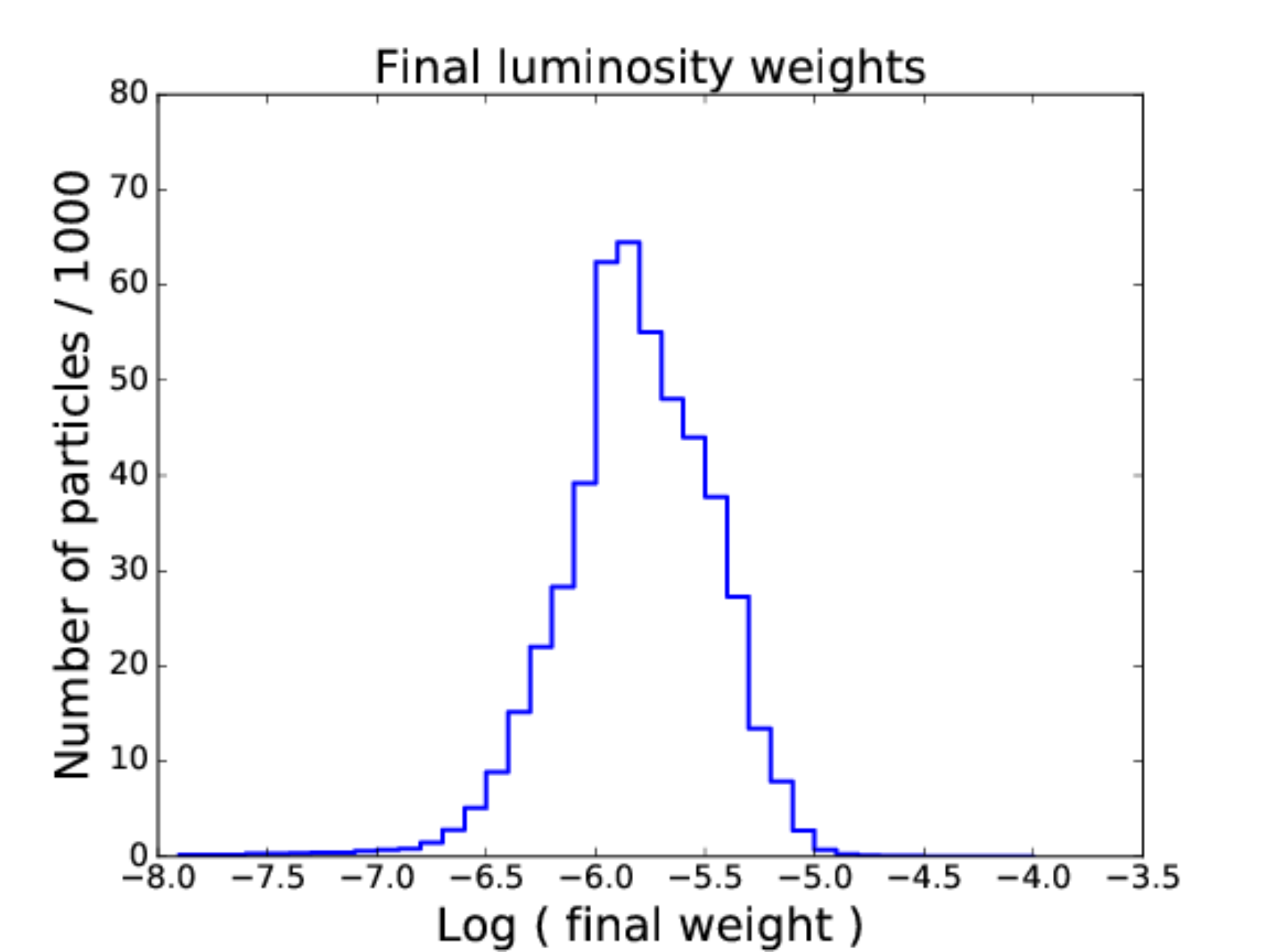} & \includegraphics[width=50mm]{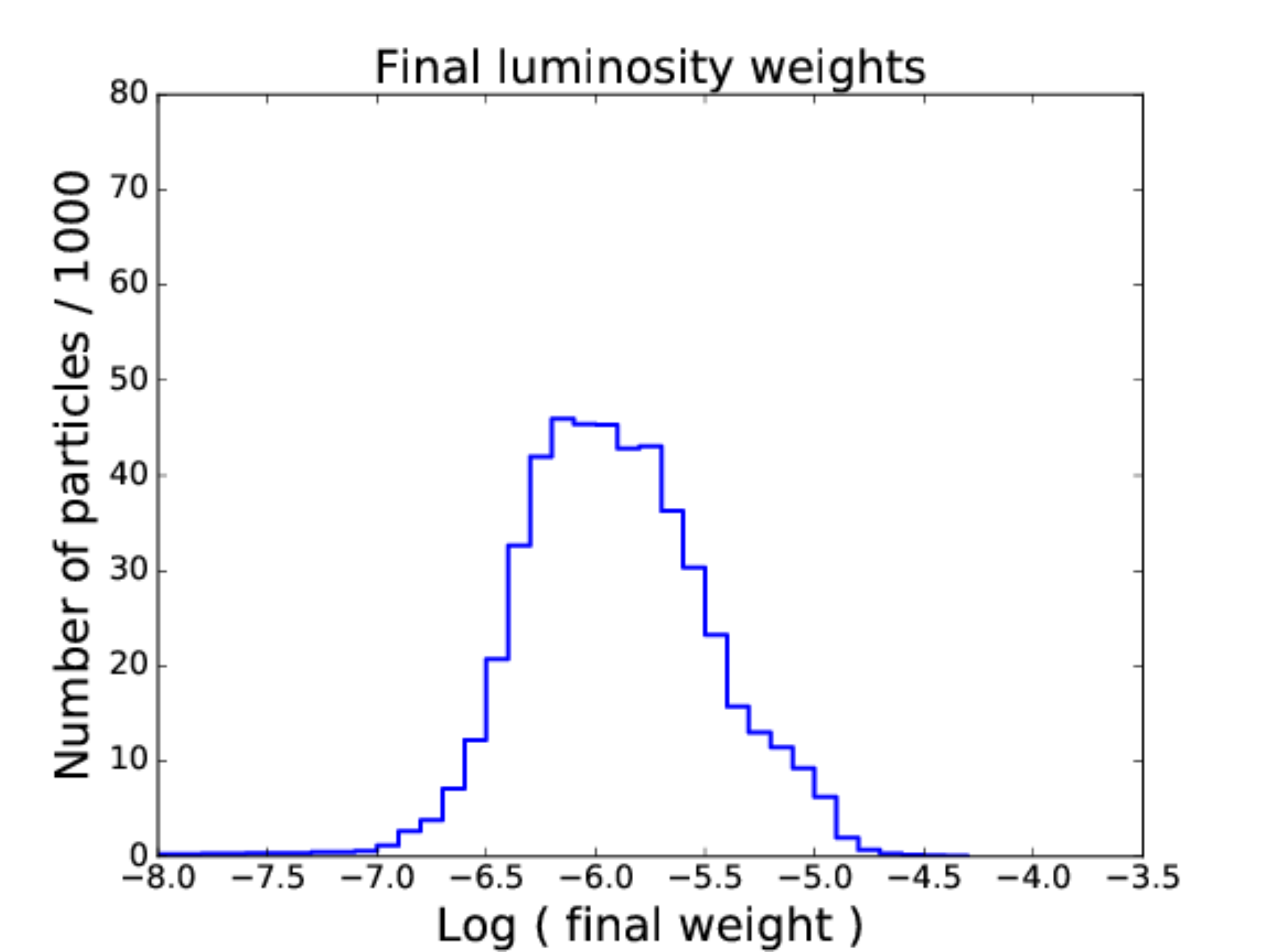} \\
 
  & NGC4551 & \\
 & \includegraphics[width=50mm]{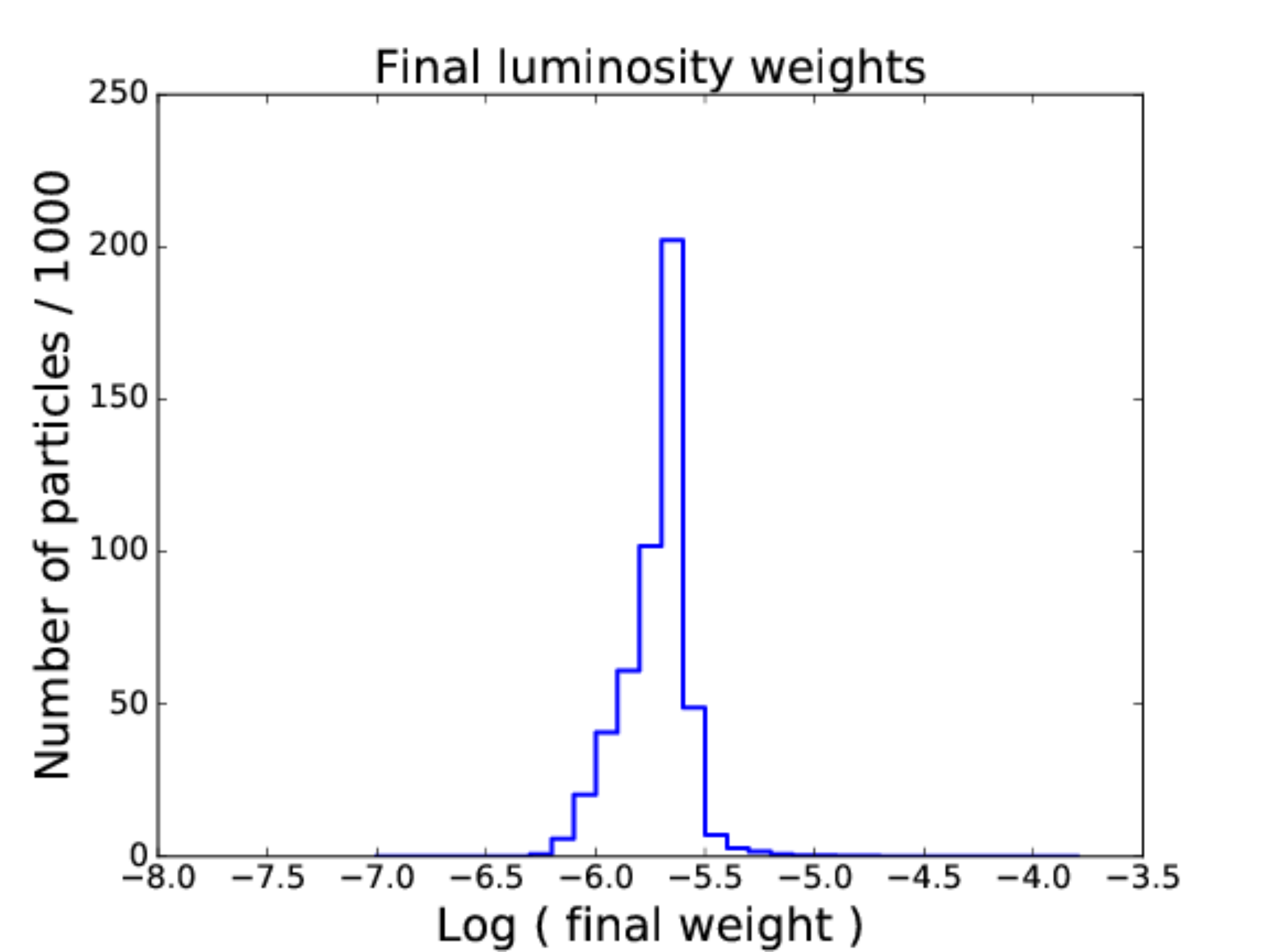} & \\

\end{tabular}

\medskip
\caption{Galaxy particle luminosity weights at the end of our M2M modelling runs. The first row shows the weight distribution for the three S0 galaxies and the second row, for the elliptical galaxy.  It is clear from the plots that the elliptical galaxy NGC 4551 has a narrower distribution with a higher peak than the S0 galaxies.  The reason for this is currently unclear.}
\end{figure*}

\begin{figure*}
\centering
\label{tab:eor_spec}
\begin{tabular}{ccc}

 \multicolumn{3}{c}{NGC 1248}  \\
 \includegraphics[width=50mm]{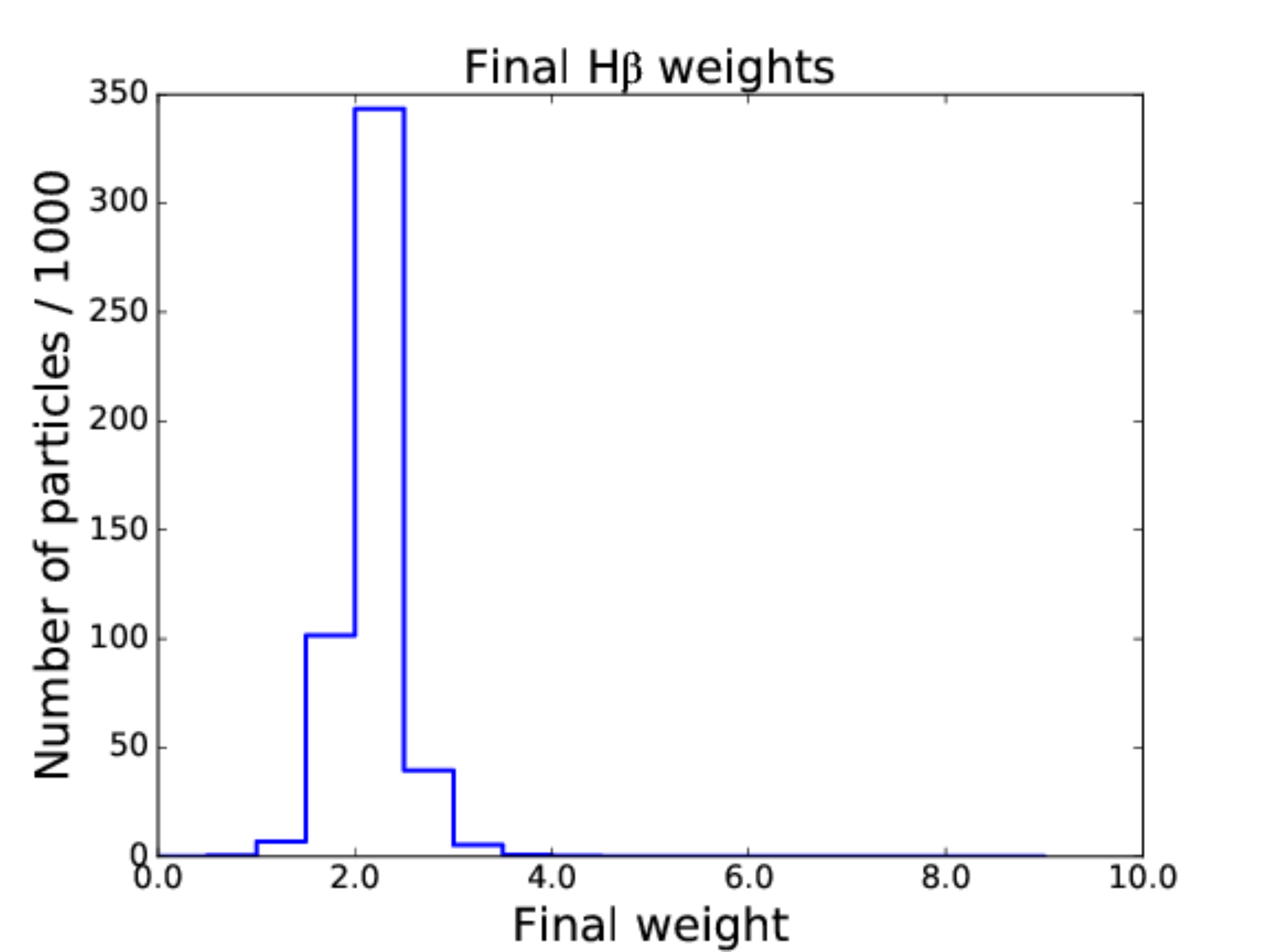} & \includegraphics[width=50mm]{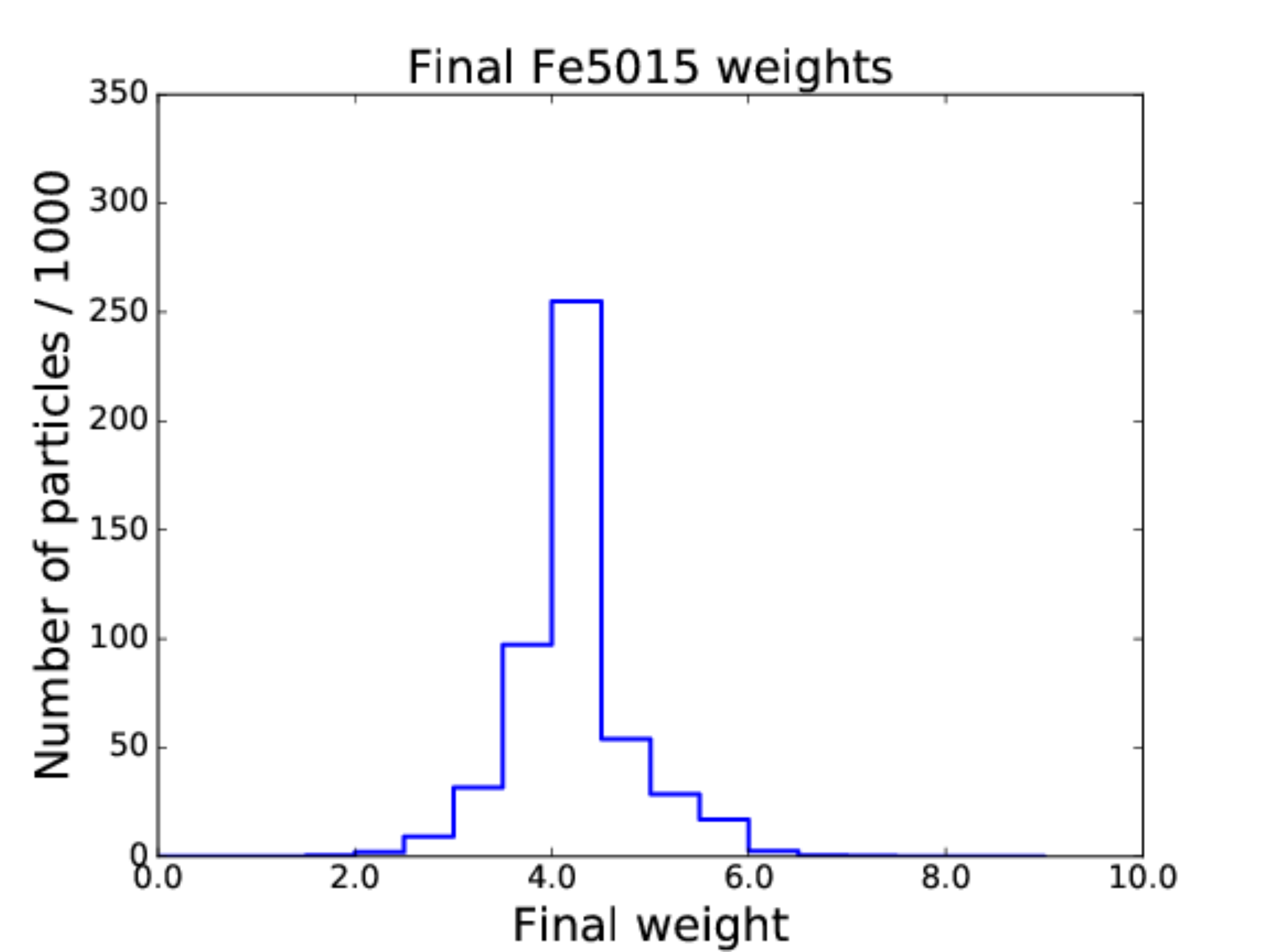} & \includegraphics[width=50mm]{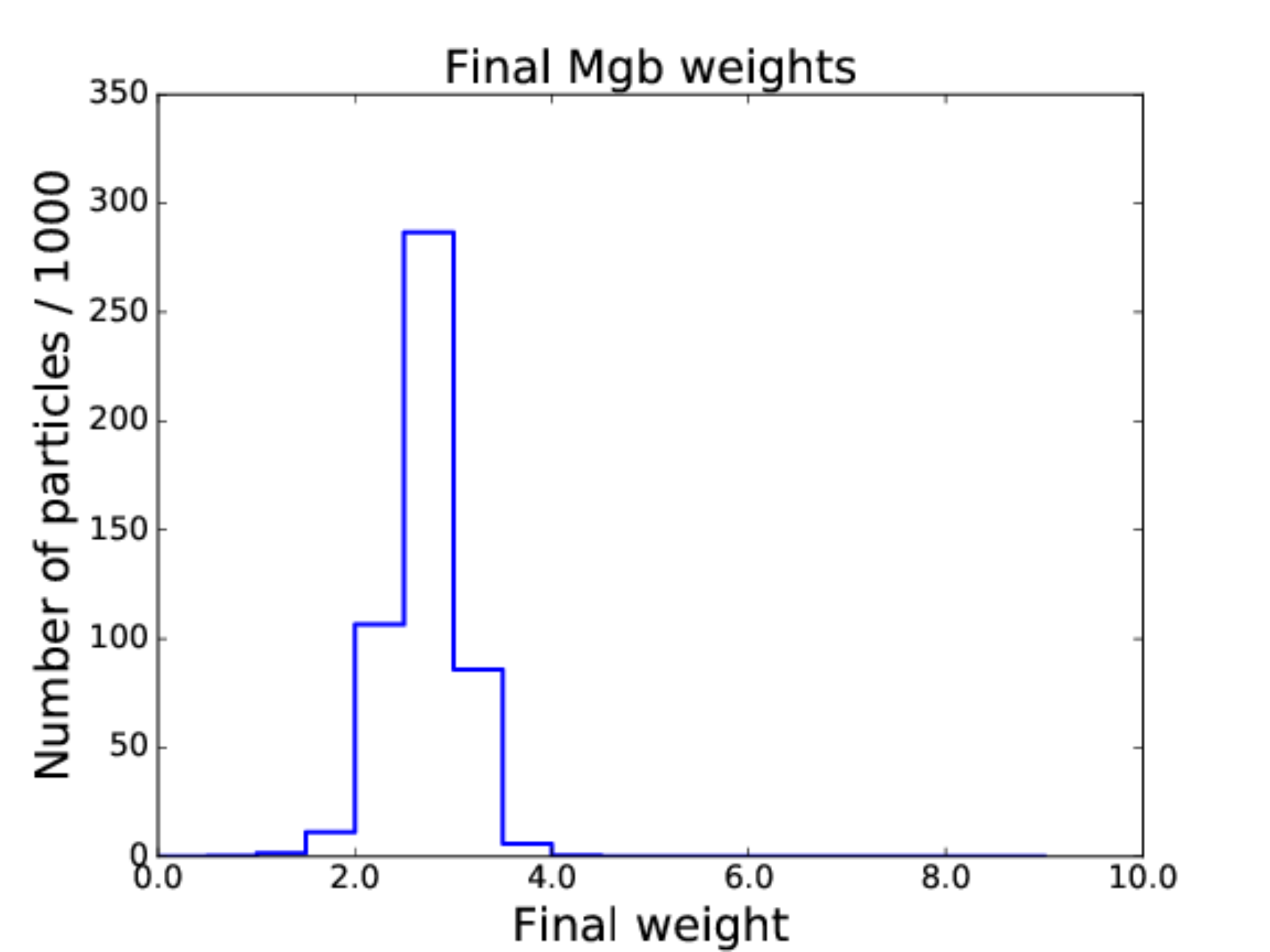}\\

 \multicolumn{3}{c}{NGC 3838} \\
 \includegraphics[width=50mm]{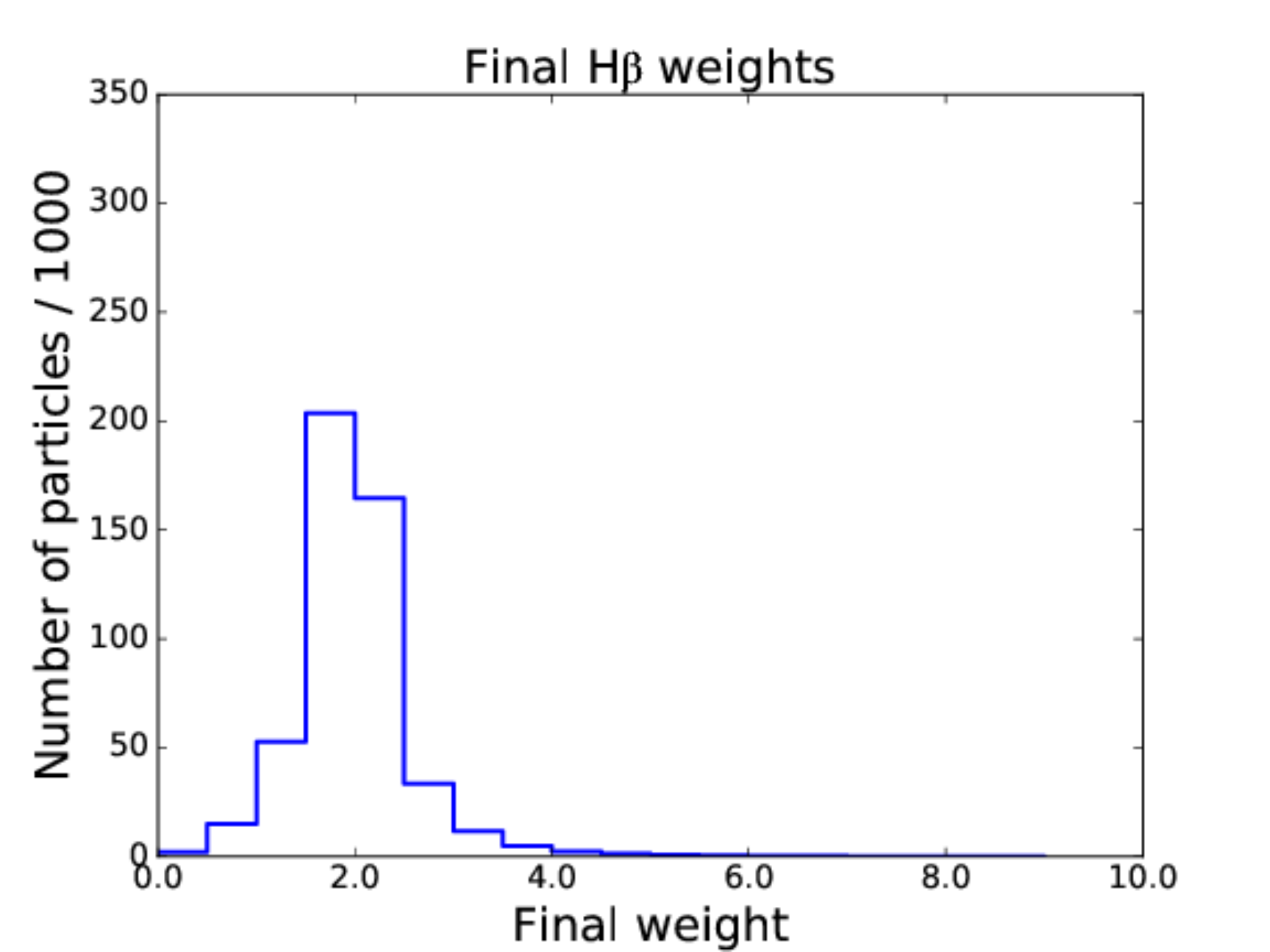} & \includegraphics[width=50mm]{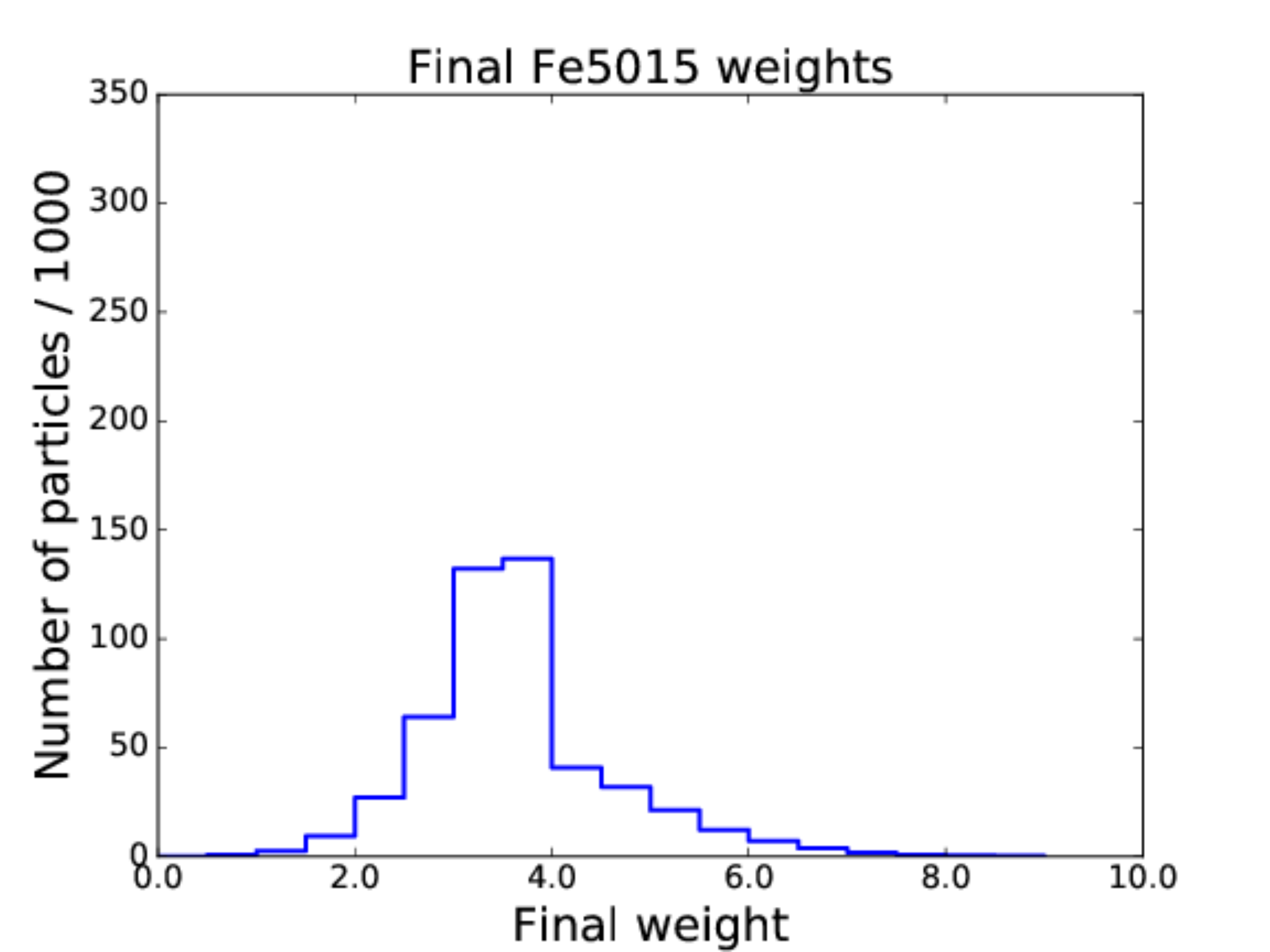} & \includegraphics[width=50mm]{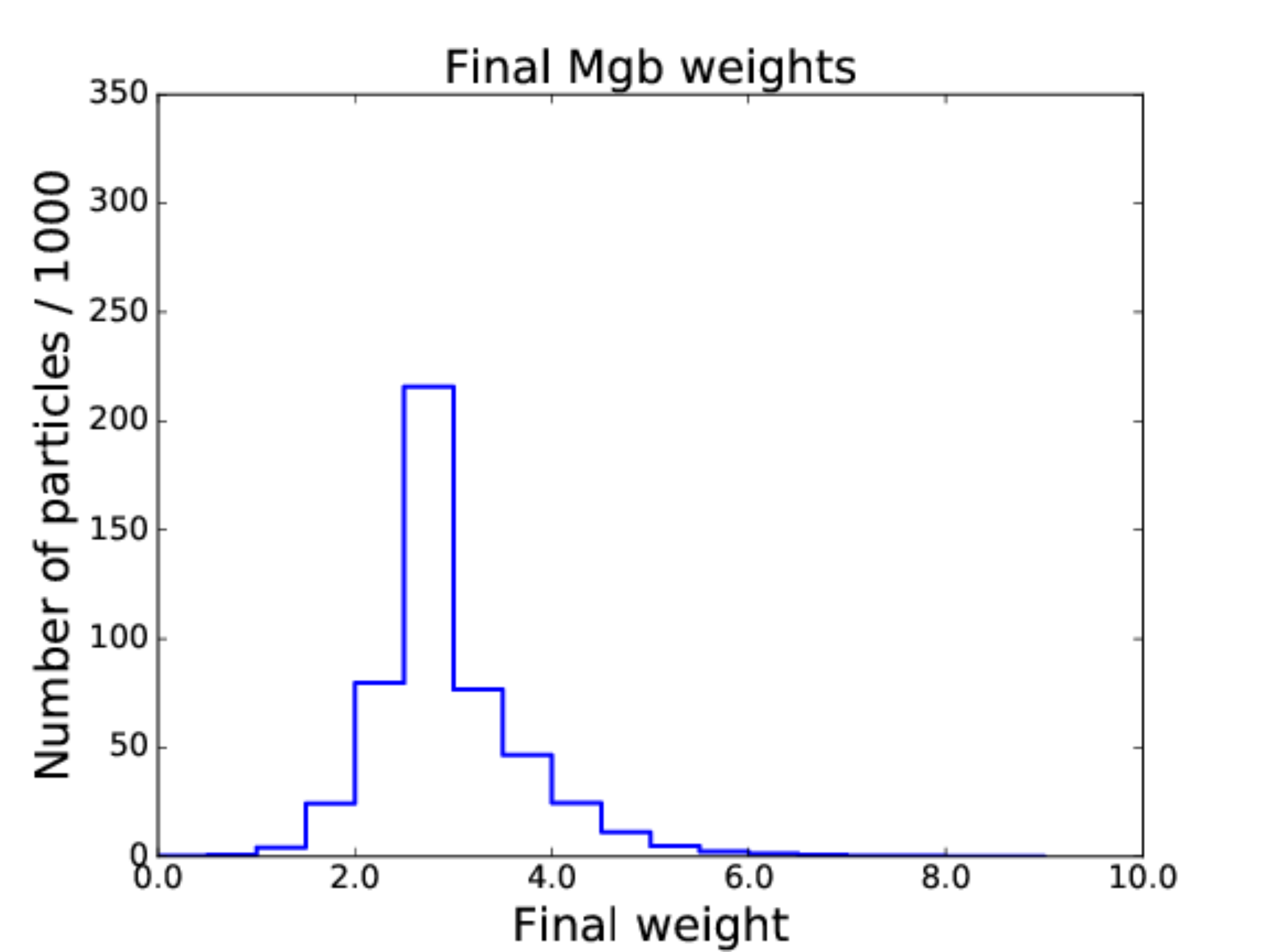}\\
 
 \multicolumn{3}{c}{NGC 4452} \\
 \includegraphics[width=50mm]{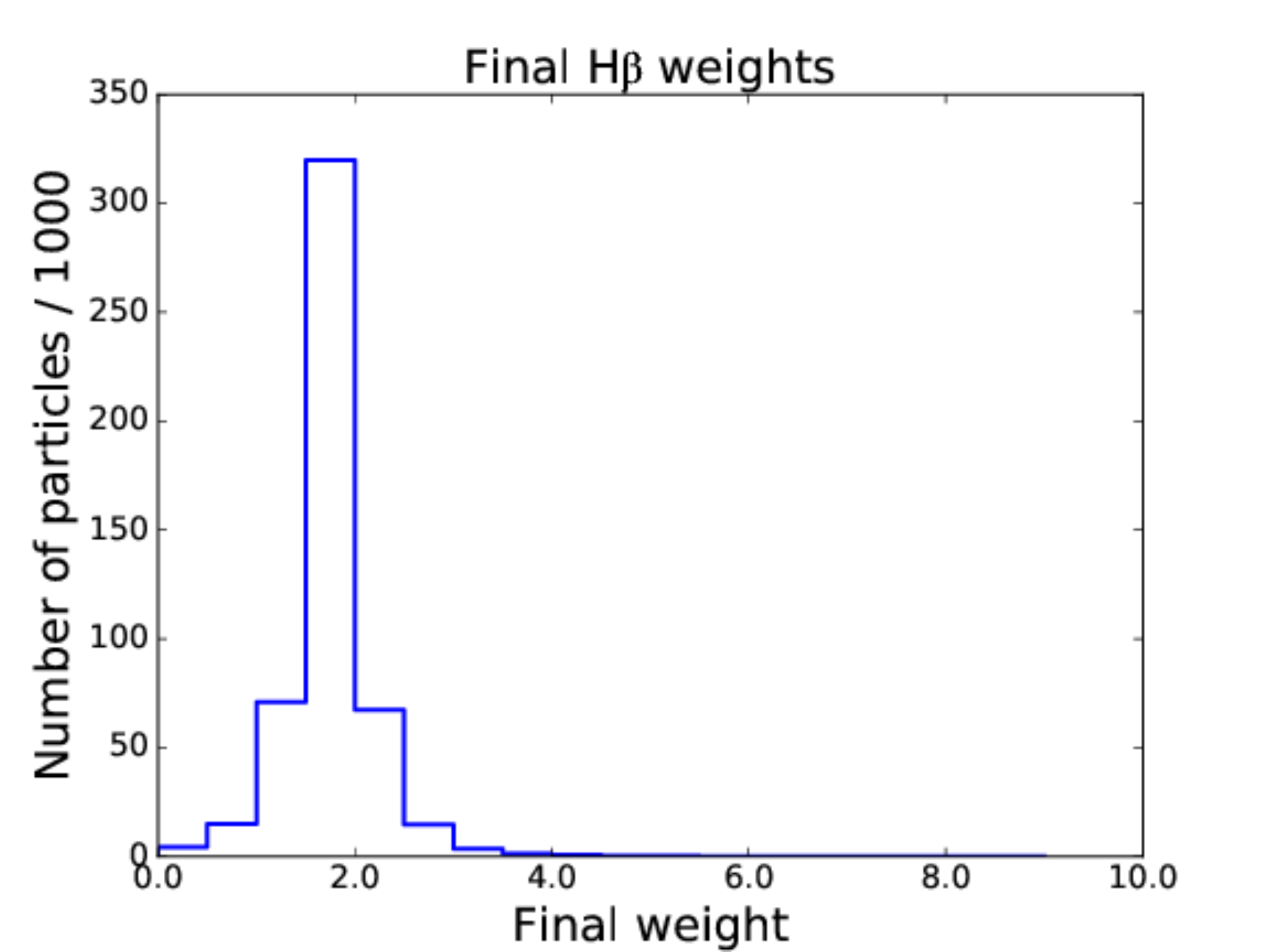} & \includegraphics[width=50mm]{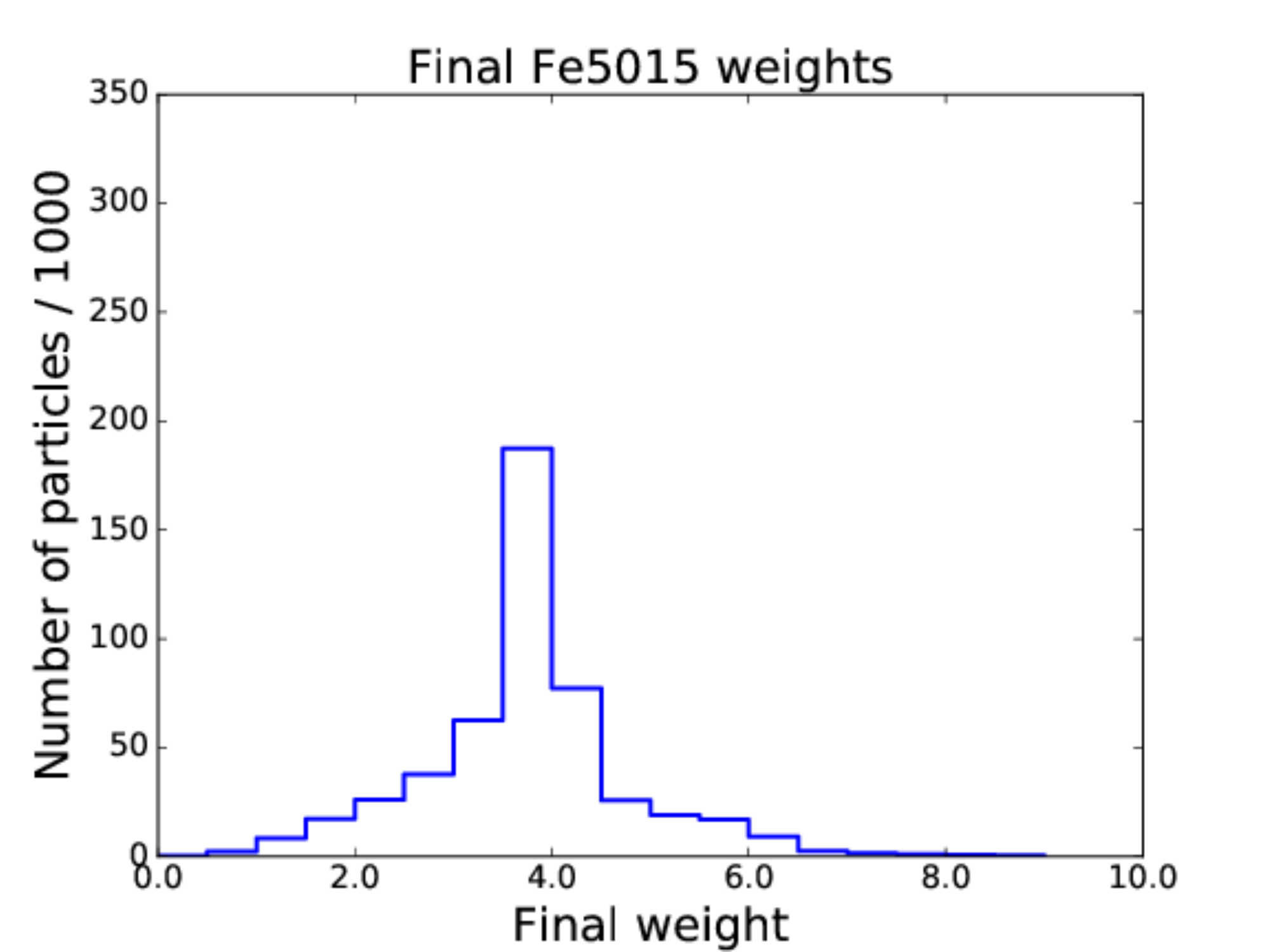} & \includegraphics[width=50mm]{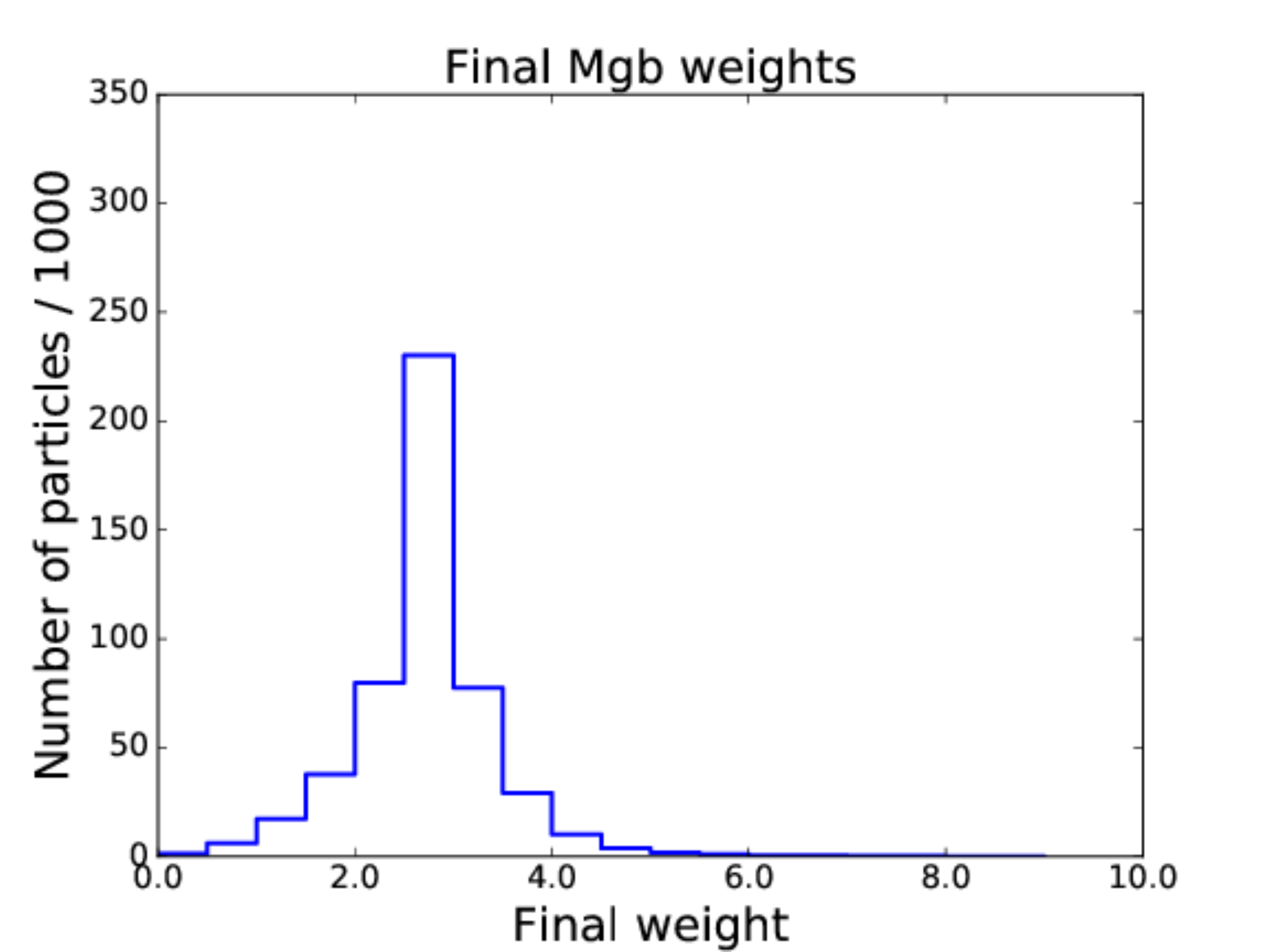}\\
 
 \multicolumn{3}{c}{NGC 4551} \\
 \includegraphics[width=50mm]{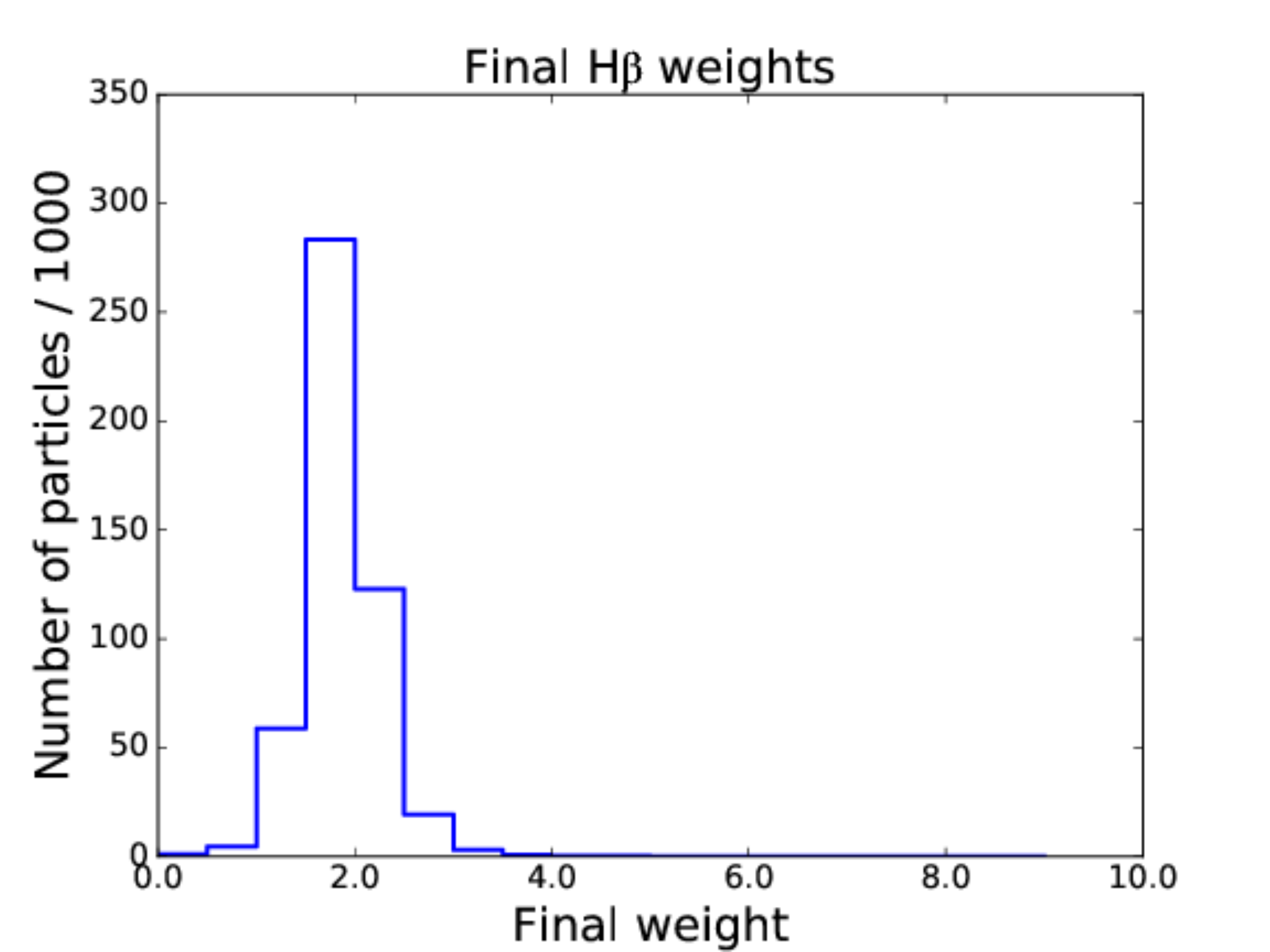} & \includegraphics[width=50mm]{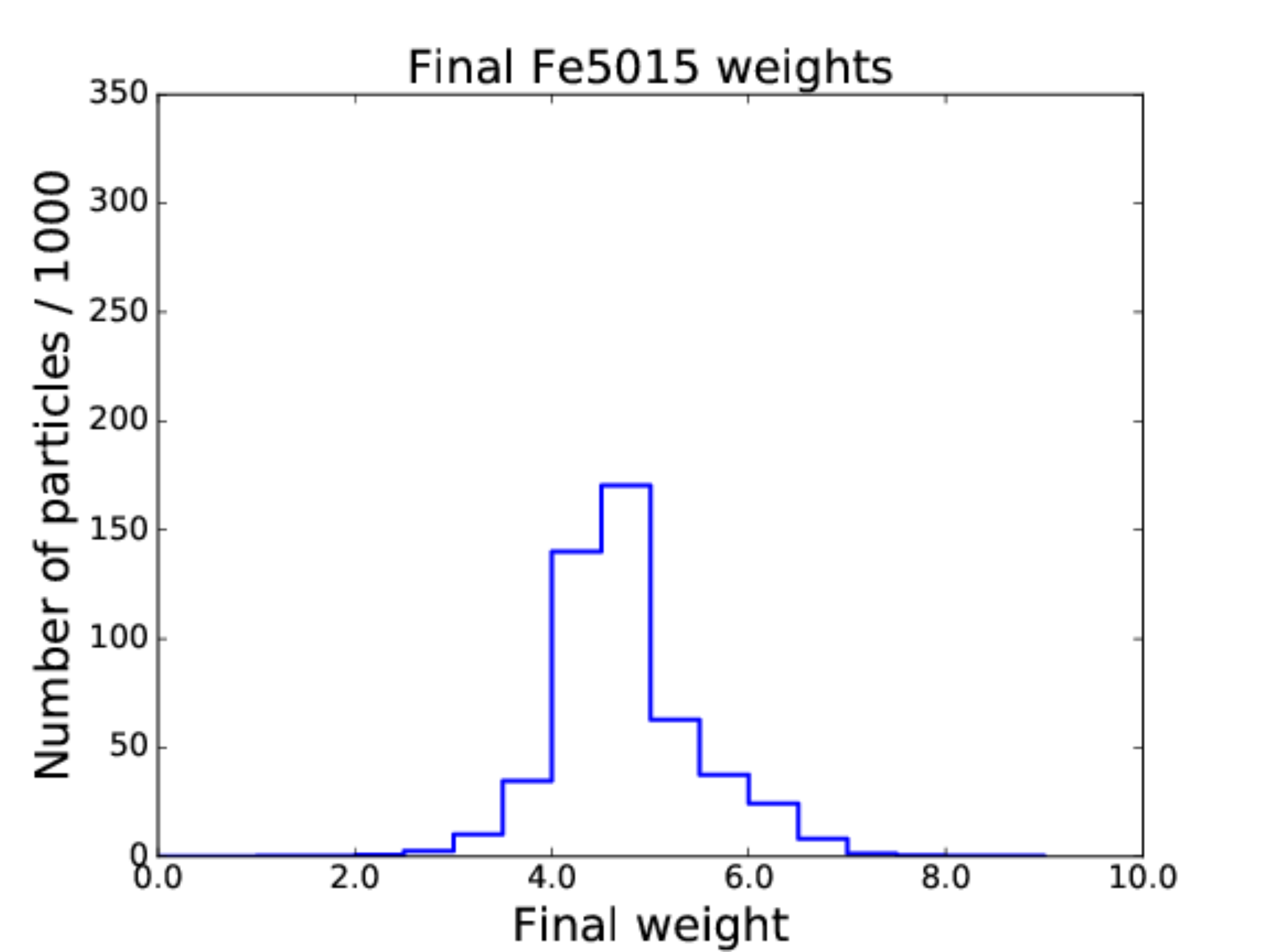} & \includegraphics[width=50mm]{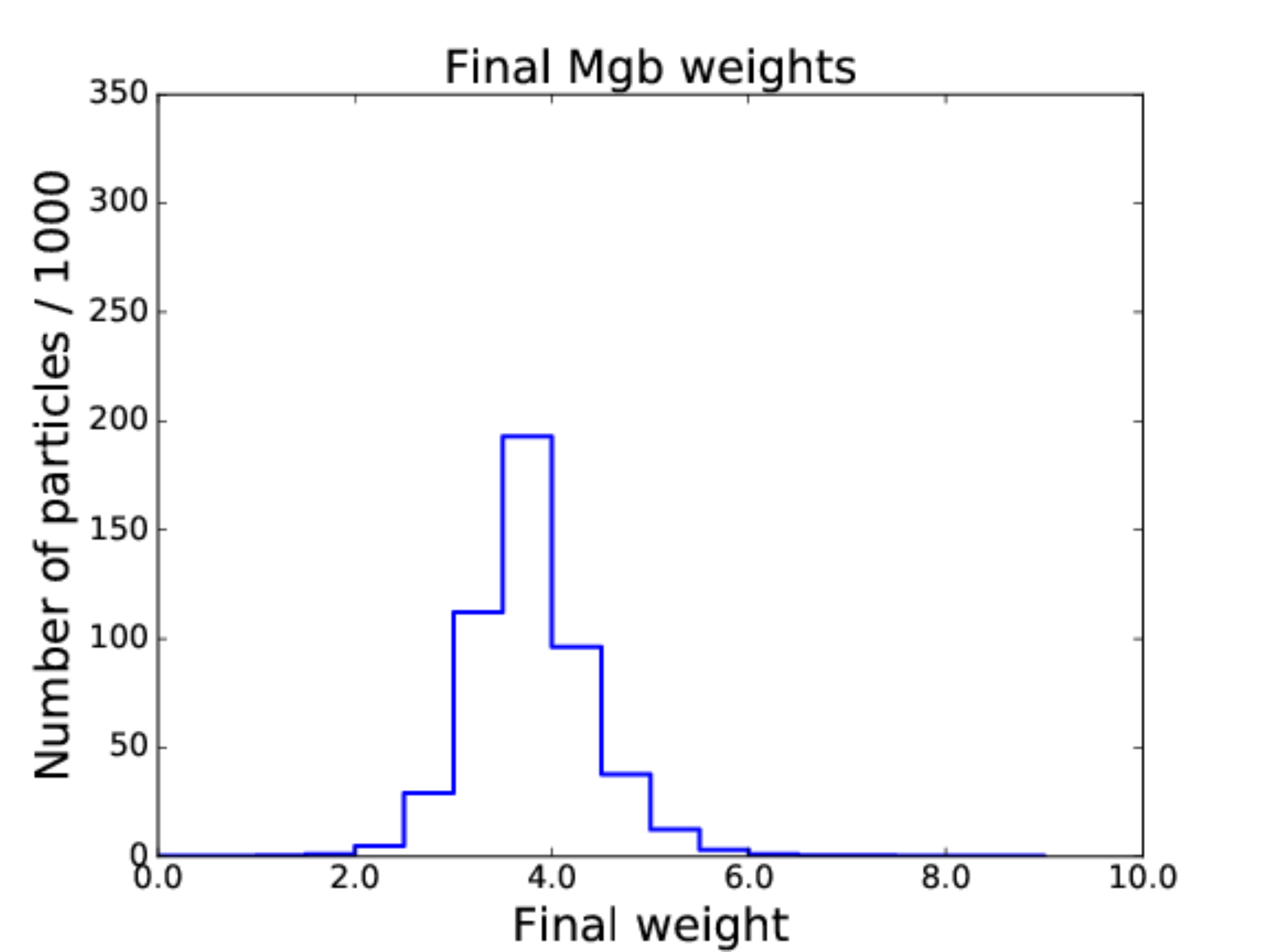}\\

\end{tabular}

\medskip
\caption{Galaxy particle spectral line weight distributions at the end of our M2M modelling runs.}
\end{figure*}

This appendix contains in Figure \ref{tab:eor_lum}.1 the end of run particle luminosity weights for all 4 galaxies.  The spectral line weights are in Figure \ref{tab:eor_spec}.2.

\bibliographystyle{raa}
\bibliography{ms2016-0175}

\end{document}